\documentclass[11pt,a4paper]{scrartcl}  
 
\usepackage[applemac]{inputenc}    
 
\usepackage{enumitem}

\newcommand{\bi}{\begin{itemize}}
\newcommand{\ei}{\end{itemize}}
\newcommand{\be}{\begin{enumerate}}
\newcommand{\ee}{\end{enumerate}}
\newcommand{\+}{\item}

\usepackage{soul}

\usepackage[]{hyperref} 
\usepackage{url}
\usepackage{colortbl} 
\hypersetup{pdftex,bookmarks,breaklinks,colorlinks,linktocpage,pdfnewwindow=true}

\usepackage[]{natbib}
\setcitestyle{aysep={}}
\usepackage{amsfonts,amsmath,amssymb,amsthm,amscd} \usepackage{cancel}

\usepackage[british]{babel}  
\usepackage[babel]{csquotes}  

\usepackage{setspace} 
\usepackage[]{units}

\definecolor{schwarz}{gray}{0}

\hypersetup{ 
linkcolor=blue, 
citecolor=blue, 
urlcolor=blue  
} 

\hypersetup{ 
pdftitle	={A Stronger Bell Argument for (Some Kind of) Parameter Dependence}, 
pdfsubject	={}, 
pdfcreator	={Paul M. Näger},  
pdfproducer ={}, 
pdfkeywords ={
Philosophy of physics; Quantum theory; Quantum entanglement; Bell's theorem; Bell inequalities; Quantum non-locality; Outcome dependence; Parameter dependence
} 
} 

\usepackage{booktabs}

\usepackage{rotating}
\usepackage{pdflscape}
\usepackage{multirow} 
\usepackage{longtable}

\usepackage[]{mathtools}

\usepackage{totpages} 
\usepackage[headsepline]{scrpage2}
\usepackage{epsfig}

\usepackage{hyphenat}
\usepackage[shortcuts]{extdash}

\graphicspath{
{graphics/}
{../graphics/}
}

\newcounter{mycounter}

\usepackage{thmtools, thm-restate}
\usepackage{
nameref,%\nameref
cleveref,% \cref
}

\newcounter{theoremcounter}[section]

\declaretheorem
[name=Corollary, refname={Corollary,Corollaries}, Refname={Corollary,Corollaries}]
{corollary} 
\declaretheorem
[
name=Lemma,
refname={Lemma,Lemmas}, 
Refname={Lemma,Lemmas},
parent=theoremcounter    
] 
{lemma}

\usepackage{tabularx}
 
\def \zv {\boldsymbol}
 
\usepackage{comment}

\usepackage{xcolor}

\definecolor{hellgruen}{rgb}{0.7,1,0.7}
\definecolor{banane}{rgb}{1,1,0.4}
\definecolor{helllila}{rgb}{1,0.75,1} 

\usepackage{soulutf8} 

\usepackage[ 	textsize=footnotesize  %scriptsize, small, normalsize
			,bordercolor=white
			,backgroundcolor=hellgruen
			,linecolor=hellgruen
			,colorinlistoftodos
			,disable
			]{todonotes}
			
\newcommand{\note}[2]{\todo[]{#1}\sethlcolor{hellgruen}\hl{#2}}

\newenvironment{Kommentar}{\color{black}}{\normalcolor}
\newcommand{\bki}{\begin{Kommentar}}
\newcommand{\eki}{\end{Kommentar}}

\definecolor{gruen}{RGB}{64,128,0}
\newenvironment{Kommentar2}{\color{black}}{\normalcolor}

\allowdisplaybreaks

\begin{document}

\title{
A Stronger Bell Argument for \\ (Some Kind of) Parameter Dependence 
}
\author{Paul M. Näger \\
\small \emph{Department of Philosophy, WWU Münster}\\
\small \emph{Domplatz 23, 48143 Münster, Germany}\\
\small \href{mailto:paul.naeger@wwu.de}{\nolinkurl{paul.naeger@wwu.de}}  
} 

\date{\normalsize 4th July 2018 \\
\small{Version 5.52} 
\\[0.4cm] forthcoming in: \\ \emph{Studies in the History and Philosophy of Modern Physics}
}

\maketitle  

\begin{abstract}

It is widely accepted that the violation of Bell inequalities excludes local theories of the quantum realm. This paper presents a new derivation of the inequalities from non-trivial non-local theories and formulates a stronger Bell argument excluding also these non-local theories. Taking into account all possible theories, the conclusion of this stronger argument provably is the strongest possible consequence from the violation of Bell inequalities on a qualitative probabilistic level (given usual background assumptions). Among the forbidden theories is a subset of outcome dependent theories showing that outcome dependence is not sufficient for explaining a violation of Bell inequalities. Non-local theories which can violate Bell inequalities (among them quantum theory) are rather characterised by the fact that at least one of the measurement outcomes in some sense (which is made precise) probabilistically depends both on its local as well as on its distant measurement setting (‘parameter’). When Bell inequalities are found to be violated, the true choice is not ‘outcome dependence or parameter dependence’ but between two kinds of parameter dependences, one of them being what is usually called ‘parameter dependence’. Against the received view established by Jarrett and Shimony that on a probabilistic level quantum non-locality amounts to outcome dependence, this result confirms and makes precise Maudlin's claim that some kind of parameter dependence is required.
\end{abstract}

\newpage 
\tableofcontents 
\newpage

\section{Introduction} 
\label{sec:intro}

Bell's argument (\citeyear{Bell1964,Bell1971,Bell1975}) 
establishes a mathematical no-go theorem for theories of the micro-world. 
In its standard form, it derives that theories which are local (and fulfill certain auxiliary assumptions) 
cannot have correlations of arbitrary strength between events which are space-like separated. An upper bound for the correlations is given by the famous Bell inequalities. Since certain experiments with entangled quantum objects have results which violate these inequalities (\enquote{EPR/B correlations}), it concludes that the quantum realm cannot be described by a local theory. Any correct theory of the quantum realm must involve some kind of non-locality (\enquote{quantum non-locality}). 
This result is one of the central features of the micro-realm. It is the starting point for extensive debates concerning the nature of quantum objects and their relation to space and time.

Since Bell's first proof (\citeyear{Bell1964}) the theorem has evolved considerably towards stronger forms: There has been a sequence of improvements which derive the inequalities from increasingly weaker assumptions. 
The main focus has been on getting rid of those premises which are commonly regarded as auxiliary assumptions: 
\citet{Clauser1969} derived the theorem without assuming perfect correlations; \citet{Bell1971} abandoned the assumption of determinism; \citet{Grasshoff2005} and \citet{Portmann2007} showed that possible latent common causes do not have to be \emph{common} common causes.%
\footnote{{}
However, the debate about common common causes versus separate common causes is to some degree still undecided \citep[see][]{Hofer-Szabo2008}. 
}
What all  these different derivations share is that they assume one or another form of locality. Locality seems to be the central assumption in deriving the Bell inequalities---and hence it is the assumption that is assumed to fail when one finds that the inequalities are violated.

In this paper I shall present a strengthening of Bell's theorem which relaxes the central assumption: One does not have to assume locality in order to derive the Bell inequalities. Certain forms of non-locality, which we shall call \enquote{weakly non-local} suffice: 
An outcome may depend on the other outcome or on the distant setting---as long as it does not depend on \emph{both} settings, it still implies that the Bell inequalities hold. 
As a consequence, the violation of the Bell inequalities also excludes those weakly non-local theories. 
So it does not require {any} kind of non-locality, but a very \emph{specific} one: 
At least one of the outcomes must depend probabilistically on both settings. 
While previous strengthenings of Bell's theorem secured that a certain auxiliary assumption is {not} the culprit, our derivation here for the first time strengthens the \emph{conclusion} of the theorem. Formulating the stronger  argument and deriving the new conclusion will make up a first part of this paper (\autoref{sec:bell}). In \autoref{sec:strong-poss-cons} I discuss some immediate consequences of the new argument. 

In a second part (\autoref{sec-analysis}), we shall probabilistically analyze this new conclusion in a similar way as \citet{Jarrett1984} famously analyzed the result of the standard Bell argument as \enquote{outcome dependence or parameter dependence}. The result of the new analysis will differ considerably from Jarrett's. Especially it will make explicit that some kind of parameter dependence cannot be avoided, while outcome dependence is irrelevant for the question whether Bell inequalities can be violated. 

A third part (\autoref{sec:cons}) is dedicated to comparing the result 
from \autoref{sec-analysis} 
with existing positions and to discussing possible consequences. 
It will turn out that, while correct in a strictly literal sense, 
Jarrett's classic analysis is misleading; the received view, which is based on that analysis and holds that quantum non-locality on a probabilistic level is outcome dependence, 
is not an appropriate characterization; and Maudlin's information theoretic result \citeyearpar{Maudlin1994}, that there must be some dependence between an outcome and its distant parameter, is confirmed and made precise in probabilistic terms. This will also resolve the persistent tension between Jarrett's and Maudlin's opposing positions in favor of the latter. 

Note that in this paper 
we restrict our investigations to the qualitative {probabilistic} level, 
i.e. we investigate which probabilistic dependences and independences the violation of Bell inequalities in EPR/B experiments implies.
Especially, here we refrain from further considerations concerning the compatibility of the resulting non-local dependences with relativity or 
the existence of certain non-local physical or metaphysical connections
(e.g. a non-separability)
because that would require further assumptions and non-trivial reasoning  
(\enquote{correlation is not causation}).% 
\footnote{{}
Elsewhere I have shown what the probabilistic results derived in this paper imply for the causal structure of the experiments \citep{Naeger2013b}. 
}

\section{Strengthening Bell's argument} 
\label{sec:bell}

\subsection{EPR/B experiments and the standard Bell argument}
\label{sec:experiments}

\begin{figure}[htbp]
\noindent \centering
\begin{minipage}[t]{\linewidth}
\epsfig{file=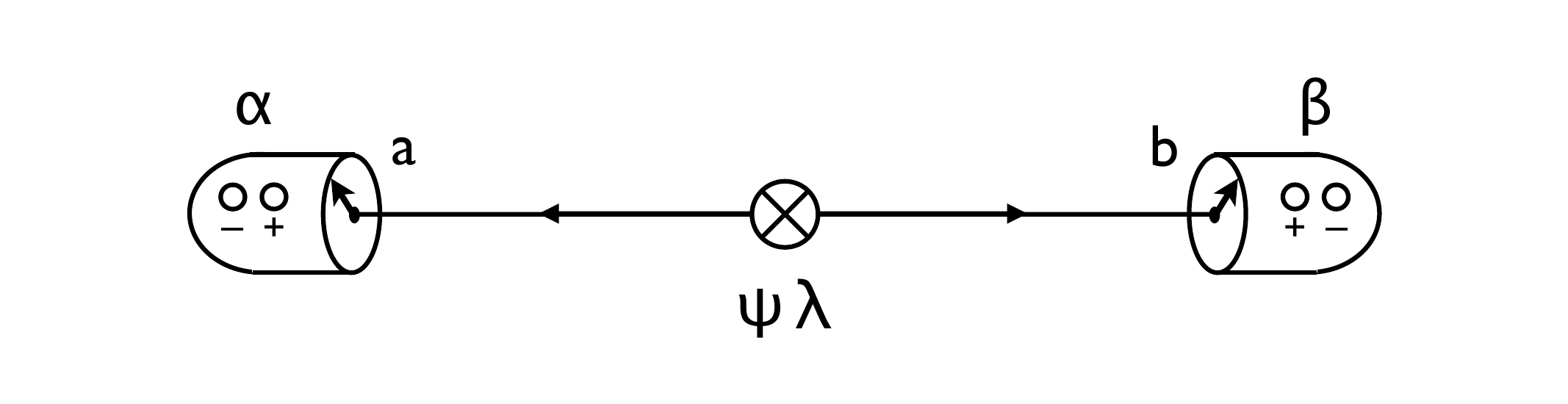,width=\linewidth} \caption{EPR/B setup} \label{epr-exp}
\end{minipage}
\end{figure}

We consider a usual EPR/B setup  
with space-like separated polarization measurements of an ensemble of photon pairs in an entangled quantum state $\zv \psi = \psi_0$ (\citealt*{Einstein1935,Bohm1951,Clauser1974}; see \autoref{epr-exp}). 
Possible hidden variables of the photon pairs are called $\zv \lambda$, so that the \emph{complete} state of the particles at the source is $(\zv \psi, \zv \lambda)$. 
Since in this setup the state $\zv \psi$ is the same in all runs, it will not explicitly be noted in the following (one may think of any probability being conditional on one fixed state $\zv \psi = \psi_0$). 
We denote Alice's and Bob's measurement setting as $\zv a$ and $\zv b$, respectively, and the corresponding (binary) measurement results as $\zv \alpha$ and $\zv \beta$.     
On a probabilistic level, the experiment is described by the joint probability distribution%
\footnote{{}
\citet[46--7]{Butterfield1992} points to the potential problem that the settings might not have well-defined probabilities if the experimenters freely choose them (and explains how to describe EPR/B experiments by a restricted probability distribution in such cases). 
Assuming, as we do here, that the probability distribution of the settings is well-defined, however, is no substantial restriction because free choice of the settings is not a necessary requirement for relevant EPR/B experiments. Rather, what is mandatory is that the measurement directions are chosen \emph{independently}. In fact, in contemporary EPR/B experiments, the settings are typically determined by independently operating random number generators,  securing that the settings have a well-defined probability distribution. 
} 
 $P(\alpha \beta a b \lambda) := P(\zv \alpha = \alpha, \zv \beta = \beta, \zv a = a, \zv b = b, \zv \lambda = \lambda)$ of these five random variables.%
\footnote{{} 
While the outcomes are discrete variables and the settings can be considered to be discrete (in typical EPR/B experiments there are two possible settings on each side), the hidden state may be continuous or discrete. In the following I assume $\zv \lambda$ to be discrete, but all considerations can be generalized to the continuous case.}
We shall consistently use bold symbols ($\zv \alpha, \zv \beta, \zv a, …$) for random variables and normal font symbols ($\alpha, \beta, a, …$) for the corresponding values of these variables. We use indices to refer to \emph{specific} values of variables, e.g. $\alpha_- = -$ or $a_1 = 1$, 
which provides useful shorthands, e.g. $P(\alpha_- \beta_+ a_1 b_2 \lambda) := P(\zv \alpha = -, \zv \beta = +, \zv a = 1, \zv b = 2, \zv \lambda = \lambda)$. Expressions including probabilities with non-specific values of variables, e.g. $P(\alpha|a) = P(\alpha)$, are meant to hold for all values of these variables (if not otherwise stated). 

Containing the hidden states $\zv \lambda$, which are by definition not measurable, the total distribution is empirically not accessible (\enquote{hidden level}), i.e. purely theoretical. Only the marginal distribution $P(\alpha \beta a b)$, which does not involve $\zv \lambda$,  is empirically accessible and is determined by the results of actual measurements in EPR/B experiments (\enquote{observable level}).
A statistical evaluation of a series of many runs with similar preparation procedures yields that the outcomes are strongly correlated given the settings and the quantum state.% 
\footnote{{} A correlation of the outcomes given the settings and the quantum state means 
$\exists \, \alpha, \beta, a, b\!\!\!: \quad P(\alpha \beta | a b) ≠ P(\alpha | ab) P(\beta | ab)$. 
} 
For instance, in case the quantum state is the Bell state $\psi_0 = (|+\rangle |+ \rangle + |-\rangle |-\rangle)/\sqrt{2}$ (and the settings are chosen with equal probability $\frac{1}{2}$) the correlations read:  
\begin{align}
\tag{Corr}
\label{Corr}
P(\alpha \beta | a b ) & = P(\alpha|\beta a b ) P(\beta) =  \begin{cases}  \cos^2 (a-b) \cdot \frac{1}{2} & \text{if } \alpha = \beta\\
 \sin^2 (a-b) \cdot \frac{1}{2} & \text{if } \alpha ≠ \beta
\end{cases}
\end{align}
These famous EPR/B correlations between space-like separated measurement outcomes were first measured in a convincing way by \citet{Aspect1982}, were confirmed 
over large distances \citep{Ursin2007}, and have recently been demonstrated also under closure of all major loopholes 
\citep{Hensen2015,Giustina2015,Shalm2015}. 
All these findings are correctly predicted by quantum mechanics: 
Involving only empirically accessible variables, the quantum mechanical probability distribution essentially agrees with the empirical one.

Since according to \eqref{Corr} one outcome depends on both the other space-like separated outcome as well as on the distant (and on the local) setting, the observable part of the probability distribution  (or the quantum mechanical distribution, respectively) clearly is non-local. 
Bell's idea \citeyearpar{Bell1964} was to show that EPR/B correlations are so extraordinary that even if one allows for hidden states $\zv \lambda$ one cannot restore locality: Given EPR/B correlations the \emph{theoretical} probability distribution (including possible hidden states) must be non-local as well. Hence, \emph{any possible} probability distribution which might correctly describe the experiment must be non-local.   

This \enquote{Bell argument for quantum non-locality}, as I shall call it, 
proceeds by showing that the empirically measured EPR/B correlations violate certain inequalities (\enquote{Bell inequalities}). 
It follows that at least one of the assumptions in the derivation of the inequalities must be false. Indeterministic generalizations \citep{Bell1971,Clauser1974,Bell1975} of Bell's original deterministic derivation \citeyearpar{Bell1964} employ two probabilistic assumptions, \enquote{local factorization}\footnote{{} \enquote{Local factorization} is my term. Bell calls (\ref{local-causality}) \enquote{local causality}, some call it \enquote{Bell-locality}, but most often it is simply called \enquote{factorization} or \enquote{factorizability} (introduced by \citealt{Fine1980}). Bell's terminology already suggests a causal interpretation, which I would like to avoid in this paper, and the latter two names are too general since, as I shall show, there are other forms of the hidden joint probability which legitimately might be said to \enquote{factorize}; hence the qualification \enquote{local}.} 
\begin{equation}
\label{local-causality}
\tag{LF}
P(\alpha \beta | a b \lambda) = P(\alpha |a \lambda) P(\beta|b \lambda)
\end{equation}
and \enquote{measurement independence}
\begin{equation}
\label{MI}
\tag{MI}
P(\lambda | a b) = P(\lambda).
\end{equation}
While there are suggestions to explain the violation of the Bell inequalities by a failure of measurement independence,%
\footnote{{}
A failure of measurement independence can be realized by different kinds of models: conspiratorial or superdeterministic models, simulation or prism models \citep[][]{Fine1982}, 
models with backwards causation \citep[e.g.][]{Price1994,Corry2015} or with non-locality \citep[][]{SanPedro2012}. 
}
the main route in the debate has been to assume that it holds; and 
in order to focus on the factorization condition (and possible modifications to it), this will also be one of the basic assumptions throughout this paper. 
If measurement independence holds, 
local factorization fails, 
implying that any correct theory of the quantum realm must involve an irreducibly non-local statistical dependence.

Here and in the following I shall presuppose the Wigner-type derivation of Bell inequalities 
\citep{Wigner1970,vanFraassen1989}, 
which, besides measurement independence and local factorization, 
requires the assumption that there are perfect correlations 
between the outcomes for a certain relative angle of the measurement settings, e.g. for parallel settings given the quantum state $\psi_0$:% 
\footnote{{} 
From the indicated values of the joint probability distribution one can derive 
\begin{align}
P(\alpha_\pm | \beta_\pm a_i b_i) &= 1 & P(\alpha_\pm | \beta_\mp a_i b_i) &= 0 & P(\beta_\pm | \alpha_\pm a_i b_i) &= 1   & P(\beta_\pm | \alpha_\mp a_i b_i) &= 0,
\end{align}
which makes the perfect correlation explicit.  
}
\begin{align}
\label{PCorr-def}
\tag{PCorr}
\forall i: \quad P(\alpha_\pm  \beta_\pm | a_i b_i) &= \frac{1}{2}   & P(\alpha_\pm  \beta_\mp | a_i b_i) &= 0 
\end{align}
While the additional assumption at first sight seems to be a disadvantage, this derivation, as we shall see, will turn out to be the 
most powerful one allowing to derive Bell inequalities from the widest range of probability distributions. For this purpose we shall also need the further similar fact, which is not required for the original Wigner-Bell derivation, that in typical EPR/B experiments there are perfect anti\-/correlations for perpendicular settings:%
\begin{align}
\label{PACorr-def} 
\notag
\forall i: \quad P(\alpha_\pm \beta_\mp | a_i b_{i_\bot}) &= \frac{1}{2} & P(\alpha_\pm \beta_\pm | a_i b_{i_\bot}) &= 0\\
P(\alpha_\pm \beta_\mp | a_{i_\bot} b_i) &= \frac{1}{2}  & P(\alpha_\pm \beta_\pm | a_{i_\bot} b_i) &= 0 \tag{PACorr} 
\\(i_\bot := i+90^\circ)   \notag
\end{align} 

For my following strengthening of the standard Bell argument it is important to have a clear account of its logical structure:  

\setlength{\leftmargini}{2cm}
 \begin{enumerate}
\+[(P1)] There are EPR/B correlations: (Corr) 
\+[(P2)] \label{corr->BI} EPR/B correlations violate Bell inequalities: $\text{(Corr)} \rightarrow \neg(\text{BI})$ %\\
\end{enumerate}

\setlength{\leftmargini}{2cm}
\begin{enumerate}
\+[(P3)] \label{QM-PCorr} EPR/B correlations include perfect correlations: 
 (Corr) $\rightarrow$ (PCorr) 
\end{enumerate}

\setlength{\leftmargini}{2cm}
\begin{enumerate}
\+[(P4)] Bell inequalities can be derived from measurement independence, perfect correlations and local factorization: $(\text{MI}) \wedge (\text{PCorr}) \wedge (\text{LF}) \rightarrow (\text{BI})$ %\\
\end{enumerate}

\setlength{\leftmargini}{2cm}
\begin{enumerate}
\+[(P5)] Measurement independence holds: $(\text{MI})$ \\
\hrule \hrule 
\+[(C1)] Local factorization fails: $\neg(\text{LF})$ \hfill (from P1--P5)
\end{enumerate}

The core idea of my critique concerning this argument is that its conclusion 
can be made considerably stronger, providing a tighter, more informative probabilistic constraint for quantum non-locality.
Specifically, I shall show that it is premise (P4), the premise concerning the derivation of the inequality, which can be made stronger. 
The idea of the strengthening is to weaken the antecedent in (P4), 
i.e. the assumptions to derive the inequalities. 
While former improvements have concentrated on relaxing assumptions \emph{except} the locality condition, here I shall try to find \emph{weaker alternatives to local factorization}, which 
(jointly with other usual assumptions) also imply that Bell inequalities hold.  
Since local factorization is the weakest possible form of \emph{local} distributions, it is clear that such alternatives have to involve a kind of \emph{non-locality}, i.e. what I am trying to show in the following is that we can derive Bell inequalities from certain non-local probability distributions. 
This will make the overall argument stronger for it will allow for the conclusion that not only local theories but also those non-local ones that imply the inequalities are ruled out.

\subsection{Classification scheme for possible theories} 
\label{sec:classes} 

\def \mystar {^\star}

What alternatives to local factorization are there, that might serve to derive Bell inequalities? 
Local factorization is a specific product form of the \enquote{hidden joint probability} (of the outcomes), as I shall call $P(\alpha \beta | a b \lambda)$.%
\footnote{{} \enquote{Hidden} because the probability is conditional on the hidden state $\lambda$ and thus is not empirically accessible.} 
In general, according to the product rule of probability theory, any hidden joint probability can equivalently be written as a product, 
\begin{align}
\label{min-fact-alpha}
P(\alpha \beta | a b \lambda) & = P(\alpha | \beta b a \lambda) P(\beta |a b \lambda) \\
\label{min-fact-beta}
&= P(\beta | \alpha a b \lambda) P(\alpha | b a   \lambda) 
\end{align} 
(if according to the underlying probability distribution the involved conditional probabilities are well-defined). Since there are two product forms, one whose first factor is a conditional probability of $\zv \alpha$ and one whose first factor is a conditional probability of  $\zv \beta$, for the time being let us restrict our considerations to the product form \eqref{min-fact-alpha} until in  \autoref{sec:compl-part} we shall transfer the results to the other form \eqref{min-fact-beta}. 

We stress that 
the product form \eqref{min-fact-alpha} of the hidden joint probability 
{holds in general}, i.e. for all probability distributions (for each set of values for which the conditional probabilities are well-defined). 
According to probability distributions with appropriate independences, however, 
the factors on the right-hand side of the equation reduce in the sense  that certain variables in the conditionals can be left out. If, for instance, outcome independence holds, $\zv \beta$ can disappear from the first factor, and the joint probability is said to \enquote{factorize}. \emph{Local} factorization further requires that the distant settings in both factors disappear, i.e. that so called \enquote{parameter independence} holds. 
Prima facie, any combination of variables in the two conditionals in (\ref{min-fact-alpha}) seems to constitute a distinct product form of the hidden joint probability. Restricting ourselves to \emph{irreducibly hidden} joint probabilities, i.e. requiring $\zv \lambda$ to appear in both factors, there are $2^5=32$ \emph{combinatorially} possible forms (since any of the three variables in the first conditional and any of the two variables in the second conditional \emph{besides $\zv\lambda$} can or cannot appear). Columns II--VI in \autoref{table-classes} show these conceivable forms: \enquote{1} denotes appearance of a variable in the product form, \enquote{0} means its non-appearance.  
We label these product forms by (F$_{1}^\alpha$) to (F$_{32}^\alpha$) (the superscript $\alpha$ is due to the fact that we have used \eqref{min-fact-alpha} instead of \eqref{min-fact-beta}).

\def \breitem {-0pt} 

\begin{table}[htbp]
\begin{small} 
\caption{Classes of probability distributions}
\label{table-classes} 
\begin{center}
\renewcommand{\arraystretch}{1.35}
\begin{tabular}{l c c @{\extracolsep{-7pt}}
c c c @{\extracolsep{-7pt}} c c @{\extracolsep{-5pt}}
c c @{\extracolsep{-7pt}} c  @{\extracolsep{3pt}} | @{\extracolsep{5pt}}   c@{\extracolsep{5pt}} c @{\extracolsep{5pt}} l l}
\hline \hline 
& \tiny I &  & \tiny II & \tiny III & \tiny IV & & & \tiny V & \tiny VI & &  \tiny VII &  \tiny VIII &  \tiny IX &  \tiny X \\[0cm]  
& \multicolumn{10}{l|}{$(\text{H}_i^\alpha)$: \hspace{0cm} $P(\alpha \beta | a b  \lambda) = …$}  &PCorr&nPCorr & & $\Diamond$(H$_{i^\prime}^\beta$)
\\
& $i$ & $P(\alpha|$ & $\beta$ & $b$ & $a$ & $\lambda)$ &  $\cdot \hspace{0.1cm} P(\beta|$ & $a$ & $b$ & $\lambda) \hspace{0.1cm}$ & $\Box$(BI) & $\Box$(BI) & Notes & $i^\prime$
\\ \hline \hline

\multirow{14}{*}{\begin{sideways} strong non-locality$^\alpha$ \end{sideways}} & 1 & &  1 & 1 & 1 & & & 1 & 1 & & 0 & 0 & & 1–5,7,10,15,16    \\ \cline{2-15}

& 2 & &  1 & 1 & 1 & & & 1 & 0 & & 0 & 0 & & 1–3,5,7,16    \\ \cline{2-15}

& 3 & &  1 & 1 & 1 & & & 0 & 1 & & 0  & 0  & QM$_\text{p}$ & 1–4,7,15 \\ \cline{2-15}

& 4 & &  1 & 1 & 0 & & & 1 & 1 & & --- & 0 & & 1,3,4  \\ \cline{2-15}

& 5 & &  1 & 0 & 1 & & & 1 & 1 & & --- & 0 & & 1,2,5  \\ \cline{2-15}

& 6 & &  0 & 1 & 1 & & & 1 & 1 & & 0 & 0  & Bohm$_\text{s}$  & 6 
\\ \cline{2-15}

& 7 & &  1 & 1 & 1 & & & 0 & 0 & & 0 & 0  & QM$_\text{m}$ & 1,2,3,7\\ \cline{2-15}

& 8 & & 0 & 1 & 1 &&& 1 & 0 & & 0 & 0  & & 11\\ \cline{2-15}

& 9 & & 0 & 1 & 1 &&& 0 & 1 & & 0 & 0  & Bohm$_{\beta<a}$ & 12
 \\ \cline{2-15}

& 10 && 1& 0 & 0 &&& 1 & 1 & & --- & 0  & & 1 \\ \cline{2-15}

& 11 && 0 & 1 & 0 &&& 1 & 1 & & 0 & 0 && 8  \\ \cline{2-15}

&12 && 0 & 0 & 1 &&& 1 & 1 & & 0 & 0 &  Bohm$_{\alpha<b}$ & 9 %($t_\alpha \! \!< \! t_b$)
 \\ \cline{2-15}

& 13 && 0 & 1 & 1 &&& 0 & 0 & & 0 & 0 && 14  \\ \cline{2-15}

& 14 && 0& 0 & 0 &&& 1 & 1 & & 0 & 0 &&13  \\ \hline \hline

\multirow{14}{*}{\begin{sideways} weak non-locality$^\alpha$ \end{sideways}} & 15 & &  1 & 1 & 0 & & & 1 & 0 & & --- & 1 & & 1,3 \\ \cline{2-15}

& 16 & &  1 & 0 & 1 & & & 0 & 1 & &--- & 1& &1,2 
\\ \cline{2-15}

& 17 & &  1 & 0 & 1 & & & 1 & 0 & &--- & --- & & 18,19,24\\ \cline{2-15}

& 18 & &  1 & 1 & 0 & & & 0 & 1 & & ---& ---& & 17,20,21 \\ \cline{2-15}

& 19 & &  1 & 1 & 0 & & & 0 & 0 & & ---& ---& & 17,20 \\ \cline{2-15}

& 20 & &  1 & 0 & 1 & & & 0 & 0 & & ---& ---& & 18,19  \\ \cline{2-15}

& 21 & &  1 & 0 & 0 & & & 1 & 0 & & ---& ---& & 18 \\ \cline{2-15}

& 22 & &  0 & 1 & 0 & & & 1 & 0 & & 1 & 1& & 22\\ \cline{2-15}

& 23 & &  0 & 0 & 1 & & & 1 & 0 & & ---& ---& & 25\\ \cline{2-15}

& 24 & &  1 & 0 & 0 & & & 0 & 1 & & ---& ---& &17 \\ \cline{2-15}

& 25 & &  0 & 1 & 0 & & & 0 & 1 & & ---&  ---& & 23\\ \cline{2-15}

& 26 & &  1 & 0 & 0 & & & 0 & 0 & & ---& ---& & 26 \\ \cline{2-15}

& 27 & &  0 & 1 & 0 & & & 0 & 0 & & ---& --- & & 28\\ \cline{2-15}
 
& 28 & &  0 & 0 & 0 & & & 1 & 0 & & ---& ---& & 27\\ \hline \hline

\multirow{4}{*}{\begin{sideways} locality$^\alpha$ \end{sideways}} & 29 &  &  0 & 0 & 1 & & & 0 & 1 & & 1 & 1 & local fact. & 29
\\ \cline{2-15}

& 30 & &  0 & 0 & 1 & & & 0 & 0 & & ---& ---& & 31\\ \cline{2-15}

& 31 & &  0 & 0 & 0 & & & 0 & 1 & & ---& ---& & 30\\ \cline{2-15}

& 32 & &  0 & 0 & 0 & & & 0 & 0 & & ---& ---& & 32\\ \hline \hline

\multicolumn{2}{l}{Analysis:}  & & $\neg(\text{OI}_1)$ & $\neg(\text{PI}^\alpha_1)$ & $\neg(\text{LPI}^\alpha_1)$ & & & $\neg(\text{PI}^\beta_2)$ & $\neg(\text{LPI}^\beta_2)$   %\parbox{1.5cm}{logically possible}
\\ \hline \hline

\end{tabular}
\end{center}
\end{small}
\end{table}%

The specific product form of the hidden joint probability 
is {an essential feature} of the 
probability distributions of EPR/B experiments. For, as we shall see, it not only  determines whether a probability distribution can violate Bell inequalities, but also carries unambiguous information about which variables of the experiment are probabilistically independent of another. 
Therefore, it is natural to use the product form of the hidden joint probability in order to classify the 
probability distributions. 
We can say that each product form of the hidden joint probability 
 constitutes a \emph{class of probability distributions} in the sense that probability distributions with the same form 
 (but different numerical weights of the factors) belong to the same class. 
In order to make the assignment of probability distributions to classes unambiguous let us require that each probability distribution belongs only to that class which corresponds to its \emph{simplest} product form, i.e. to the form with the minimal number of variables appearing in the conditionals according to the distribution in question. So each class is defined by a characteristic product form (F$_i^\alpha$) and a minimality condition for that form; we 
label the classes by (H$_{1}^\alpha$) to (H$_{32}^\alpha$). 
{Since there are also probability distributions whose product form is not well-defined for all values of the variables, we further define that a distribution fulfills a product form if and only if the distribution obeys the form for all values of the variables for which the involved conditional probabilities are defined.}%  
\footnote{{} 
For instance, 
assume that for a certain probability distribution all three  probabilities in the equation 
\begin{equation}
\label{H16-2}
P(\alpha \beta | a b \lambda) = P(\alpha | \beta a \lambda) P(\beta | b \lambda)
\end{equation}
are well-defined for most values of the variables and that for these values the equation holds
(as it is required for distributions of class (H$^\alpha_{16}$)). For the remaining values, however, it is possible that the distribution yields $P(\beta a \lambda) = 0$, entailing that the conditional probability $P(\alpha | \beta a \lambda) := P(\alpha \beta a \lambda) / P(\beta a \lambda)$ is not well defined for these values and, hence, neither is the fact whether the distribution in question fulfills Equation \eqref{H16-2} for these values. 
According to our definition, however, we count such distributions as fulfilling (F$^\alpha_{16}$) because for all values for which the probabilities in the defining Equation \eqref{H16-2} are defined the equation holds. Hence, if there is no stronger product form (with less variables) which the distribution in question obeys, 
we include the distribution in class (H$^\alpha_{16}$). 
Since in \autoref{sect-analysis-classes} we shall analyze classes in terms of pairwise conditional (in-)dependences we should remark that the assignment of such partially defined probability distributions to classes that we have introduced here fits well with this analysis.

Note that only the first factor of a product form can be undefined and only if it involves the distant outcome in its conditional. For all probabilities involving at most $a b \lambda$ in their conditional---such as the hidden joint probability $P(\alpha \beta | a b \lambda)$ as well as the second factor in the product form, $P(\beta | a b \lambda)$---are always well-defined because 
$P(a b \lambda)$ never vanishes: Due to measurement independence it reduces to $P(a)P(b)P(\lambda)$ and if any of the single probabilities equals 0 for any variable value, say $P(\lambda_1) = 0$, one can always restrict the distribution to not include this value. 

It is never the case that a product form is completely unspecified, i.e. there are always some variable values for which conditional probabilities of a certain form are well-defined. For it is impossible that $P(\beta X \lambda)$ (with $X \subseteq \{a,b\}$), vanishes for all values of the involved variables: 
Suppose $P(\beta_+ X \lambda) = 0$, then $P(\beta_+ | X \lambda) = 0$, but then $P(\beta_- | X \lambda)=1$ and hence $P(\beta_- X \lambda) = P(X \lambda) ≠ 0$  (and vice versa for $P(\beta_{-} X \lambda)=0$).

Finally, if according to a probability distribution a product form $P(\alpha \beta | a b \lambda) = P(\alpha | \beta X \lambda) P(\beta | Y \lambda)$ (with $X, Y \subseteq \{a,b\}$) is undefined in its first factor for some variable values (due to $P(\beta X \lambda) = 0$), weaker factorization forms including more variables in the conditional of the first factor are neither well-defined because $P(\beta X \lambda) = 0$ implies $P(\beta a b \lambda) = 0$.
} 
We call such latter distributions \enquote{partially defined} in contrast to \enquote{well-defined} ones.

This scheme of classes is comprehensive: Any probability distribution of the EPR/B experiment must belong to one of these 32 classes. In this systematic overview, the class constituted by local factorization is (H$_{29}^\alpha$). 
Furthermore, if we allow that there might be no hidden states~$\zv \lambda$, the quantum mechanical distribution (as textbook quantum theory and GRW theory imply it) 
as well as the empirical distribution (which as far as we know coincide, but see our discussion of perfect (anti\=/)correlations in \autoref{sec:perf-correlations}) belong to class (H$_7^\alpha$) (if the photon state $\zv \psi$ is \emph{maximally} entangled, noted by \enquote{QM$_\text{m}$}) or to (H$_3^\alpha$), respectively (if $\zv \psi$ is \emph{partially} entangled, noted by \enquote{QM$_\text{p}$}).% 
\footnote{{}
The typical case for EPR/B experiments is to prepare a maximally entangled quantum state (e.g. $|\psi\rangle = \sqrt{p} |+\rangle |+\rangle + \sqrt{1-p} |-\rangle |-\rangle $ with $p=\frac{1}{2}$), because one wants to have a maximal violation of the Bell inequalities.  
The slightest deviation from maximal entanglement ($p≠\frac{1}{2}$), however, breaks the symmetry of the state. The probability distribution of such \emph{partially} entangled states 
shows an additional probabilistic dependence on the local setting in the second factor; hence, they fall in class~(H$_3^\alpha$). For an overview of the dependences and independences in the quantum mechanical probability distribution of maximally and partially entangled states see \citet[Table~1]{Naeger2015}. 
} 
The de-Broglie-Bohm theory falls under class (H$_6^\alpha$), 
when the measurements are performed simultaneously (in the sense that both measurement devices are installed before the detector at the respective other side registers; not in the strict sense that the registering of the measurement outcomes $\zv \alpha$ and $\zv \beta$ have to be simultaneous),%
\footnote{{}
Such temporal ordering between space-like separated events is, of course, only possible when there is a  preferred frame of reference, which Bohm's theory presupposes. 
}
and we label the corresponding probability distribution by \enquote{Bohm$_\text{s}$} (the index standing for \enquote{\emph{s}ymmetrical time ordering}). Otherwise, when the $\zv \beta$-measurement completes before the measurement device at A 
is installed 
(labelled by \enquote{Bohm$_{\beta<a}$}), the theory falls in class  (H$_9^\alpha$); and when the $\zv \alpha$-measurement is over before the 
measurement device at B 
is arranged (labelled by \enquote{Bohm$_{\alpha<b}$}), we have class (H$_{12}^\alpha$).%
\footnote{{}
\citet{Dewdney1987} 
show that according to Bohmian mechanics in a typical EPR/B setup, where both measurement devices with fixed measurement settings are in place right from the start, 
each outcome depends on both settings and both initial states (represented by $\lambda$), but not on the other outcome. 
If, however, one of the measurement devices, say the device at A, 
is only installed after the 
measurement on the other side, $\zv \beta$, has been completed 
it is clear that $\zv \beta$ does not depend on $\zv a$ (since $\zv a$ only comes into play later than $\zv \beta$ and is chosen independently of $\zv \beta$). This is why the statistics of the Bohmian description crucially depends on the time order. 
}
Similarly, any other theory of the quantum realm has its unique place in one of the classes.
Note that our scheme also contains classes that do not seem physically plausible. It is important, however, to include these classes into our investigation because in the end we aim to show that the argument that we are now going to develop on the basis of this  scheme, is the strongest possible argument on a qualitative probabilistic level---which requires not to have overlooked any logical possibility (see Paragraph (4) in \autoref{sec:strong-poss-cons}).

One crucial advantage of such an abstract classification is that it simplifies matters insofar we can now derive features of %irreducible
\emph{classes} of probability distributions and can be sure that these features hold for all members of a class, i.e. for all theories whose probability distributions fall under the class in question.
The feature that we are most interested in is, of course, which of these classes (given measurement independence) are consistent with the empirical probability distribution of EPR/B experiments. 
There are two hurdles: A class needs to be compatible with perfect (anti\=/)correlations (or at least with nearly perfect (anti\=/)correlations) and it may not imply Bell inequalities.

\subsection{The hurdle of perfect (anti\=/)correlations}
\label{sec:perf-correlations}

One can show that many of the classes are straightforwardly impossible if measurement independence, perfect correlations and perfect anti\-/correlations hold. 
(\enquote{Straightforward} here means that the impossibility is not demonstrated via a Bell inequality, but in a more direct way.)
Precisely the claim is: 

\begin{restatable}{theorem}{theorema}  
\label{th:dir-pcorr}
A class of probability distributions forms an inconsistent set with measurement independence, perfect correlations and perfect anti\-/correlations 
if and only if (i) its defining product form involves at most one of the settings or (ii)~its defining product form involves both settings and the first factor of this product form involves the distant outcome and at most one setting. 
\end{restatable} 
\vspace{-0.8cm}  \hfill (Proof in Appendix~\ref{sec:proof-theorem1.1})

\begin{corollary}
\label{cor:dir-pcorr}
A class of probability distributions 
forms a consistent set with measurement independence, perfect correlations and perfect anti\-/correlations if and only if  $\neg$(i) its defining product form  
involves both settings and $\neg$(ii) in case the distant outcome appears in the first factor of  its defining product form, also both settings appear in that factor. 
\end{corollary}

The consequences of the theorem for the status of the different classes are represented in column~VII of \autoref{table-classes}. All classes marked by \enquote{---} are inconsistent, while all classes with a number (\enquote{0} or \enquote{1}) are consistent (we shall explain the meaning of these numbers below). 
The inconsistent classes divide into two subgroups, corresponding to which condition for inconsistency, (i) or (ii) (cf. \autoref{th:dir-pcorr}), is fulfilled: 
\begin{itemize} 
\+[] Inconsistency due to condition (i): $\{(\text{H}_{17}^\alpha),...,(\text{H}_{32}^\alpha)\} \backslash \{(\text{H}_{22}^\alpha), (\text{H}_{29}^\alpha)\}$ 
\+[] Inconsistency due to condition (ii):  $\{(\text{H}_{4}^\alpha), (\text{H}_{5}^\alpha), (\text{H}_{10}^\alpha), (\text{H}_{15}^\alpha), (\text{H}_{16}^\alpha)\}$
\end{itemize}

We emphasize that the consistency and inconsistency claims of classes with the background assumptions have asymmetric consequences on the level of single probability distributions. On the one hand, a class being inconsistent with the background assumptions means that \emph{every} probability distribution of that class forms an inconsistent set with the assumptions. It is the general product form defining the class which is in conflict with the assumptions, hence all members of the class are. 
The same, mutatis mutandis, however, is not true of the consistent classes. 
A class being consistent with the background assumptions does {not} mean that {every} probability distribution of that class is consistent with the assumptions. 
Rather, by the laws of logic, it just means that \emph{at least one} probability distribution of a class is consistent with the assumptions, showing that the general product form of that class is {not per se} in conflict with them.  This is what consistency of a class means (when we define inconsistency in the natural way as just stated). 
This definition of consistency is perfectly compatible with the fact that there are distributions in a consistent class that are inconsistent with the assumptions due to their specific \emph{numerical values}. For instance, one can easily imagine distributions falling under class (H$^\alpha_7$) that, at parallel settings, involve correlations that are weaker than perfect. These distributions are obviously not consistent with the background assumptions, although their general product form is. 
Hence, we have to keep in mind that being consistent with the background assumptions on the level of classes, which is the level the present analysis proceeds on, is just a necessary condition for the distributions in that class to be consistent with the assumptions.

Quantum mechanics predicts perfect (anti\=/)correlations, 
but they are empirically not confirmed beyond doubt, 
because usual measurements typically show a certain deviation from perfectness. 
Though it might seem reasonable to assume that they nevertheless do hold (because the experimental deviations from perfectness might be attributed to measurement errors and non-ideal detectors), it has become usual in the discussion about Bell's theorem to avoid the strong assumption of perfectness: Either one does not make any reference to the correlations at parallel (or perpendicular) settings altogether (which we cannot do here), 
or one assumes only \emph{nearly} perfect correlations (nPCorr)   
and \emph{nearly} perfect anti\-/correlations (nPACorr) (e.g. for parallel or perpendicular settings, respectively, given $\psi_0$):% 
\footnote{{} 
The conditions in \eqref{nPCorr-def} entail 
\begin{align*}
P(\alpha_\pm | \beta_\pm a_i b_i) &= 1-2\delta_{ii}, & P(\alpha_\mp | \beta_\pm a_i b_i), &= 2\delta_{ii}, & P(\beta_\pm | \alpha_\pm a_i b_i) &= 1-2\delta_{ii},   & P(\beta_\mp | \alpha_\pm a_i b_i) &= 2\delta_{ii}, 
\end{align*}
revealing the nearly perfect correlations (and mutatis mutandis for the conditions in \eqref{nPACorr-def1}). 
}
\begin{align}
\label{nPCorr-def} \tag{nPCorr}
\forall i: && P(\alpha_\pm \beta_\pm | a_i b_i) &= \frac{1}{2}-\delta_{ii} & P(\alpha_\pm \beta_\mp | a_i b_i) &= \delta_{ii}  \\
\label{nPACorr-def1} 
\tag{nPACorr} 
&& P(\alpha_\pm \beta_\mp | a_i b_{i_\bot}) &= \frac{1}{2}-\delta_{i i_\perp} & P(\alpha_\pm \beta_\pm | a_i b_{i_\bot}) &= \delta_{i i_\perp}\\
\label{nPACorr-def2} \notag %\tag{nPACorr}
&&P(\alpha_\pm  \beta_\mp | a_{i_\bot} b_i) &= \frac{1}{2}-\delta_{i_\perp i} & P(\alpha_\pm  \beta_\pm | a_{i_\bot} b_i) &= \delta_{i_\perp i}  
\end{align}

Clearly, nearly perfect (anti\=/)correlations are a weaker assumption than perfect ones, and one can show that fewer classes are inconsistent with the former. 
Precisely: 

\begin{restatable}{theorem}{theoremb}
\label{th:dir-npcorr}
A class of probability distributions forms an inconsistent set with measurement independence, nearly perfect correlations and nearly perfect anti\-/correlations if and only if (i) its defining product form involves at most one of the settings. 
\end{restatable} 
\vspace{-0.3cm}  
\hfill (Proof in Appendix~\ref{sec:proof-theorem2.1})

\begin{corollary}
\label{cor:dir-npcorr}
A class of probability distributions forms a consistent set with measurement independence, nearly perfect correlations and nearly perfect anti\-/correlations if and only if $\neg$(i) its defining product form involves both settings. 
\end{corollary}

The consequences of these claims are represented in column~VIII of \autoref{table-classes}. 
Unlike \autoref{th:dir-pcorr}, \autoref{th:dir-npcorr} does not rule out classes fulfilling condition~(ii), so in the case of nearly perfect (anti\=/)correlations more classes are consistent.  

In sum, both cases exclude a number of non-local theories which are not ruled out by Bell's original theorem. 
We should note, however, that the exclusion here neither requires to derive a Bell inequality (it is more direct, as can be seen in the proofs), nor does it make sense to try to derive a Bell inequality for classes that are impossible due to a more direct conflict with the empirical probability distribution.
For this reason, \Cref{th:dir-pcorr,th:dir-npcorr} are not in a literal sense a strengthening of the Bell argument. But since the aim of Bell's argument is to exclude certain theories of the micro-realm one might say that they are \emph{amendments} to the argument which strengthen its conclusion.

\subsection{The hurdle of violating Bell inequalities} 

We now turn to the second hurdle, which requires that classes need to be able to violate Bell inequalities. Classes which imply the inequalities are ruled out via the Bell argument---so which of the consistent classes do entail the inequalities?%
\footnote{{} 
Classes that are inconsistent even with nearly perfect (anti-)correlations, $(\text{H}^\alpha_{17})$--$(\text{H}^\alpha_{32})\backslash\{(\text{H}^\alpha_{22}), (\text{H}^\beta_{29})\}$, trivially obey Bell inequalities because their product only involves one of the settings. 
Class ($\text{H}_{17}^\alpha$), for instance, whose product form does not involve the setting $\zv b$, 
makes the empirical joint probability independent of $\zv b$: 
$P(\alpha \beta | a b) = \sum_{\lambda}  P(\alpha | \beta a \lambda) P(\beta | a \lambda) P(\lambda) = P(\alpha \beta | a)$. 
Inserting this empirical joint probability, 
which does not depend on $\zv b$, 
into a usual Bell-Wigner inequality \eqref{BI-usual}, yields $P(\alpha_- \beta_+ | a_1) ≤  P(\alpha_- \beta_+ | a_1) + P(\alpha_- \beta_+ | a_2)$ and thus reveals that in this case the inequality holds trivially, because it 
has lost its functional dependence on $\zv b$. 
}
As we have announced at the outset of the paper, besides the well-known class constituted by local factorization there are non-local classes that allow a derivation. 
We start with the case of strictly perfect (anti\=/)correlations:

\begin{restatable}{theorem}{theoremc}
\label{th:ind-pcorr}
Given measurement independence, perfect correlations and perfect anti\hyp{}correlations, a consistent class (i.e. a class that fulfills $\neg$(i) and $\neg$(ii)) 
implies Bell inequalities if and only if (iii) each factor of its defining product form involves at most one setting. 
\end{restatable} 
\vspace{-0.8cm}  \hfill (Proof in Appendix~\ref{sec:proof-theorem1.2})

\begin{corollary}
Given measurement independence, perfect correlations and perfect anti\hyp{}correlations, a consistent class (i.e. a class that fulfills $\neg$(i) and $\neg$(ii)) does not imply Bell inequalities if and only if $\neg$(iii) at least one factor of its defining product form involves both settings. 
\end{corollary}

The consequences of these results for the status of the different classes are represented  in column~VII of \autoref{table-classes}. 
The heading of the column, \enquote{$\Box$(BI)}, means \emph{necessarily,  Bell inequalities hold}. 
Recall that the dashes \enquote{---} in this column represent classes that are inconsistent with the strictly perfect (anti\=/)correlations; in these cases the question whether Bell inequalities are implied does not make sense. 
The numbers in the column indicate whether a certain product form implies Bell inequalities (\enquote{1}) or does not imply them (\enquote{0}) (according to \autoref{th:ind-pcorr}). 
Clearly, all classes that are marked either by \enquote{---} or \enquote{1} 
are impossible if measurement independence and perfect (anti\=/)correlations hold.

The two classes 
marked by \enquote{1}, i.e. $(\text{H}_{22}^\alpha)$ and $(\text{H}_{29}^\alpha)$, can explicitly be shown to imply a Bell inequality. That $(\text{H}_{29}^\alpha)$, local factorization, implies the inequalities is familiar; less known is the fact that the non-local class
\begin{align}
P(\alpha \beta | a b \lambda) = P(\alpha | b \lambda) P(\beta | a \lambda) \tag{H$_{22}^\alpha$} 
\end{align}
implies the inequalities as well \citep[cf.][sec.~3.3]{Seevinck2008}.
The latter class is the symmetrical counterpart to local factorization, compared to which the settings are interchanged, such that each outcome depends on its distant setting. 
For this reason the derivation of the Bell inequalities runs very similarly as for local factorization (just swap the settings in the original proof).
\note{*leichte Redundanz zu Beweis im Anhang}{}

On the other hand, the theorem also says that any consistent class that violates~(iii), 
can be shown \emph{not} to imply the Bell inequalities. 
Here we have a similar asymmetry between the level of classes and that of distributions as in the case of (in)consistency with perfect \mbox{anti-}correlations (see \autoref{sec:perf-correlations}). 
Since a class implying Bell inequalities (given the background assumptions) means that \emph{every} probability distribution having the product form in question obeys the inequalities, the claim that a class does {not} imply the inequalities (given the background assumptions) denotes the fact that \emph{there is at least one probability distribution in that class that violates the inequalities}. 
Therefore, not implying Bell inequalities emphatically does {not} mean that {every} probability distribution in a class violates the inequalities. 
For this reason, given just the product form of one of the classes violating (iii)
one cannot decide whether Bell inequalities hold; whether they do in these cases depends on the \emph{numerical} features of the probability distribution in question. In this sense, one might reasonably say that probability distributions of these classes \emph{can} violate Bell inequalities.

Let us now turn to the case that only \emph{nearly} perfect (anti\=/)correlations hold: 
\begin{restatable}{theorem}{theoremd}
\label{th:ind-npcorr}
Given measurement independence, nearly perfect correlations and nearly perfect anti\-/correlations, a consistent class (i.e. a class that fulfills $\neg$(i)) implies Bell inequalities if and only if (iii) each factor of its defining product form involves at most one setting. 
\end{restatable}  
\vspace{-0.8cm}  \hfill (Proof in Appendix~\ref{sec:proof-theorem2.2}) 

\begin{corollary}
Given measurement independence, nearly perfect correlations and nearly perfect anti\-/correlations, a consistent class (i.e. a class that fulfills $\neg$(i)) does not imply Bell inequalities if and only if $\neg$(iii) at least one factor of its defining product form involves both settings. 
\end{corollary}

The consequences of these claims are represented in column~VIII of \autoref{table-classes}. 
Since nearly perfect (anti\=/)correlations are a considerably weaker requirement than that of strictly perfect ones, 
there are more consistent classes.
Those classes that (compared to strictly perfect (anti-)correlations) become consistent and do not fulfill condition~(iii), namely (H$_{4}^\alpha$), (H$_{5}^\alpha$) and (H$_{10}^\alpha$), can be shown to be able to violate Bell inequalities;
the other classes that become consistent, 
(H$_{15}^\alpha$) and (H$_{16}^\alpha$), imply Bell inequalities. 
Let me emphasize that 
\begin{align}
P(\alpha \beta | a b \lambda) = P(\alpha | \beta a \lambda) P(\beta | b \lambda)  \tag{H$_{16}^\alpha$}
\end{align}
(as well as the related class (H$_{15}^\alpha$) with swapped settings) 
is the weakest and most surprising class which implies Bell inequalities according to \autoref{th:ind-npcorr}. 
While this class is inconsistent with perfect (anti\=/)correlations, it is consistent with nearly perfect ones, but then implies Bell inequalities. 
As local theories, theories in class (H$^\alpha_{16}$)
do not produce correlations that are strong enough to violate  Bell inequalities. 
Demonstrating this fact is the central step in the proof for \autoref{th:ind-npcorr}.
This so far unnoticed implication 
is remarkable, because (H$^\alpha_{16}$) differs from local factorization in that it involves the distant outcome $\zv \beta$ in the first factor on the right hand side, which makes it a  product form that involves a dependence between the \emph{non-locally} related outcomes---and such product forms have been  believed to be able to violate the inequalities.  
In our analysis in \autoref{sec-analysis} we shall see that the fact that (H$^\alpha_{16}$) cannot violate  the inequalities has far reaching consequences for the status of theories that are usually called \enquote{outcome dependent}.

Since local factorization is a stronger condition than (H$^\alpha_{16}$)
the derivation of Bell inequalities from
class (H$^\alpha_{16}$) (given measurement independence and nearly perfect (anti\=/)correlations)
can easily be modified to show the derivation of Bell inequalities from 
local factorization (given measurement independence and nearly perfect (anti\=/)correlations). 
In this way our proof en passant shows the remarkable fact that one can \emph{derive a Wigner-Bell inequality without strictly perfect correlations} (so far the latter have been regarded to be a necessary assumption for deriving that type of Bell inequality). 

The Bell inequality that follows by this new kind of proof is a generalized Wigner-Bell inequality, 
\begin{align} 
 \label{BI-dev}
\frac{P(\alpha_- \beta_+ | a_1 b_3)}{1-\epsilon_1}  & ≤ \frac{P(\alpha_- \beta_+ | a_1 b_2 \lambda) + P(\alpha_- \beta_+ | a_2 b_3 \lambda)}{(1-\epsilon_2)^2} \notag\\
&\hphantom{= \;}+  \delta \left (\frac{2}{\epsilon_1}+\frac{8}{\epsilon_2} + 4 \frac{1-\epsilon_1-\epsilon_2}{\epsilon_1^2 \epsilon_2} + \frac{1}{(1-\epsilon_2)^2} -\frac{2}{1-\epsilon_1} -4 \right )  \nonumber \\
&\hphantom{= \;} + 2\epsilon_1 \left (\frac{2}{\epsilon_2}-1 \right), 
\end{align}
that differs from a usual Wigner-Bell inequality, 
\begin{equation}
 \label{BI-usual} P(\alpha_- \beta_+ | a_1 b_3) ≤ P(\alpha_- \beta_+ | a_1 b_2) + P(\alpha_- \beta_+ | a_2 b_3), 
\end{equation}
by certain correction terms involving 
the deviation from perfect correlations and perfect anti\-/correlations $\delta$ as well as two parameters $\epsilon_1$ and $\epsilon_2$, which can be freely chosen inside the limits $\epsilon_1 ≤ \frac{1}{2}$ and $\epsilon_2 < 1-\epsilon_1$; especially they can be chosen such that the inequality is best violated (for the meaning of these parameters, see the proof of Theorem 4). 
It is easy to see that in the border case $\delta\rightarrow 0$ the generalized Wigner-Bell inequality agrees with the usual one  (if we also assume $\epsilon_1 \rightarrow 0$ and $\epsilon_2 \rightarrow 0$ in such a way that the correction terms vanish). 
One can further show (see the proof of \autoref{th:ind-npcorr}) that the generalized inequality is violated by the usual statistics of EPR/B experiments, if at least 99.9999258\% of the runs with parallel settings as well as those with perpendicular settings turn out to be perfectly correlated and perfectly anti-correlated, respectively. 
This defines the above condition of nearly perfect (anti\=/)correlations more precisely: Only in worlds where the fraction of perfectly (anti-)correlated runs 
exceeds the indicated threshold, 
all theories from class (H$^\alpha_{16}$) are ruled out.

This quantitative limit reveals a final possibility to avoid the implication:
One might hint to the fact that in actual experiments far less than 99.9999258\% of the entangled objects show perfect \mbox{(anti-)}cor\-re\-la\-tions. This indeed shows that the question whether 
theories from class (H$^\alpha_{16}$) can hold or not is not yet decided  empirically beyond doubt.  
Let me stress, however, that the main aim in this paper is not to decide this empirical and quantitative question, but the conceptual and qualitative one, namely whether it is possible to amend Bell's argument for a stronger conclusion, ruling out even certain non-local theories. 

That said, I can add that I think that there are good reasons not to take the mentioned empirical discrepancy to undermine the argument against 
theories from class (H$^\alpha_{16}$). 
First, the derivation of the inequality \eqref{BI-dev} uses certain rather rough estimations, which contribute to the fact that the degree of perfectness that is required for a violation to take place is high. Improved future derivations, which include more precise (and expectedly more complicated) estimations, might lower that degree considerably. 
Second, the past has shown that experimental physicists have continuously been increasing the fraction of measured perfectly (anti-)correlated pairs of entangled objects, by using more and more sophisticated experimental techniques. So it is to be expected that the empirically confirmed degree of perfect correlation will increase in the future as well. Finally, quantum mechanics predicts perfect correlations and at present there is no further, independent evidence (besides the fact that experiments do not yield strictly perfect (anti\=/)correlations) to doubt that quantum mechanics is wrong; for this reason, it seems reasonable to assume that the deviation from perfectness in experiments is due to experimental imperfections.

In the above theorems, condition (iii), that a class does not involve more  
than one setting in each factor of its product form, is the essential characteristic (in terms of the product form) to tell apart classes that imply Bell inequalities, (iii), from those that do not, $\neg$(iii).
Let us introduce some appropriate terminology: 

\setlength{\leftmargini}{1cm}
\begin{itemize}
\+[] {Local$^\alpha$} classes:  (H$_{29}^\alpha$)--(H$_{32}^\alpha$) 
\vspace{-0.1cm} 

\begingroup
\leftskip0.5cm \rightskip0.8cm 
Each factor only contains variables that are time-like (or light-like) separated to the respective outcome. 
\par
\endgroup

\+[] {Weakly non-local$^\alpha$} classes:  ($\text{H}_{15}^\alpha$)--($\text{H}_{28}^\alpha$) 
\vspace{-0.1cm} 

\begingroup
\leftskip0.5cm \rightskip0.8cm 
At least one of the factors involves variables that are space-like separated to the respective outcome, but none of the factors involves both settings. \par
\endgroup

\+[] {Strongly non-local$^\alpha$} classes: (H$_{1}^\alpha$)--(H$_{14}^\alpha$) 
\vspace{-0.1cm} 
 
\begingroup
\leftskip0.5cm \rightskip0.8cm 
At least one of the factors involves both settings. $\neg$(iii)
\par
\endgroup

\end{itemize}

Strongly non-local$^\alpha$ classes are just those classes that fulfill criterion $\neg$(iii) not to imply the Bell inequalities, while the remaining classes fulfilling criterion~(iii) have been divided into local$^\alpha$ and weakly non-local$^\alpha$ ones. 
With these new concepts we can summarize the central consequence of Theorems~\ref{th:dir-pcorr}\==\ref{th:ind-npcorr} as:
\begin{corollary}
\label{cor:sum}
Given measurement independence and strictly or nearly perfect (anti\=/)correlations 
every local$^\alpha$ and weakly non-local$^\alpha$ class is forbidden, 
either because it is inconsistent with measurement independence and the strictly or nearly perfect (anti\=/)correlations or (if it is consistent) because it necessarily obeys Bell inequalities. 
\end{corollary}

As opposed to what the standard discussion suggests, this corollary stresses the fact that besides local classes \emph{even certain non-local classes}, namely the weakly non-local$^\alpha$ ones, \emph{are ruled out} by the empirical statistics of EPR/B experiments (if measurement independence holds). 
We have found that 18 (21 in the case of \emph{strictly} perfect (anti\=/)correlations) of the 32 logically possible classes are forbidden, 
among them 14 (17 in the case of \emph{strictly} perfect (anti\=/)correlations) \emph{non-local} classes. 
It is a central result of this investigation that among the forbidden non-local classes is the class (H$^\alpha_{16}$), which includes a dependence between the distant outcomes. In the case of strictly perfect correlations it is forbidden because it is inconsistent with the correlations and measurement independence, and when nearly perfect (anti\=/)correlations hold, it is consistent but implies Bell inequalities.

\subsection{Complementary classification scheme}   
\label{sec:compl-part}

Our argument up to this point has been based on the partition of probability distributions in \autoref{table-classes}, which we found by 
writing the hidden joint probability according to the general product rule \eqref{min-fact-alpha} and conceiving all logically possible product forms. 
We can, however, as well write the hidden joint probability according to the second general product rule \eqref{min-fact-beta}, and similar considerations as above lead us to a similar table, whose classes, (H$_1^\beta$)--(H$_{32}^\beta$), differ to those in \autoref{table-classes} in that both the outcomes and the settings are swapped in the defining product forms. For instance, class (H$_{16}^\beta$) is defined by the product form $P(\alpha \beta | a b \lambda) = P(\beta | \alpha b \lambda) P(\alpha | a \lambda)$ in contrast to (H$_{16}^\alpha$), which is constituted by  $P(\alpha \beta | a b \lambda) = P(\alpha | \beta a \lambda) P(\beta | b \lambda)$. Note that this new classification is a different partition of the possible probability distributions, which reasonably might be called \enquote{complementary partition}. Any probability distribution must fall in exactly one of the classes  (H$_1^\alpha$)--(H$_{32}^\alpha$) \emph{and} in exactly one of the classes  (H$_1^\beta$)--(H$_{32}^\beta$).

Which of these new classes (H$_{i}^\beta$) is compatible with measurement independence and strictly or nearly perfect (anti\=/)correlations? And which implies Bell inequalities? 
The answer simply is that \emph{Theorems~1--4 literally apply to the these new classes as well}. For the theorems are formulated in a way that generally applies to classes defined by product forms and the proofs can be adjusted mutatis mutandis. 
Consequently, the theorems also imply for the new partition:
\begin{corollary}
\label{cor:sum-beta}
Given measurement independence and strictly or nearly perfect (anti\=/)correlations 
every local$^\beta$ and weakly non-local$^\beta$ class is forbidden, 
either because it is inconsistent with measurement independence and the strictly or nearly perfect (anti\=/)correlations or (if it is consistent) because it necessarily obeys Bell inequalities. 
\end{corollary} 

How do these two partitions of  possible probability distributions relate? 
Due to logical restrictions a probability distribution from a certain class (H$_i^\alpha$) cannot fall into any class (H$_{i^\prime}^\beta$), i.e. not any combination of classes$^\alpha$ and classes$^\beta$ is logically possible. 

\begin{restatable}{theorem}{theoreme} 
\label{th:combi}
Each class $(\text{\normalfont H}_i^\alpha)$ is consistent with those and only those classes $(\text{\normalfont H}_i^\beta)$ that are indicated in column~X of \autoref{table-classes}. 
\end{restatable}  
\vspace{-0.8cm}  \hfill (Proof in Appendix~\ref{sec:proof-theorem-combi}) 
\vspace{0.4cm}
 
\noindent (The heading \enquote{$\Diamond (\text{H}_{i^\prime}^\beta)$} of column~X means \enquote{(H$_{i^\prime}^\beta)$ that possibly hold}).
For instance, probability distributions falling in 
class (H$_7^\alpha$) can only belong to  either of the classes  (H$_1^\beta$), (H$_2^\beta$), (H$_3^\beta$) or (H$_7^\beta$). (Systems with maximally entangled quantum states exclusively fall into the combination of classes  $(\text{H}_7^\alpha) \wedge (\text{H}_7^\beta)$.)  In total, there are 65 possible combinations of classes$^\alpha$ with classes$^\beta$. This provides a much more fine-grained qualitative partition of the distributions than just considering the classes$^\alpha$. 

The overview reveals that local$^\alpha$ classes can only be combined with local$^\beta$ classes, but some strongly non-local$^\alpha$ classes (viz. $(\text{H}_{1}^\alpha)$, $(\text{H}_{2}^\alpha)$ and $(\text{H}_{3}^\alpha)$) can be combined with weakly non-local$^\beta$ ones (viz. $(\text{H}_{15}^\alpha)$ or $(\text{H}_{16}^\alpha)$, respectively) and vice versa. Since a distribution from a weakly non-local class necessarily implies Bell inequalities, it is clear that the complementary partition can provide important additional information that is relevant for dividing the distributions into those that can and those that cannot violate Bell inequalities. Therefore, in the following we shall indicate into which \emph{combined} class a distribution falls. 
Note that the fact that there are distributions from strongly non-local classes that do not violate Bell inequalities does not contradict anything we have said so far (we have said that at least one but not necessarily all  distributions in a strongly non-local class violates Bell inequalities); here we learn that we can qualitatively characterize  \emph{some} of them by the complementary partition (though there are further strongly non-local distributions that do not violate Bell inequalities for numeric reasons and cannot be captured by qualitative features.) 

Let us finally agree to say that a probability distribution is \enquote{local} (without qualification by $\alpha$ or $\beta$) if it is local$^\alpha$ \emph{or} local$^\beta$, i.e. $\bigvee_{i = 29}^{32} (\text{H}_i^\alpha)  \vee \bigvee_{i = 29}^{32} (\text{H}_i^\beta)$ holds; 
it is \enquote{weakly non-local} 
if it is weakly non-local$^\alpha$  \emph{or} weakly non-local$^\beta$, i.e. $\bigvee_{i = 15}^{28} (\text{H}_i^\alpha)  \vee \bigvee_{i = 15}^{28} (\text{H}_i^\beta)$ is true; and it is \enquote{strongly non-local} if it is strongly non-local$^\alpha$ \emph{and} strongly non-local$^\beta$, i.e. $\bigvee_{i = 1}^{14} (\text{H}_i^\alpha)  \wedge  \bigvee_{i = 1}^{14} (\text{H}_i^\beta)$ holds. 
Especially, a distribution which is weakly non-local$^\alpha$ and strongly non-local$^\beta$, e.g. a distribution belonging to classes $(\text{H}^\alpha_{16})$ and $(\text{H}^\beta_1)$, counts as weakly non-local.

\subsection{A stronger Bell argument}
\label{sec:strength-Bell}

It is clear that each set of corresponding theorems (\ref{th:dir-pcorr} and \ref{th:ind-pcorr} as well as \ref{th:dir-npcorr} and \ref{th:ind-npcorr}) can be used to strengthen Bell's argument. On the other hand, it is not clear which of these available new arguments should be considered to be the strongest. (The first set results in an argument that, compared to the argument resulting from the second set, requires the stronger assumption of strictly perfect correlations (weakening the argument), but allows for a stronger conclusion, because it rules out even some of the strongly non-local$^\alpha$ classes). 
Here we restrict our discussion to the argument resulting from the second set, because it avoids the controversial assumption of strictly perfect (anti\=/)correlations.  
(The argument from the first set can be formulated mutatis mutandis.) 

\setlength{\leftmargini}{2cm}
 \begin{enumerate}
\+[(P1)] There are EPR/B correlations: (Corr) 
\+[(P2)] \label{corr->BI} EPR/B correlations violate Bell inequalities: $\text{(Corr)} \rightarrow \neg(\text{BI})$ %\\
\end{enumerate}

\setlength{\leftmargini}{2cm}
\begin{enumerate}
\+[(P3$^\prime$)] \label{QM-PCorr} EPR/B correlations include nearly perfect correlations and nearly perfect anti\-/correlations: 
 (Corr) $\rightarrow$ (nPCorr) $\wedge$ (nPACorr) 
\end{enumerate}

\setlength{\leftmargini}{2cm}
\begin{enumerate}
\+[(P6)] \label{prem-incons-2} Those local$^\alpha$, weakly non-local$^\alpha$, local$^\beta$ and weakly non-local$^\beta$ classes that involve at most one setting in their product form are inconsistent with measurement independence, nearly perfect correlations and nearly perfect anti\-/correlations: 
\begin{equation*}
(\text{MI}) \wedge (\text{nPCorr}) \wedge (\text{nPACorr}) \rightarrow \quad\bigwedge_{\mathclap{\substack{i = 17, \dots, 32 \\ \hphantom{i = }\backslash \{22,29\}}}}^{} \neg (\text{H}_i^\alpha)  \quad  \wedge \quad \bigwedge_{\mathclap{\substack{i = 17, \dots, 32 \\ \hphantom{i = }\backslash \{22,29\}}}}^{} \neg (\text{H}_i^\beta) 
\end{equation*}
\end{enumerate}

\setlength{\leftmargini}{2cm}
\begin{enumerate}
\+[(P4$^{\prime}$)] Bell inequalities can be derived from measurement independence, nearly perfect correlations, nearly perfect anti\-/correlations and any local$^\alpha$, weakly non-local$^\alpha$, local$^\beta$ or weakly non-local$^\beta$ class of probability distributions that involves both settings in its product form: 
\begin{equation*}
\bigg[ (\text{MI}) \wedge (\text{nPCorr}) \wedge (\text{nPACorr}) \wedge \Big( \bigvee_{\mathclap{\substack{i = 15,16,\\ \hphantom{i =} 22,29}}} \; (\text{H}_i^\alpha) \quad \vee \quad  \bigvee_{\mathclap{\substack{i = 15,16,\\ \hphantom{i =} 22,29}}} \; (\text{H}_i^\beta) \Big) \bigg] \rightarrow (\text{BI})
\end{equation*}
\end{enumerate}

\setlength{\leftmargini}{2cm}
\begin{enumerate}
\+[(P5)] Measurement independence holds: $(\text{MI})$ \\
\hrule \hrule 
\+[(C1$^{\prime}$)] \textbf{Failure of locality and weak non-locality}: All local$^\alpha$, weakly non-local$^\alpha$, local$^\beta$ and weakly non-local$^\beta$ classes fail: 
\end{enumerate} 
\begin{equation}
\left( \bigwedge_{i = 15}^{32} \neg (\text{H}_i^\alpha)  \wedge \bigwedge_{i = 15}^{32} \neg (\text{H}_i^\beta) \right) 
\notag
\end{equation}\\

Compared to the original Bell argument (\autoref{sec:experiments}) there are three substantial changes, which strengthen the argument. 
A first change concerns the fact that everywhere in the argument we have relaxed  controversial strictly perfect correlations to uncontroversial nearly perfect correlations (in premisses (P3) and (P4) of the original argument). This is a strengthening in the sense that the argument makes weaker assumptions. 
At the same places in the argument where nearly perfect correlations occur we have additionally introduced nearly perfect \emph{anti}\-/correlations. This might seem as a weakening of the argument; in fact, however, it is a neutral move,  
because it is uncontroversial that the nearly perfect anti\-/correlations follow from the EPR/B correlations (as the nearly perfect correlations do; see premise (P3$^\prime$)), and these EPR/B correlations have already been assumed in the original argument (premise (P1)). 

A second strengthening of the argument stems from introducing a completely new premise (P6), which states the content of \autoref{th:dir-npcorr}, that certain classes are not compatible with measurement independence, nearly perfect correlations and nearly perfect anti\-/correlations. 
Given that measurement independence and nearly perfect (anti\=/)correlations are assumed anyway (or derive from usual assumptions), it is clear that these classes will be ruled out by the overall argument. In this sense, (P6) provides a genuine strengthening of the conclusion of the theorem. 
\bki Deriving a direct contradiction between the background assumptions and certain classes without involving a Bell inequality, premise (P6) has no counterpart in the original argument and rather has the status of an amendment---however, an amendment that naturally fits in.\eki{}
Note that assuming the additional premise (P6) does not weaken the argument because it can be proven mathematically (see the proof of \autoref{th:dir-npcorr}). 

A third modification, \bki indeed the central strengthening\eki{}, consists in the adaption of premise (P4) to \autoref{th:ind-npcorr}, which says that one can derive Bell inequalities not only from local factorization but from all those local$^\alpha$, weakly non-local$^\alpha$, local$^\beta$ and weakly non-local$^\beta$ classes that are consistent given measurement independence and nearly perfect (anti\=/)correlations. 
Accordingly, we have replaced local factorization in the antecedent by the disjunction of  these product forms. 
This makes the antecedent of (P4$^\prime$) weaker than that in (P4) and, hence, the argument stronger. Since the overall Bell argument is a modus tollens argument to the negation of that premise, this modification also strengthens the conclusion of the theorem.

Making these changes has a considerable effect on the conclusion of the Bell argument.
While the original result, the failure of local factorization, implied that all local$^\alpha$ and local$^\beta$ classes fail (because the other local classes are specializations of local factorization), the new result additionally excludes all weakly non-local$^\alpha$ and weakly non-local$^\beta$ classes---and thus clearly is stronger. 
 
Stating which classes are excluded, the conclusion formulated here is a negative one. 
But it is easy to turn it into a positive statement: 
since our scheme of logically possible classes is comprehensive, the failure of all local$^{\alpha}$ and weakly non-local$^\alpha$ classes is equivalent to the fact that one of the strongly non-local$^\alpha$ classes, (H$_1^\alpha$)--(H$_{14}^\alpha$), holds. Analogously, if a probability distribution is neither  local$^{\beta}$ nor weakly non-local$^\beta$ it must be strongly non-local$^\beta$, i.e. belong to one of the classes (H$_1^\beta$)--(H$_{14}^\beta$). 
Therefore, equivalently to (C1$^{\prime}$) we can say:

\begin{enumerate}
\+[(C1$^{\prime\prime}$)] 
\textbf{Strong non-locality}: 
One of the strongly non-local$^\alpha$ classes and one of the strongly non-local$^\beta$ classes holds:
\begin{align} 
\left( \bigvee_{i = 1}^{14} (\text{H}_i^\alpha)  \wedge \bigvee_{i = 1}^{14} (\text{H}_i^\beta)  \right) \quad \quad 
\notag
\end{align}

\end{enumerate}
This is the positive conclusion of the stronger Bell argument in terms of classes. 
(Recall that due to logical restrictions not any strongly non-local$^\alpha$ class is compatible with any strongly non-local$^\beta$ class, see \autoref{table-classes} column~X.)

\subsection{Discussion I: Immediate consequences}   
\label{sec:strong-poss-cons}
\newcounter{count2}
\stepcounter{count2}

(\arabic{count2})\stepcounter{count2}  Let us first shortly summarize our results so far: The strengthening of the Bell argument rests on the insight that the members of a range of non-local theories, which we have called \emph{weakly non-local}, either are inconsistent with measurement independence and nearly perfect correlations or imply Bell inequalities (as do local theories). 
For instance, it is impossible to violate Bell inequalities even if a dependence on the distant outcome holds as in the product form (H$^\alpha_{16}$), $P(\alpha \beta | a b \lambda) = P(\alpha | \beta a \lambda) P(\beta | b \lambda)$.   
Consequently, the empirical violation of the inequalities does rule out local theories (which is well known from the original argument) \emph{and} these weakly non-local ones (which is one central result of this paper). 
Showing that the violation of Bell inequalities excludes more theories than the standard Bell argument suggests, \emph{the new argument has a stronger conclusion than the original one}. 

The remaining theories, which are compatible with a violation of Bell inequalities, are called \enquote{strongly non-local}; a list of their product forms is given by (H$_{1}^\alpha$)--(H$_{14}^\alpha$) in \autoref{table-classes} (and the corresponding complementary classes (H$_{1}^\beta$)--(H$_{14}^\beta$)).
They are characterized by the fact 
that \emph{at least one of the factors in the product form involves both settings in its conditionals}, i.e. at least one of the outcomes must depend probabilistically (or functionally, respectively) on both settings. Without such a dependence between an outcome and both settings Bell inequalities cannot be violated. 
Before we will examine the required dependences in more detail (\Cref{sec-analysis,sec:cons}), we shall now discuss some immediate consequences of these results.

(\arabic{count2})\stepcounter{count2} 
The fact that certain non-local theories imply Bell inequalities first of all illustrates   
that \emph{Bell inequalities are not locality conditions} in the sense that, if a probability distribution obeys a Bell inequality, it must be local.
In the discussion, Bell inequalities are so closely linked to locality that one could have this impression. 
Of course, Bell's argument never really justified that view, for the logic of the standard Bell argument is that local factorization 
(given measurement independence and perfect correlations) 
is merely \emph{sufficient} (and not necessary) for Bell inequalities. 
The association between Bell inequalities and locality might have arisen from the fact that for a long time local factorization had been the \emph{only} product form which has been shown to imply Bell inequalities. Given only this information, it was at least possible  (though unproven) that the holding of Bell inequalities implies locality.  
However, since we have shown that some weakly non-local classes in general imply Bell inequalities and since one can easily find examples of strongly non-local distributions that conform to Bell inequalities, 
it has become explicit that this is not true. 
Not all probability distributions obeying Bell inequalities are local.% 
\footnote{{} 
Note that this result is not in conflict with Fine's insight  \citeyearpar{Fine1982a} that an empirical probability distribution obeying a Bell inequality is equivalent with the existence of at least one \emph{local} hidden probability distribution that implies the empirical distribution in question  (\enquote{local stochastic hidden variable model}). 
Notwithstanding, my claim that not every hidden probability distribution which 
obeys a Bell inequality is local can nevertheless be true because besides the local hidden probability distribution there can be non-local hidden probability distributions that imply the empirical distribution in question. 
Combining the two insights, for any non-local distribution, which implies Bell inequalities, there exists a local distribution such that the two share the same empirical distribution. Since weakly non-local distributions necessarily imply Bell inequalities this makes explicit that they cannot even come closer to violating the Bell inequalities than local distributions do. 
}

(\arabic{count2})\stepcounter{count2} 
The conclusions of the new Bell argument, which we have derived, are considerably stronger than those of previous versions. 
We have shown that the violation of Bell inequalities not only excludes local theories but also weakly non-local$^\alpha$ and weakly non-local$^\beta$ ones. 
In contrast, the conclusion of the standard Bell argument only forbids local theories and allows for \emph{all} non-local ones, including the weakly non-local classes that we have shown to imply Bell inequalities. 
In this sense, {the usual constraint} following from the standard Bell argument, 
\emph{is inappropriately weak}. 
While this is not to say that the standard argument is logically incorrect, it does mean that its conclusion is not as tight as it could be. 
We should keep in mind that any argument based on this standard conclusion, especially Jarrett's analysis, proceeds from a mixture of classes that can violate Bell inequalities with classes that imply them---and therefore might yield misleading results. 

(\arabic{count2})\stepcounter{count2}
The same is not true of our new result: All classes that it allows, all strongly non-local classes, can violate Bell inequalities. 
For this reason it is impossible to strengthen Bell's argument in such a way as to rule out more classes of probability distributions than we have ruled out here. 
In this sense, we can say that if our considerations have been correct and the typical background assumptions hold (measurement independence and nearly perfect (anti\=/)correlations), by our systematic approach we can be sure that the conclusions from the new Bell argument are the \emph{strongest possible consequences of the violation of Bell inequalities on a qualitative probabilistic level}. 
Note that this is not to say that further classes might not be ruled out due to other criteria, maybe due to their incompatibility with relativity or the like. 
The label \enquote{qualitative probabilistic} indicates that we have only referred to classes of probability distributions defined by their probabilistic dependences and independences without referring to quantitative features (or to qualitative sub-probabilistic features, see fn.~\ref{fn-subprob}).  

It might be interesting to make explicit how we have arrived at this strong conclusion. Especially, our considerations in this paper have two important features that preclude future strengthenings of the argument to rule out more classes. First, the central methodological procedure of our argument was to consider \emph{all logically possible} classes of probability distributions. Hence, any probability distribution that conceivably might describe an EPR/B experiment must fall under one of the classes in our systematic overview (cf. \autoref{table-classes}). 
For this reason, we can be sure that we have not overlooked any probability distribution for the EPR/B experiment. There simply are no probability distributions left that might bring in surprise; we have captured them all. 

A second important feature is that our argument provides sufficient \emph{and} necessary conditions for classes to imply Bell inequalities. By stating that local classes imply Bell inequalities, former arguments typically have only provided sufficient criteria. This left open the possibility that there are further classes implying the inequalities---and, indeed, here we have found that many non-local classes, viz. the weakly non-local ones, do as well. 
On the other hand, by explicitly showing that the remaining classes, the strongly non-local classes, can violate the inequalities \bki (see the proofs of \Cref{th:ind-pcorr,th:ind-npcorr}, where we have constructed explicit examples of distributions in those classes that violate the inequalities)\eki{}, we have precluded that future arguments might show one of the strongly non-local classes to imply the inequalities as well. 
And if this argument, that proceeds on the qualitative probabilistic level of the classes and their product forms, is correct, \bki and the background assumptions we have presupposed hold\eki{}, we cannot entail a stronger claim on that level than that local and weakly non-local classes imply Bell inequalities while strongly non-local classes can violate them.

(\arabic{count2})\stepcounter{count2}
The latter claim also reveals a certain limitation of the argument presented here. It emphatically does not say that strongly non-local classes violate Bell inequalities; it only says that strongly non-local classes \emph{can} violate Bell inequalities, meaning that some of the strongly non-local distributions do violate the inequalities while others do not. 
In fact, one can explicitly find examples for  probability distributions in each of the strongly non-local$^\alpha$ classes (H$_1^\alpha$)--(H$_{14}^\alpha$) (as well as in the strongly non-local$^\beta$ classes (H$_1^\beta$)--(H$_{14}^\beta$)) which \emph{obey} Bell inequalities---and these distributions clearly could be ruled out by more precise arguments.  
However, belonging to the same class, discerning strongly non-local classes which violate the inequalities from those that obey them clearly cannot be made on a \emph{qualitative probabilistic} level. 
Any improvement of the argument must refer to the 
specific \emph{quantitative} features of the probability distribution in question (or  qualitative features on a sub-probabilistic level%
\footnote{{}
\label{fn-subprob}
\citet[sec.~3.3.2]{Seevinck2008}, generalizing an insight of \citet{Fahmi2002}, provides an example of 
a subclass of the strongly non-local$^\alpha$ class $(\text{H}^\alpha_1)$ 
that implies Bell inequalities. The defining feature of the subclass is that the factors of the product form that defines $(\text{H}^\alpha_1)$---$P(\alpha \beta | a b \lambda) = P(\alpha | \beta a b \lambda) P(\beta | a b \lambda)$---have the specific functional form $P(\alpha | \beta a b \lambda) = f(\alpha, a,\lambda) x(\beta, b,\lambda)$ and $P(\beta | a b \lambda) = g(\beta, b,\lambda) y(a,\lambda)$, and with these it is easy to derive Bell inequalities. 
Note that specifying the conditional probabilities 
in this way leaves the product form and hence the central (in\=/)dependences untouched because the functions $f, g, x, y$ are not themselves probabilities, but rather determine values of probabilities. 
Therefore, it seems appropriate to say that indicating the functional form of conditional probabilities is \enquote{a qualitative characterization on a \emph{sub}-probabilistic level} and that the possibility of such additional characterizations does not speak against my claim that I have provided the strongest possible consequences of the violation of Bell inequalities on a qualitative probabilistic level. 
In contrast to the probabilistic level, which is connected to laws and causation and thereby to the question of locality, it seems unclear, however, whether this sub-probabilistic level has a physical or metaphysical meaning. 
}%
),
so there is no general claim that can be made on the basis of the mere product form; the product form of any strongly non-local class alone does not determine whether Bell inequalities hold or fail.%

It follows that the consequence of my stronger Bell argument, that the quantum world can only be described correctly by a theory falling under  a strongly non-local class, is only a \emph{necessary} condition for violating Bell inequalities; it is not a sufficient one. (Note the difference between conditions for \emph{violating} Bell inequalities and conditions for \emph{not implying} them; we have provided necessary and sufficient conditions for the latter but only necessary ones for the former.) Sufficient criteria to violate Bell inequalities would have to involve conditions for the \emph{strength} of the correlations. 
A common measure for how strong a correlation is, is mutual information, so information theoretic works which derive numerical values for how much mutual information has to be given in order to violate Bell inequalities, provide an answer to that question (cf. \citealt[ch.~6]{Maudlin1994}  and \citealt{Pawlowski2010}). 
These are important works, which can further sharpen the constraints for quantum non-locality following from EPR/B experiments. 
Such quantitative improvements, however, do not count against my claim here that the conclusion of my new stronger Bell argument  
captures the strongest possible consequences of the violation of Bell inequalities on a \emph{qualitative probabilistic} level.

\section{Analyzing the conclusions}   
\label{sec-analysis} 

Having strengthened Bell's argument to a more informative conclusion, we now have to make precise what this new, stronger constraint for quantum non-locality amounts to. 
\citet{Jarrett1984} proved that the standard probabilistic constraint for quantum non-locality following from the usual Bell argument, the failure of local factorization, is equivalent to \enquote{outcome dependence or parameter dependence}. 
The very idea of Jarrett's analysis is that 
a \emph{complex} dependence condition (the failure of local factorization)
can be analysed by 
\emph{pairwise} statistical dependences (outcome dependence and parameter dependence). Our new constraint for quantum non-locality, the failure of local and weakly non-local product forms, 
is a complex dependence condition as well. 
So we can apply Jarrett's idea to our new case and understand \enquote{analysis} as providing an expression in terms of pairwise probabilistic dependences which is equivalent to the new constraint. 
Providing an analysis of the new stronger constraint will make explicit the differences to the usual constraint. 

I first recall shortly Jarrett's analysis (\autoref{sec:Jarretts-anal}) and 
introduce an appropriate set of independences, which can serve as analysantia of the new constraint (\autoref{terminology}). Then I shall develop an analysis for each of the classes (H$_i^\alpha$) (\autoref{sect-analysis-classes})
and subsequently of the new probabilistic constraint for quantum non-locality (\autoref{qnl-pd}).

\subsection{Jarrett's analysis}
\label{sec:Jarretts-anal}

\citet{Jarrett1984} had the idea that one can be more explicit about the probabilistic nature of quantum non-locality by analyzing the probabilistic statement local factorization (\ref{local-causality}) in terms of pairwise conditional probabilistic  independences. 
By a \enquote{pairwise conditional probabilistic independence} I mean the fact that a random variable $\zv x$ is independent of another $\zv y$ given a conjunction of further variables $\zv z$. This is said to be true if and only if
\begin{equation}
\label{independence} \forall \; x,y,z : \quad P(x|yz) = P(x|z),
\end{equation}
where $x,y,z$ denote the values of the variables $\zv x, \zv y, \zv z$.  
The independence is noted as $I(\zv x, \zv y | \zv z)$. If, however, \eqref{independence} fails, because there is at least one set of values for which $P(x|yz)≠P(x|z)$, the variables $\zv x$ and $\zv y$ are called \enquote{dependent given $\zv z$}, and this probabilistic dependence is noted as $\neg I(\zv x, \zv y | \zv z)$. 

Jarrett uses three pairwise independences: \enquote{outcome independence} is defined as $I(\zv \alpha, \zv \beta | \zv a \zv b \zv \lambda)$ 
and \enquote{parameter independence} as a conjunction of two independences,  \linebreak[4] \mbox{$I(\zv \alpha, \zv b | \zv a \zv \lambda) \wedge I(\zv \beta, \zv a | \zv b \zv \lambda)$}. (Originally, Jarrett denotes these independences as \enquote{completeness} and \enquote{locality} respectively, but we shall use the  now established names.) 
Jarrett proved mathematically that 

\begin{enumerate}
\+[(P7)] Local factorization is equivalent to the conjunction of outcome independence and parameter independence:  
\begin{equation}
(\text{LF}) \leftrightarrow I(\zv \alpha, \zv \beta | \zv a \zv b \zv \lambda) \wedge I(\zv \alpha, \zv b | \zv a \zv \lambda) \wedge I(\zv \beta, \zv a | \zv b \zv \lambda)
\end{equation} 
\end{enumerate}

From (C1), the conclusion of the standard Bell argument that local factorization fails, and (P7) he concluded that 
\begin{enumerate}
\+[(C2)] Outcome dependence or  parameter dependence holds:  
\begin{equation}
\neg I(\zv \alpha, \zv \beta | \zv a \zv b \zv \lambda) \vee \neg I(\zv \alpha, \zv b | \zv a \zv \lambda) \vee \neg I(\zv \beta, \zv a | \zv b \zv \lambda)
\end{equation}
\end{enumerate}

\noindent which is the \emph{analysis of the probabilistic constraint following from the standard Bell argument} (\enquote{Jarrett's analysis}).

\subsection{Different kinds of parameter independences}
\label{terminology}

Aiming to analyze the new probabilistic constraint for quantum non-locality we first have 
to get an overview which concepts can play the role of the analysantia. In \autoref{table-independences} I introduce 
those nine pairwise independences which will be relevant. 
Among the relevant independences we find usual outcome independence, $I(\zv \alpha, \zv \beta | \zv a \zv b \zv \lambda)$, as well as $I(\zv \alpha, \zv b | \zv a \zv \lambda )$, one independence of the conjunction which is  usually called \enquote{parameter independence}. Here we see a first problem with the standard names: How shall we call the latter if its conjunction with $I(\zv \beta, \zv a | \zv b \zv \lambda )$ is called \enquote{parameter independence}? My table introduces new terminology, which tries to stay as close to the standard names as possible, but obviously further qualifications are needed. 
My suggestion is to continue to use the name \enquote{parameter independence} for all independences between an outcome and its distant parameter (i.e. setting), but to add the outcome in question, namely \enquote{$\alpha$-parameter independence} or \enquote{$\beta$-parameter independence} respectively. Further differentiation in the nomenclature is required by the fact that there is another $\alpha$-parameter independence in the table, $I(\zv \alpha, \zv b | \zv \beta \zv a \zv \lambda )$, which differs from the one already mentioned in the 
conditional variables (it additionally includes the outcome~$\zv \beta$).  Such independences of the same type but with different conditional variables are different independences and are in general logically independent of another: One can hold or not irrespective of whether the other does or does not. (One can show that only for more than two pairwise independences logical restrictions appear, see the semi-graphoid axioms \enquote{contraction} and \enquote{intersection} in \citealt[11]{Pearl2000}.) I discern them by indices, e.g. the former is called \enquote{$\alpha$-parameter independence$_2$}, the latter \enquote{$\alpha$-parameter independence$_1$}. 
Of course, there are further $\alpha$-parameter independences (namely those conditional on $\zv \beta \zv \lambda$ and $\zv \lambda$), which, however, do not play any role for the analysis here. 

Similarly to \enquote{parameter independences} I define \enquote{\emph{local} parameter independences} (see \autoref{table-independences}), which instead of the independence of an outcome on its distant parameter (e.g. $\zv \alpha$ on $\zv b$) claim the independence of an outcome on its \emph{local} parameter (e.g. $\zv \alpha$ on $\zv a$). 
Besides these new names I have also introduced short labels for each independence, which we will mainly use in the following.  

\renewcommand{\arraystretch}{1.4}
\begin{table}[htbp]
\caption{Definition of conditional independences}
\begin{center}
\begin{tabular}{>{$} l <{$}ccc}
\text{independence} & standard name & new name  & label\\ \hline \hline
I(\zv \alpha, \zv \beta | \zv a \zv b \zv \lambda )  & outcome independence & outcome independence$_1$ & (OI$_1$)\\ \hline
I(\zv \alpha, \zv b | \zv \beta \zv a \zv \lambda ) & -- & $\alpha$-parameter independence$_1$ & $(\text{PI}_1^\alpha)$ \\
I(\zv \alpha, \zv b | \zv a \zv \lambda )  & [part of] parameter ind. & $\alpha$-parameter independence$_2$ & $(\text{PI}_2^\alpha)$ \\
I(\zv \beta, \zv a | \zv \alpha \zv b \zv \lambda ) & -- & $\beta$-parameter independence$_1$ & $(\text{PI}_1^\beta)$ \\
I(\zv \beta, \zv a | \zv b \zv \lambda )  & [part of] parameter ind. & $\beta$-parameter independence$_2$ & $(\text{PI}_2^\beta)$ \\ \hline
I(\zv \alpha, \zv a | \zv \beta \zv b \zv \lambda ) & -- & $\alpha$-local parameter independence$_1$ & $(\text{LPI}_1^\alpha)$ \\
I(\zv \alpha, \zv a | \zv b \zv \lambda ) &  -- & $\alpha$-local parameter independence$_2$ & $(\text{LPI}_2^\alpha)$ \\
I(\zv \beta, \zv b | \zv \alpha \zv a \zv \lambda ) & -- & $\beta$-local parameter independence$_1$ & $(\text{LPI}_1^\beta)$ \\
I(\zv \beta, \zv b | \zv a \zv \lambda ) &  -- & $\beta$-local parameter independence$_2$ & $(\text{LPI}_2^\beta)$ 
\end{tabular}
\end{center}
\label{table-independences}
\end{table}%

Having introduced these new concepts we are now in a position to clearly see one of the sources of  confusion in the standard discussion. \enquote{Outcome dependence or parameter dependence} does \emph{not necessarily} mean that if you accept outcome dependence you can avoid parameter dependence in the sense of \emph{any kind} of dependence of an outcome on its distant parameter (conditional on whatever variables). The slogan just says that in this case you can avoid parameter dependence in the usual sense of $\neg(\text{PI}^\alpha_2) \vee \neg(\text{PI}^\beta_2)$,
 while other kinds of parameter dependences like $\neg$(PI$^\alpha_1$) might still hold. Indeed the analysis of the new constraint will yield that at least one of the two parameter dependences $\neg$(PI$^\alpha_1$) and $\neg$(PI$^\beta_2$) \emph{must} hold  (as well as at least one of $\neg(\text{PI}^\beta_1)$ and $\neg(\text{PI}^\alpha_2)$). Parameter dependence in this broader sense cannot be avoided but will turn out to be  a necessary condition for violating the Bell inequalities.

\subsection{Analysis of the classes} 
\label{sect-analysis-classes} 

With these pairwise independences we can now attempt to analyse each class of probability distributions. For the analysis of the classes (H$_i^\alpha$) in Table \autoref{table-classes} we shall need five independences from \autoref{table-independences} (the other four independences plus outcome independence$_1$ are only required for the analysis of the classes (H$_j^\beta$); see below). 
We have noted the corresponding dependences in the bottom line of \autoref{table-classes}, i.e. each dependence is associated with one of the columns II--VI. The idea is that the dependence holds in a class if the column of that class contains \enquote{1}. Otherwise, i.e. if it contains \enquote{0}, the corresponding independence holds.  
The result of this analysis is stated by the following theorem: 
\begin{restatable}{theorem}{theoremf} 
\label{th:analysis}
Each class $(\text{\normalfont H}_i^\alpha)$ in \autoref{table-classes} is equivalent to the conjunction of 
the specific pattern of 
dependences and independences (see the bottom line of the table, labelled \enquote{Analysis}) indicated by 1's or 0's, respectively, in columns II--VI. 
\end{restatable}
\vspace{-0.3cm} \hfill (Proof in Appendix \ref{sec:proof-theorem3}) 
\vspace{0.4cm} 

The theorem means that each pattern of dependences and independences corresponds to exactly one of the classes, e.g. 
\begin{equation}
(\text{H}_7^\alpha) \Leftrightarrow 
\neg(\text{OI}_1) \wedge \neg(\text{PI}^\alpha_1) \wedge \neg(\text{LPI}^\alpha_1) \wedge (\text{PI}^\beta_2) \wedge (\text{LPI}^\beta_2).
\end{equation}
One can see from the table that \emph{each of the five independences corresponds to exactly one of the five variables in the conditionals of the factors}: If a certain independence \emph{holds}, the corresponding variable does \emph{not} appear (and vice versa), and if a certain independence \emph{fails}, the corresponding variable \emph{does appear} (and vice versa). Specifically, if (OI$_1$) holds, the first factor of the hidden joint probability does not involve the other outcome $\zv \beta$ (and vice versa), and if it does not, the first factor includes it (and vice versa). Similarly, ($\text{PI}^\alpha_1$) and ($\text{LPI}^\alpha_1$) correspond to the distant and the local parameter in the \emph{first} factor respectively, while ($\text{PI}^\alpha_2$) and ($\text{LPI}^\alpha_2$) are linked to the distant and the local parameter  in the \emph{second} factor respectively. So the holding or failure of each of the five independences has a very well defined impact on the product form of the hidden joint probability (and vice versa), and the conjunction of \emph{all} independences which hold according to a certain probability distribution determines its product form, i.e. its  class (and vice versa).

We should note that each factorization condition is equivalent with the pattern of independences (without the dependences). In contrast, we have defined \emph{classes} by a factorization condition \emph{and} the assumption that the factorization condition in question is minimal, i.e. cannot be reduced by dropping further variables; and in order to capture the minimality claim for a given class also the corresponding dependences are required (for details see the proof of \autoref{th:analysis}).%
\footnote{{}
This explains why our analysis of the 
class defined by local factorization,
$(\text{H}_{29}^\alpha) \Leftrightarrow (\text{OI}_1) \wedge (\text{PI}^\alpha_1) \wedge \neg(\text{LPI}^\alpha_1) \wedge (\text{PI}^\beta_2) \wedge \neg(\text{LPI}^\beta_2)$, involves dependences besides independences, though Jarretts analysis of local factorization (as being equivalent to  $(\text{OI}_1) \wedge (\text{PI}^\alpha_2) \wedge (\text{PI}^\beta_2)$)
only refers to independences. 
Furthermore, in Jarrett's analysis $(\text{PI}^\alpha_1)$ 
is replaced by $(\text{PI}^\alpha_2)$ (as compared to the analysis from \autoref{table-classes}); but since $(\text{OI}_1)$ holds, the replacement is equivalent: One can easily prove that $(\text{OI}_1) \wedge (\text{PI}^\alpha_1) \Leftrightarrow (\text{OI}_1) \wedge (\text{PI}^\alpha_2)$. 
} 

Mutatis mutandis, one finds the analysis of the classes $(\text{H}_i^\beta)$: 
\begin{corollary}
\label{cor:analysis}
Each class $(\text{\normalfont H}_i^\beta)$ is equivalent to the conjunction of 
the specific pattern of dependences and independences 
indicated by 1's or 0's, respectively, in columns II--VI of each line $i$ in \autoref{table-classes}, which then denote { $\neg(\text{\normalfont OI}_1)$,  $\neg(\text{\normalfont PI}^\alpha_1)$, $\neg(\text{\normalfont LPI}^\alpha_1)$, $\neg(\text{\normalfont PI}^\beta_2)$, $\neg(\text{\normalfont LPI}^\beta_2)$}. 
\end{corollary}

\noindent 
For instance: $(\text{H}_7^\beta) \Leftrightarrow \neg(\text{OI}_1) \wedge \neg(\text{PI}^\beta_1) \wedge \neg(\text{LPI}^\beta_1) \wedge (\text{PI}^\alpha_2) \wedge (\text{LPI}^\alpha_2)$.

\subsection{Analysis of the stronger conclusion}  
\label{qnl-pd}

We can now use the analysis of the single classes to analyze the new, stronger conclusion of the Bell argument. This will provide us with sufficient and necessary conditions for a class being able to violate Bell inequalities. 

We had found that quantum non-locality is the failure of all local$^\alpha$, weakly non-local$^\alpha$, local$^\beta$ and weakly non-local$^\beta$  classes (C1$^{\prime}$)
and that these classes are characterized by the fact that their constituting product forms involve \emph{at most one setting (parameter) in each of its factors}. 
Let us first give an analysis of the local$^\alpha$ and weakly non-local$^\alpha$ classes. By our analysis of the single classes (H$^\alpha_1$)--(H$^\alpha_{32}$) 
the first factor of the defining product form involves at most one parameter if and only if at least one of the independences $\alpha$-parameter independence$_1$, $(\text{PI}^\alpha_1)$, or $\alpha$-local parameter independence$_1$, $(\text{LPI}^\alpha_1)$, holds. 
Similarly, at most one parameter appears in the second factor if and only if $\beta$-parameter independence$_2$, $(\text{PI}_2^\beta)$, or $\beta$-local parameter independence$_2$, $(\text{LPI}_2^\beta)$, hold. 
So we have found the following equivalence:

\begin{enumerate}
\+[(P7$^\prime$a)] \label{prem:anal-alpha} The disjunction of local$^\alpha$ and weakly non-local$^\alpha$ classes is equivalent to 
the fact that $\zv \alpha$ is independent$_1$ of at least one parameter and $\zv \beta$  is independent$_2$ of at least one parameter: 
\end{enumerate}
\begin{equation}
\label{analysis-non-local-alpha}
\Big( \bigvee_{i = 15}^{32} (\text{H}_i^\alpha) \Big) \leftrightarrow \bigg[ \Big((\text{PI}^\alpha_1) \vee (\text{LPI}^\alpha_1)\Big) \wedge \Big((\text{PI}^\beta_2) \vee (\text{LPI}^\beta_2)\Big) \bigg] \notag
\end{equation}

In a very similar way 
one can find an analysis for the $\beta$-classes 
(remember the table which is symmetric to \autoref{table-classes} in swapping the outcomes and the parameters and apply all considerations mutatis mutandis):
\begin{enumerate}
\+[(P7$^\prime$b)] \label{prem:anal-beta} The disjunction of local$^\beta$ and weakly non-local$^\beta$ classes is equivalent to the fact that $\zv \beta$ is independent$_1$ of at least one parameter and $\zv \alpha$  is independent$_2$ of at least one parameter:
\end{enumerate}
\begin{equation}
\label{analysis-non-local-beta}
\Big( \bigvee_{i = 15}^{32} (\text{H}_i^\beta) \Big) \leftrightarrow \bigg[ \Big((\text{PI}^\beta_1) \vee (\text{LPI}^\beta_1)\Big) \wedge \Big((\text{PI}^\alpha_2) \vee (\text{LPI}^\alpha_2)\Big) \bigg] \notag
\end{equation}

Since according to the conclusion of the stronger Bell argument (C1$^{\prime}$)  the disjunction of all local$^\alpha$, weakly non-local$^\alpha$, local$^\beta$ and weakly non-local$^\beta$ classes \emph{fails},  the negation of the disjunction of (P7$^\prime$a) and (P7$^\prime$b) finally yields the analysis of (C1$^{\prime}$):

\begin{enumerate}
\+[(C2$^\prime$)] 
\textbf{Probabilistic Bell contextuality}: $\zv \alpha$ depends$_1$ on both parameters or $\zv \beta$ depends$_2$ on both parameters and  $\zv \beta$ depends$_1$ on both parameters or $\zv \alpha$ depends$_2$ on both parameters:
\end{enumerate}
\vspace{-0.35cm}
\begin{align}
\label{analysis-qnl} 
& \bigg[ \Big(\neg (\text{PI}^\alpha_1) \wedge \neg(\text{LPI}^\alpha_1)\Big) \vee \Big(\neg(\text{PI}^\beta_2) \wedge \neg(\text{LPI}^\beta_2)\Big)\bigg]  \notag \\
 \wedge & \bigg[ \Big(\neg (\text{PI}^\beta_1) \wedge \neg(\text{LPI}^\beta_1)\Big) \vee \Big(\neg(\text{PI}^\alpha_2) \wedge \neg(\text{LPI}^\alpha_2)\Big) \bigg] \notag 
\end{align}

While the conclusion (C1$^{\prime}$) of the stronger Bell argument was in terms of classes, 
here we have the equivalent expression, the analysis, in terms of pairwise independences. 
It is a non-trivial logical expression whose meaning and implications are not immediately  obvious. A first understanding might be attained by making explicit how this analysis of the conclusion (C1$^{\prime}$) is also an analysis of the equivalent conclusion (C1$^{\prime\prime}$), which says that the 
conjunction of strongly non-local$^\alpha$ and strongly non-local$^\beta$ classes holds.
These classes were characterized by the fact that at least one of the factors in each product form must involve both parameters and this is exactly what (C2$^\prime$) says: The first term in the first disjunction, $\neg (\text{PI}^\alpha_1) \wedge \neg(\text{LPI}^\alpha_1)$ (\enquote{$\alpha$-double parameter dependence$_1$}), guarantees a dependence on both parameters in the first factor of the product forms (H$^\alpha_i$), the second term in the first disjunction, $\neg(\text{PI}^\beta_2) \wedge \neg(\text{LPI}^\beta_2)$ (\enquote{$\beta$-double parameter dependence$_2$}), implies a similar fact for the second factor of these  forms,
and analogously, the second disjunction entails a dependence on both parameters in at least one of the factors of the product forms (H$^\beta_i$) (and vice versa).

So the analysis involves double parameter dependences for each outcome in two different forms, either conditional on all other variables (double parameter dependence$_1$) or conditional on all other variables excluding the other outcome (double parameter dependence$_2$). 
By the logic of the expression there are four possible combinations: 
\begin{align}
(\text{C2}^\prime) \Leftrightarrow &\bigg[ \neg (\text{PI}^\alpha_1) \wedge \neg(\text{LPI}^\alpha_1) 
\wedge  \neg (\text{PI}^\beta_1) \wedge \neg(\text{LPI}^\beta_1) \bigg] \notag \\
\vee 
& \bigg[ \neg (\text{PI}^\alpha_1) \wedge \neg(\text{LPI}^\alpha_1) 
\wedge \neg(\text{PI}^\alpha_2) \wedge \neg(\text{LPI}^\alpha_2) \bigg]  \notag \\
\vee
& \bigg[ 
\neg (\text{PI}^\beta_1) \wedge \neg(\text{LPI}^\beta_1) \wedge 
\neg(\text{PI}^\beta_2) \wedge \neg(\text{LPI}^\beta_2) \bigg] \notag \\
\vee 
& \bigg[ 
\neg(\text{PI}^\alpha_2) \wedge \neg(\text{LPI}^\alpha_2) 
\wedge \neg(\text{PI}^\beta_2) \wedge \neg(\text{LPI}^\beta_2) \bigg] \notag 
\end{align}
This makes explicit that either one of the outcomes is both double parameter dependent$_{1}$ and double parameter dependent$_{2}$, 
or there are mixed cases in which both outcomes are double parameter dependent (each in one of the two forms). 
The Bell argument does not say which of these four possibilities is correct, but there is one fact that one cannot avoid in any of these cases:
\begin{enumerate}
\+[(C3)]  
\textbf{Double parameter dependence}: At least one of the outcomes depends probabilistically on both parameters (in at least one of the forms double parameter dependence$_1$ or double parameter depen\-dence$_2$). 
\end{enumerate}
(Note that (C2$^\prime$) implies (C3), but not vice versa.) 
So we have found two results: the precise probabilistic analysis (C2$^\prime$) of the new stronger conclusion (C1$^\prime$)
and a general feature of and deriving from that analysis (C3). 
Since the conclusion is a necessary condition for EPR/B correlations (if measurement independence and nearly perfect (anti\=/)correlations hold),  
double parameter dependence of at least one of the outcomes, which is implied by quantum non-locality, is a \emph{necessary} condition for EPR/B correlations as well: Whenever we find that EPR/B correlations hold, 
double parameter dependence (C3) must hold as well. 
So given that measurement results in our world yield EPR/B correlations 
(and assuming measurement independence), we can be sure that at least one of the outcomes depends both on the local as well as on the distant parameter. 

On the other hand, since here we have derived an analysis of a conclusion following from the violation of Bell inequalities, 
neither the analysis (C2$^\prime$) nor its consequence double parameter dependence (C3), is \emph{sufficient} for the violation of Bell inequalities. 
If, according to a certain probability distribution, an outcome depends on both parameters in the sense of (C2$^\prime$) the correlations between the two wings \emph{might} be strong enough to violate Bell inequalities---but they need not be (see paragraph (4) in \autoref{sec:strong-poss-cons}). 
However, we also know (from that section) that the conclusion of the argument is  \emph{sufficient} for a class \emph{to be able} to violate Bell inequalities, in the sense that if a class fulfills the conditions mentioned in the conclusion, there is at least one probability distribution in that class which violates the inequalities.  Hence, the analysis (C2$^\prime$) is also a sufficient condition for a class to be able to violate Bell inequalities (but its implication (C3) is not).

\section{Discussion II}
\label{sec:cons}

\subsection{Pitfalls of Jarrett's analysis} 
\label{sec:Jarrett}

Similar to how Jarrett analysed the conclusion of the standard Bell argument by pairwise (in\=/)dependences, 
we have analysed  the conclusion of the stronger Bell argument. 
The essential difference between his analysis and the one presented here is that the analysandum of the former,
the failure of local factorization, as opposed to the analysandum of the latter, the failure of local and weakly non-local classes, is a considerably weaker concept. From a logical point of view, there is nothing wrong about Jarrett's analysis, and understood literally, it is perfectly right. However, the situation with the different dependences is confusing and it is not easy to understand it correctly, so let me point to the pitfalls of the situation. 

\setcounter{mycounter}{1} 
(\arabic{mycounter}\stepcounter{mycounter}) The main message of our new result is that given EPR/B correlations and measurement independence \emph{one cannot avoid some kind of dependence between at least one of the outcomes and both parameters (C3)}. This is a necessary condition for the violation of Bell inequalities according to my new analysis. 
Jarrett's analysis, in contrast, 
does not bring out this essential requirement: 
From his result \enquote{outcome dependence or parameter dependence} one cannot see that, necessarily, there must be some kind of double parameter dependence.
This seems to be due to the fact that he analyses a weaker concept.

(\arabic{mycounter}\stepcounter{mycounter}) \label{par:Jarr2}
Jarrett's result might even seem to contradict our new result if one understands it to claim 
that one can avoid \emph{any} dependence between an outcome and its distant setting when outcome dependence holds. 
This, however, is not what it literally says (and if this interpretation were the correct reading of Jarrett's result, it would indeed be plainly wrong). 
For \enquote{parameter dependence} here does not mean \emph{any} kind of parameter dependence but a very specific kind, namely parameter dependence$_2$, and 
saying that one can avoid this specific kind does not mean that there is no dependence of the outcomes on their distant parameters at all. 
Our presentation of different kinds of parameter dependences 
(see \autoref{table-independences}) has made explicit that parameter dependence$_2$ is only one among several kinds, all of which might hold if parameter dependence$_2$ fails. 
So in a careful literal reading Jarrett's result does not contradict our result  (C2$^\prime$), that one can avoid parameter dependence$_2$ only if  parameter dependence$_1$ holds. 

However, if one is not aware of the different kinds of parameter dependences (and in the debate so far other kinds than parameter dependence$_2$ do not play a significant role)%
\footnote{{}
\citet{Maudlin1994} points to parameter dependence$_1$ and possible conceptual confusions, but he does not explain in detail the relation with parameter dependence$_2$ and outcome dependence. 
}
the slogan \enquote{outcome dependence or parameter dependence} is 
\emph{liable to be misunderstood to its non-literal false sense}, that one can avoid \emph{any} kind of parameter dependence if outcome dependence holds. 
In fact, it seems that Jarrett's result summarized by Shimony's slogan has to a large extent received this infelicitous interpretation. 
There is a bunch of literature about quantum non-locality (on any level, whether causal, spatio-temporal or metaphysical) which is based on Jarrett's distinction, and which discusses in detail what outcome dependence or parameter dependence would amount to, the preferred solution being \enquote{outcome dependence without parameter dependence}. In many cases the reasoning makes only sense, if one assumes the non-literal false reading. 
To give an example, 
\enquote{outcome dependence without parameter dependence} has often been interpreted to show that there is only a non-local metaphysical connection between the 
outcomes (but no connection between an outcome and its distant setting) and that connection has been given different names 
 (\enquote{passion at-a-distance}, \citealt{Shimony1984} and \citealt{Jarrett1989}; \enquote{relational holism}, \citealt{Teller1989}; \enquote{nonseparability}, \citealt{Howard1989}). 
These scenarios tacitly assume that there is no dependence on the distant parameter at all or at least no such dependence which is relevant. 
Both these assumptions are proven wrong by our new result that, necessarily, there must be some kind of parameter dependence.

\begin{figure}[htbp]
\noindent \centering
\begin{minipage}[t]{0.65\linewidth}
\epsfig{file=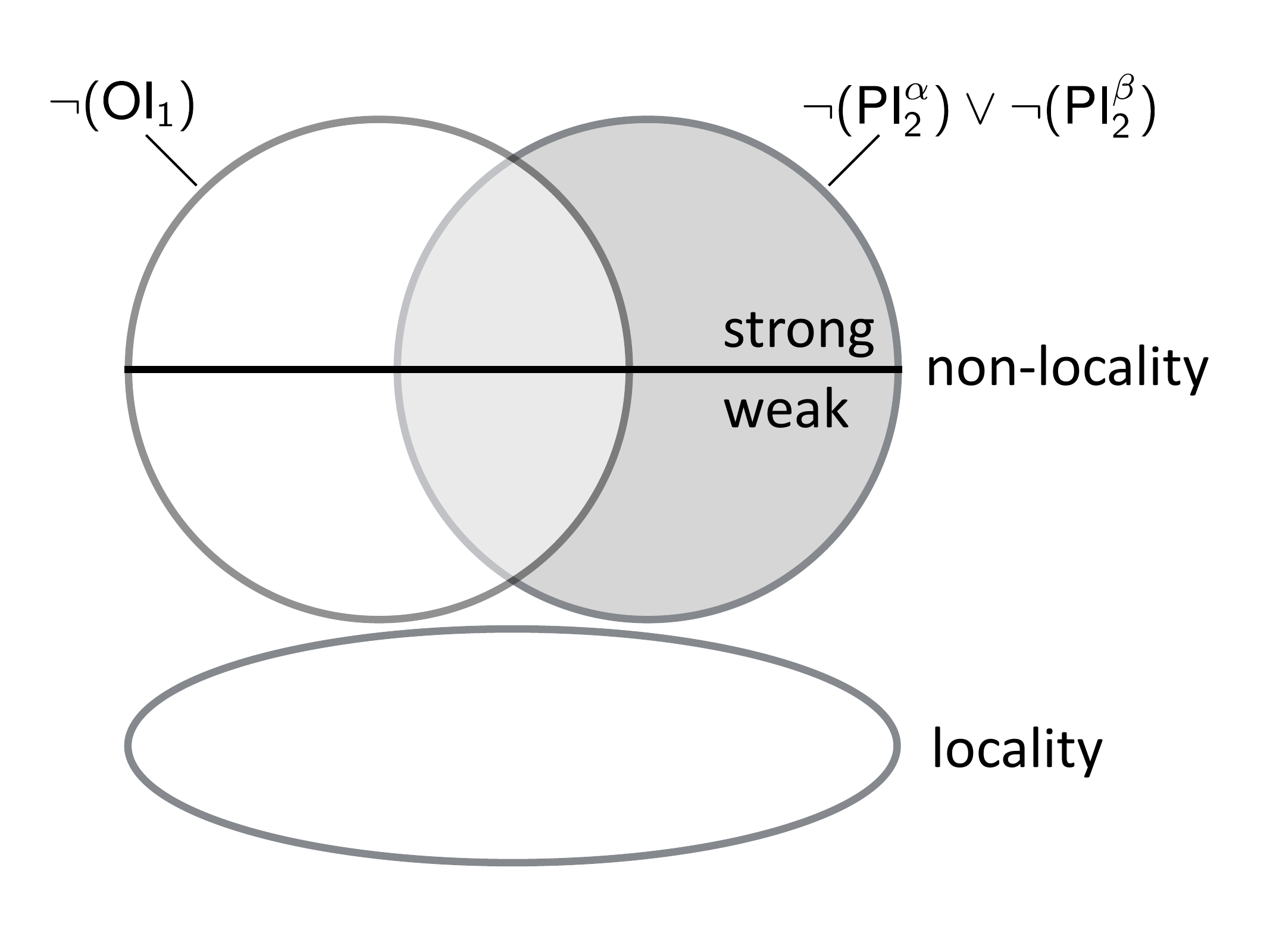,width=\linewidth}  
\end{minipage}
\caption{Outcome dependence$_1$ and parameter dependence$_2$ vs. weak and strong non-locality. 
} 
\label{image-sets}
\end{figure}

(\arabic{mycounter}\stepcounter{mycounter}) 
Another unfortunate feature of Jarrett's analysis 
is that 
it is formulated with \emph{inappropriate categories}. 
More specifically, there is no logical expression involving only his categories \enquote{outcome dependence} and \enquote{parameter dependence} that might tell apart classes that can violate Bell inequalities from classes that cannot. 
Consider the partition of the probability distributions according to the dependences in Jarrett's analysis (\autoref{Jarretts-classes}). There are four classes, which I call \enquote{Jarrett's classes} and label as $(\text{J}_1)$--$(\text{J}_4)$. 
Any of the 32 possible classes (H$^\alpha_i$) from \autoref{table-classes} (as well as any of the classes (H$^\beta_i$) from the complementary scheme) must fall into one of Jarrett's coarse-grained classes. While the local classes belong to (J4), any of the classes (J1)--(J3) includes both weakly and strongly non-local classes. So Jarrett's non-local classes 
\emph{mix classes of probability distributions which can violate Bell inequalities with such which cannot} (see \autoref{image-sets}). 
\emph{They do not cut the probability distributions at their natural joints.} 
The central reason for this seems to be that Jarrett's analysandum, the failure of local factorization, already mixes classes of the these different types. 

 \begin{table}[htbp]
\begin{small} 
\caption{Jarrett's classes of possible probability distributions}
\label{Jarretts-classes}
\begin{center}
\renewcommand{\arraystretch}{1.3}
\begin{tabular}{c c c l l}
\hline
Label & $\neg$(OI$_1$) & $\neg$(PI$^\alpha_2$) $\vee \neg$(PI$^\beta_2$)   & Notes &  \\ \hline \hline

($\text{J}_1$) &  1 & 1 &  & \multirow{3}{2cm}{quantum non-locality} \\ \cline{1-4}

($\text{J}_2$) & 0 & 1 & Bohm &   \\  \cline{1-4}

($\text{J}_3$) & 1 & 0  &  QM &  \\ \hline

($\text{J}_4$) & 0 & 0  & & locality\\  \hline
\end{tabular}
\end{center}
\end{small}
\end{table}

So it seems that a significant amount of the debate after Jarrett's paper which has focused on the question of the formal, physical and metaphysical differences between outcome dependence$_1$ and parameter dependence$_2$, in order to decide which of the two does hold, is misguided. \emph{\enquote{Outcome dependence or parameter dependence?} is just the wrong question if one wants to explore deeper into the nature of quantum non-locality}, because each of the two options subsumes classes of probability distributions which can and such which cannot violate Bell inequalities. Rather, the natural question, the new analysis shows, is which of the outcomes is double parameter dependent and whether it is double parameter dependent$_1$ or double parameter dependent$_2$. 

Since the matter with statistical dependences is abstract, 
let me illustrate this point by a rough analogy. To characterize quantum non-locality by saying that it is outcome dependent or parameter dependent is like characterizing those things that we call cars 
by the statement \enquote{cars have a radio or a combustion engine}. While there is nothing wrong about this statement from a logical point of view, it is not a statement that cuts the problem at its natural joints and that we should go on to work with. 
If it were the only characterization of cars that we had---like it was with Jarrett's characterization of quantum non-locality since 1984 until now---we could easily be misled about their nature. 
Clearly, a more reasonable characterization would be to say that cars have a combustion engine or an electric engine, which also brings out the essential fact that cars must have some kind of engine. And this is what we have done in this paper by pointing out that quantum non-locality either requires parameter dependence$_1$ or parameter dependence$_2$. 

(\arabic{mycounter}\stepcounter{mycounter}) A defender of Jarrett's analysis might respond, that the distinction is not as unnatural as I claim here, because it discerns theories that are incompatible with relativity, due to the fact that they imply the possibility of superluminal signaling (parameter dependent$_2$ theories), from those that are not (outcome dependent$_1$ theories). 
We have said in the introduction that we cannot discuss the matter of compatibility with relativity in this paper, but whatever the truth concerning this issue is:%
\footnote{{} While it has been contested that parameter dependent$_2$ theories even in principle allow for superluminal signaling, \citet{Arntzenius1994} has pointed out that  parameter dependent$_2$ theories should nevertheless be excluded because they yield inconsistencies in certain loop-like setups in a relativistic space-time.} 
If the distinction between parameter dependent$_2$ theories and theories that are not parameter dependent$_2$ is important for the compatibility with relativity in any way,   
my new analysis is equally suited to discern them
because it involves parameter dependence$_2$ as well.
In our analogy: The defender of the \enquote{radio or combustion engine} characterization of cars might insist that the characterization is interesting, because it allows to discern between cars that are compatible with law regulations in a zero-emission country (cars without combustion engine) and those that are not (cars with combustion engine).
However, the \enquote{electric engine or combustion engine} characterization accomplishes this as well (and additionally has the above mentioned obvious advantages).%

\subsection{Shortcomings of the received view: Outcome dependence cannot explain the violation of Bell inequalities}  
\label{sec:OD}

After these more general remarks about Jarrett's analysis and the resulting classes of probability distributions, let us turn to the most prominent of these classes. 
Under the label \enquote{outcome dependent theories} Jarrett's class (J$_3)$ has evolved as the received view of what quantum non-locality (that can be compatible with relativity) amounts to on a probabilistic level \citep{Jarrett1984,Shimony1986}. 
We have already said that
our new analysis reveals that (J$_3$) involves classes that imply the Bell inequalities---and this has some remarkable consequences. 

(\arabic{mycounter}\stepcounter{mycounter}) Let me emphasize at the outset that from a logical point of view there is nothing wrong about the received view. 
By the logic of the involved arguments (the Bell argument and arguments against parameter dependence$_2$ from relativity) it is a necessary condition for the violation of Bell inequalities and compatibility with relativity and, therefore, it is perfectly alright to include classes that imply the Bell inequalities. 
Furthermore, neither \citet{Jarrett1984} nor \citet{Shimony1984} says that (J$_3$) is intended to only involve classes that can violate the Bell inequalities. However, they  neither do say that this is \emph{not} the case, and for a long time it has been an open question 
whether (J$_3$) also is sufficient for classes of product forms to be able to violate Bell inequalities.

(\arabic{mycounter}\stepcounter{mycounter}) 
My new result in this paper provides an answer to this question: 
\emph{Outcome depen\-dence$_1$ and hence (J$_3$) is not sufficient for a class to be able to violate Bell inequalities}.  
For what is required from a theory to be able to violate Bell inequalities is that at least one of the outcomes depends on both settings (double parameter dependence), so especially a dependence on the distant parameter is required. 
This finding is corroborated by the fact that all classes in \autoref{table-classes} whose only non-local dependence is outcome dependence (i.e. a subset of the classes in (J$_3$), particularly including (H$^\alpha_{16}$)) 
imply the Bell inequalities.

My new analysis moreover shows that \emph{outcome dependence$_1$} is not only not sufficient, but \emph{does not play any role for the question whether a class can violate Bell inequalities}: 
If a class is double parameter dependent, it can violate Bell inequalities, and if it is not, it cannot. 
Outcome dependence is not part of these necessary and sufficient conditions for a class being able to violate Bell inequalities.  
In this sense, whether outcome dependence holds is irrelevant for the question whether a class can violate the inequalities.

This is emphatically not to say that outcome dependence does not hold if one finds that Bell inequalities are violated---it might or it might not. But if it does, it cannot be the only non-local dependence. Neither is this to say that outcome dependence, if it holds, does not contribute to a violation of the Bell inequalities---to the contrary: If outcome dependence holds, it will make such a contribution.%
\footnote{{}
A dependence on the distant outcome does matter when one considers not only the violation of Bell inequalities but the exact quantitative reproduction of EPR/B correlations. \citet{Pawlowski2010} have shown that  there must be information about the distant outcome and this information can either be available by a direct correlation between the outcomes (as in the case of quantum mechanics) or be revealed by a hidden variable (which, however, is not available in the case of quantum mechanics).  
} 
But it will not be a contribution that is crucial for making the difference between violating the inequalities or not.

(\arabic{mycounter}\stepcounter{mycounter}) 
While all these findings do not make the received view untenable from a logical point of view, 
they make it \emph{liable to be misunderstood} to its non-literal wrong meaning that \emph{outcome dependence is the crucial non-local dependence that explains the violation of the Bell inequalities}. 
In fact, it seems that \emph{a number of authors have understood the received view to this unsupported wrong meaning}.
For, as we have already said above in paragraph~(2)  
(\autoref{sec:Jarrett}), they have interpreted it to show that quantum non-locality is some kind of metaphysical dependence between the outcomes (a non-separability according to most authors), instead of a connection between the outcomes and their distant settings---and this interpretation would be pointless, if outcome dependence would not suffice to explain the violation of the Bell inequalities.

(\arabic{mycounter}\stepcounter{mycounter})  
While outcome dependence$_1$ {in a literal sense} (as  Shimony defined \enquote{outcome dependence}, referring to Jarrett's definition of \enquote{completeness})  
is not sufficient for violating Bell inequalities, 
a defender of \enquote{outcome dependence} 
might claim that the literal reading is a misinterpretation of the received view, as the view never was meant to require just outcome dependence$_1$ (and parameter independence$_2$). Rather, it should be interpreted \emph{non-literally} to require outcome dependence \emph{and appropriate local parameter dependences}, i.e. $\alpha$-local parameter depen\-dence$_{1/2}$ as well as $\beta$-local parameter depen\-dence$_{1/2}$. To strengthen her position she might point to the fact that it is physically plausible that these dependences hold anyway. 
However, this proposal still involves class (H$^\alpha_{16}$), defined by
$P(\alpha \beta | a b \lambda) = P(\alpha | \beta a \lambda) P(\beta | b \lambda)$, 
which, as we have shown above, implies Bell inequalities. 

The proponent of outcome dependence may respond that the interpretation of \enquote{outcome dependence} is still too weak. Rather, we should understand it  
to include outcome dependence$_1$, the just mentioned local parameter dependences \emph{and parameter dependences$_1$}, i.e. $\alpha$-parameter dependence$_1$ and $\beta$-parameter dependence$_1$.
These dependences are realized in class
(H$^\alpha_{3}$), 
$P(\alpha \beta | a b \lambda) = P(\alpha | \beta a b \lambda) P(\beta | b \lambda)$,  
which essentially describes the quantum mechanical distribution (if one neglects the dependence on the hidden variable), and in class (H$^\alpha_{1}$) (and their counterparts from the complementary partition). 
The non-local parameter dependence in the first factor, $\alpha$-parameter dependence$_1$, it might be said, is required, because from a physical perspective the probability distribution has to reflect which measurement has been carried out at the other wing. This dependence seems to be unproblematic because---unlike usual parameter dependence$_2$---it conditions on the distant outcome and hence is not in tension with relativity, because it cannot be used to send signals. 

In this strongly non-literal reading \enquote{outcome dependence} indeed \emph{is} sufficient for a class being able to violate Bell inequalities: All four classes are among those that can violate the Bell inequalities. 
However, this reading of \enquote{outcome dependence} is highly problematic. 
One reason is that this interpretation makes \enquote{outcome dependence} a highly misleading name,  
because it 
hides the crucial fact that another non-local dependence is included as well. 
More importantly, however, if we assume this interpretation of \enquote{outcome dependence}, 
the disjunction of \enquote{outcome dependence} and parameter dependence$_2$ ceases to be equivalent to the failure of local factorization---which is a central fact in the argument for the received view. While one might try to argue that the local parameter dependences are plausible and tacit background assumptions rather than a part of what \enquote{outcome dependence} means, one surely cannot say the same about the non-local parameter dependences$_1$. 

To sum up, a proponent of  the received view who wishes to defend that theories that are \enquote{outcome dependent} (and parameter independent$_2$) provide a central category in the debate about quantum non-locality, faces the following dilemma: There is {no} meaning of \enquote{outcome dependence} such that the following claims are both true: 
\setlength{\leftmargini}{1cm}
\bi
\+[(O1)] {Outcome dependence} and parameter independence$_2$ only include classes that can violate Bell inequalities. 
\+[(O2)]  {Outcome dependence} or parameter dependence$_2$ are equivalent to the failure of local factorization.  
\ei

(\arabic{mycounter}\stepcounter{mycounter}) 
There is, however, one sense in which outcome dependence is required. 
If one accepts my argument that (J$_3$) is an inappropriate characterisation of quantum non-locality but wants to carry over Jarrett's and Shimony's idea that parameter depen\-dence$_2$ is forbidden due to relativistic constraints (Jarrett claims that it is equivalent with superluminal signalling in the EPR/B setup), the new analysis (C2$^\prime$)
yields that we must have 
\begin{equation}
\label{analysis-qnl-pi2} 
\neg (\text{PI}^\alpha_1) \wedge \neg(\text{LPI}^\alpha_1) \wedge \neg (\text{PI}^\beta_1) \wedge \neg(\text{LPI}^\beta_1) 
\end{equation}
Since one can prove that $\neg(\text{PI}^x_1) \wedge (\text{OI}_1) \Rightarrow \neg(\text{PI}^x_2)$, where $x$= $\alpha, \beta$, condition \eqref{analysis-qnl-pi2} and parameter independence$_2$ jointly imply  
outcome dependence$_1$. So outcome dependence$_1$ necessarily holds when Bell inequalities are found to be violated and parameter depen\-dence$_2$ is forbidden. 
I emphasize that this fact does not speak 
in any way against my result or for the received view. For outcome dependence only follows if, by my result, it is clear that parameter independence$_2$ and the violation of Bell inequalities imply parameter dependence$_1$. 
So by the logic of the situation, outcome dependence$_1$ is not the central feature of that position, but just a consequence thereof, and, especially, it has nothing to do with explaining the violation of the Bell inequalities. 
It would, therefore, be massively misleading to describe such theories just as \enquote{outcome dependent}.

(\arabic{mycounter}\stepcounter{mycounter})
Finally, we should explain how quantum mechanics fits into this picture. According to the received view, quantum mechanics is regarded as the paradigm of an \enquote{outcome dependent (and parameter independent) theory} that violates Bell inequalities. We have made clear throughout this paper that this means that besides outcome dependence$_1$ a specific parameter independence holds, viz. parameter independence$_2$, and that this does not preclude that other kinds of parameter dependences hold. 
In fact, if the result of my analysis (C2$^\prime$) is correct, quantum mechanics can only violate Bell inequalities under these conditions if it is double parameter dependent$_1$, i.e. if $\neg(\text{PI}^\alpha_1) \wedge \neg(\text{LPI}_1^\alpha)$ and $\neg(\text{PI}_1^\beta) \wedge \neg(\text{LPI}_1^\beta)$ hold. 

It is easy to check that this is indeed the case: From the quantum mechanical probability distribution for the EPR/B experiment 
one can calculate the relevant conditional probabilities, and it turns out that the quantum mechanical product forms (here: for maximally entangled states), read 
\begin{equation}
\label{pf-qm}
P(\alpha \beta | a b)  = P(\alpha | \beta a b) P(\beta) = P(\beta | \alpha a b) P(\alpha).%
\footnote{{} 
For non-maximally entangled states the product forms read $P(\alpha \beta | a b)  = P(\alpha | \beta a b) P(\beta|b) = P(\beta | \alpha a b) P(\alpha|a)$, i.e. there is an additional dependence on the local setting in each second factor.
}
\end{equation}
The conditionals of the first factors show that, besides outcome dependence, quantum mechanics exactly involves the required dependences on the local and distant setting%
---and it is these dependences on both settings rather than the dependence between the outcomes, which is crucial for the fact that quantum mechanics violates Bell inequalities. 

The parameter dependence in quantum mechanics is not as surprising as it may seem since, according to the formalism, the measurement direction at $A$  determines the possible collapsed states at $B$ and the actual outcome at $A$ only determines in which of the (two) possible states the photon state at $B$ collapses. 

So both claims have turned out to be true: Quantum mechanics is \enquote{outcome dependent and parameter independent} and nevertheless involves a dependence on the distant parameter.

(\arabic{mycounter}\stepcounter{mycounter})  Given that the probabilistic picture has changed considerably, it remains to be investigated on the basis of the new analysis whether the (meta-)physical picture of the received view, that there is just a (meta-)physical connection between the outcomes, can still be maintained. 
Inferring physical or metaphysical relations from probabilistic facts requires careful analysis, since the transition is well known to be vulnerable to notorious fallacies (\enquote{correlation is not causation}). 
We cannot provide such an analysis here. Having said this, it might be interesting to remark that there seem to be good arguments that the current result, that a probabilistic dependence between the outcomes is too weak to explain a violation of the Bell inequalities, most plausibly entails that also an influence between the outcomes is not strong enough to account for a violation \citep[][]{Naeger2013b}.

\subsection{Resolving the Jarrett-Maudlin debate} 
\label{sect-discussion-II}

(\arabic{mycounter}\stepcounter{mycounter})
Opposed to the received view, there is another position concerning quantum non-locality, whose result seems to agree with ours. 
Maudlin (\citeyear[ch.~6]{Maudlin1994}; cf. also a recent refinement by  \citealp{Pawlowski2010}) 
proves that, in order to reproduce the EPR/B correlations, at least one of the outcomes must depend on \emph{information} about both settings. 
Since (Shannon mutual) information implies correlation,%
\footnote{{}
Shannon mutual information, which is the concept that Maudlin's and Pawlowski et al.'s considerations essentially are based on, is a measure for the strength of a correlation. 
} 
one can infer that at least one of the outcomes must depend probabilistically on both settings---and this is exactly my result (C3). 

(\arabic{mycounter}\stepcounter{mycounter})
This convergence is good news, both for Maudlin's as well as my argument here, because the two investigations use very different methods, and two different methods yielding the same result are evidence for the stability of a claim. 
On the one hand, Maudlin's approach is an information theoretic investigation proceeding from the EPR/B correlations without invoking Bell's theorem. 
In contrast, my argument in this paper approaches quantum non-locality via Bell inequalities and probabilistic analysis, i.e. it stands methodologically in the Bell-Jarrett tradition, which has started and shaped the debate. 
For these reasons it seems fair to say that my approach here confirms Maudlin's result by a different method.

(\arabic{mycounter}\stepcounter{mycounter})
We should note that our result is in one sense weaker and in one sense stronger than the information theoretic one. 
It is weaker because it is purely qualitative: It just says which probabilistic dependences are required, but it is tacit about how strong the correlations have to be in order to violate Bell inequalities. 
In \autoref{sec:strong-poss-cons} I have argued  that such qualitative results can only be necessary conditions for a violation, because having the right dependences for violating Bell inequalities does not mean that the inequalities are in fact violated. 
In contrast, sufficient criteria must involve conditions on the strength of the correlations, and the information theoretic approach derives such criteria by calculating the amount of information (the quantitative strength of the correlations) that is required in order to reproduce EPR/B correlations; these are important results.

In another sense, however, our result is also stronger than the information theoretic one. 
Maudlin's analysis just implies \emph{some kind of} dependence of an outcome on its distant parameter. But which precisely? We have seen that there are different kinds of parameter dependences, which differ in the conditional variables. Especially it cannot be an unconditional parameter dependence because that would contradict the empirical distribution. So which are the ones that are required? 
My detailed result (C2$^\prime$) makes precise which kind of parameter dependences are required. 
Present information theoretic results do not provide a similar detailed characterization. 
However, a complete list of which dependences exactly hold or fail might be important for the discussion of quantum non-locality on other levels: Causal inference \citep[cf.][]{Spirtes1993,Pearl2000}, for instance, is very sensitive to the exact pattern of dependences and independences.

(\arabic{mycounter}\stepcounter{mycounter})
My results also allow to resolve the tension between Maudlin's approach on the one hand and Jarrett's analysis and the received view on the other hand. 
While Jarrett's result seems to suggest that there is a choice to make between 
outcome dependence and parameter dependence 
and the received view holds that it is the dependence between the outcomes which is realized, Maudlin's informational approach opposes to these positions by saying that one of the outcomes must depend on information about the distant setting. 
This tension has existed unresolved for over 20 years now.

During that time the two analyses have coexisted; Maudlin's critique of Jarrett's analysis and the resulting standard position did not succeed in convincing the adherents of outcome dependence---though Maudlin did have good arguments: He realized that Jarrett's analysis can be misleading because there are different kinds of parameter dependences and outcome dependences, according to which different variables appear in the conditional \citep[ch.~4]{Maudlin1994}; he argued that Jarrett's analysis is also misleading because his informational approach unveiled that some kind of parameter dependence is unavoidable; and, hence, it is wrong to assume that outcome dependence per se can explain EPR/B correlations \citep[ch.~6]{Maudlin1994}. 
To me it is not exactly clear, why Jarrett's analysis and the received view based on it could keep on for such a long time, given Maudlin's critique. One reason might have been that Maudlin's arguments do not connect to the Bell-Jarrett methodology, such that it was hard to compare the two approaches and to see which in fact is right. 

In this paper, however, we have provided that connection. We have strengthened the Bell-Jarrett approach to our new results and these (i) confirm Maudlin's claim that there must be a dependence of at least one outcome on the distant parameter (we have furthermore derived, which exact combinations of dependences are required, \autoref{qnl-pd}), (ii) show that Jarrett's analysis is indeed misleading (we have made precise how exactly, \autoref{sec:Jarrett}) and (iii) prove that the received view is wrong (because outcome dependence is not sufficient for a class to be able to violate Bell inequalities; \autoref{sec:OD}). 
This clearly resolves the Jarrett-Maudlin controversy in favour of the latter.

(\arabic{mycounter})\stepcounter{mycounter} Finally, we may ask, \emph{why the stronger consequences of the Bell argument}, that we have derived in this paper, \emph{have been overlooked so far}. 
Obviously, it has wrongly been assumed that local factorization is the \emph{only} basis to derive Bell inequalities, and the main reason for neglecting other product forms of hidden joint probabilities might have been the fact that, originally, Bell inequalities were derived to capture consequences of a \emph{local} worldview. 
The question that shaped Bell's original work clearly was Einstein's search for a local hidden variable theory 
and his main result was that such a theory is impossible: Locality has consequences which are in conflict with the quantum mechanical distribution---one cannot have a local hidden variable theory which yields the same predictions as quantum mechanics. 
Given this historical background, the idea to derive Bell inequalities from non-local assumptions maybe was beyond interest 
because the conflict with locality was considered to be the crucial point; or maybe it was neglected because Bell inequalities were so tightly associated with locality that a derivation from non-locality sounded totally implausible.  
Systematically, however, since today it is clear that the quantum mechanical distribution is empirically correct and Bell inequalities are violated, it is desirable to draw as strong consequences as possible from the argument, which requires to check without prejudice whether some non-local classes allow a derivation of Bell inequalities as well. 
That this is indeed the case and what exactly this implies on a qualitative probabilistic level has been the topic of this paper.

\section*{Acknowledgements}

I would like to thank Frank Arntzenius, Arthur Fine, Tjorven Hetzger, Gábor Hofer-Szabó, Meinard Kuhlmann, Wayne Myrvold, Thorben Petersen, Manfred Stöckler, Nicola Vona, Adrian Wüthrich and audiences at the conference \enquote{Philosophy of Physics in Germany} (Hanover, 2010) and at the \enquote{14th Congress of Logic, Methodology and Philosophy of Science}  (Nancy, 2011) for helpful comments and discussion. I am also grateful to two anonymous referees for very valuable comments. 
Research on this paper was supported by the 
Deutsche Forschungsge\-mein\-schaft~(DFG, STO 158/14-1).

\appendix 

\section{Mathematical appendix}
\markboth{Appendix}{Appendix}

\subsection{Proof of ~\autoref{th:dir-pcorr}%1 
} 
\label{sec:proof-theorem1.1}
\stepcounter{theoremcounter}

\theorema*

\noindent The theorem is equivalent to the conjunction of the following claims: 

\begin{lemma} 
\label{lemma-1.1}
A class of probability distributions forms an inconsistent set with measurement independence, perfect correlations and perfect anti\-/correlations 
if  (i) its defining product form involves at most one of the settings. 
\end{lemma}
\begin{lemma}
\label{lemma-1.2}
A class of probability distributions forms an inconsistent set with measurement independence, perfect correlations and perfect anti\-/correlations 
if  (ii)~its defining product form involves both settings and the first factor of this product form involves the distant outcome and at most one setting. 
\end{lemma}
\begin{lemma}
\label{lemma-1.3}
A class of probability distributions forms a consistent set with  measurement independence, perfect correlations and perfect anti\-/correlations if $\neg$(i) its defining product form involves both settings and $\neg$(ii) in case the distant outcome appears in the first factor of its defining product form, also both settings appear in that factor. 
\end{lemma}

\subsubsection{Proof of \autoref{lemma-1.1}} 

Condition (i), that the defining product form involves at most one of the settings, is fulfilled by the classes $\{(\text{H}_{17}^\alpha),…,(\text{H}_{32}^\alpha)\}\backslash \{(\text{H}_{22}^\alpha),(\text{H}_{29}^\alpha)\}$. Here we have to show the inconsistency of these classes with the set of assumptions measurement independence, perfect correlations and perfect anti\-/correlations. 

Consider, for instance, 
\begin{equation}
P(\alpha \beta | a b \lambda) = P(\alpha | \beta a \lambda) P(\beta | a \lambda) = P(\alpha \beta | a \lambda), 
\tag{H$^\alpha_{17}$}
\end{equation}
which fails to involve the setting $\zv b$.%
\footnote{{} 
If for certain values $P(\beta a \lambda) = 0$ holds, the first factor of the product form, $P(\alpha|\beta a \lambda)$, 
is not well-defined. In these cases, however, the equality $P(\alpha \beta | a b \lambda) = P(\alpha \beta | a \lambda)$ follows by the fact that $P(\beta a \lambda) = 0$ implies $P(\alpha \beta | a \lambda) =0$ as well as $P(\alpha \beta | a b \lambda) = 0$.
}
It is easy to show that this product form cannot account for the perfect correlations and the perfect anti\-/correlations together. 
Take, for instance, the conditions 
\begin{align}
\label{pcorr-1-1}
P(\alpha_\pm \beta_\pm | a_i b_i) &= \frac{1}{2}  & 
P(\alpha_\pm \beta_\pm | a_i b_{i_\bot}) &= 0, 
\end{align}
that are part of the perfect correlations and anti\-/correlations, respectively. 
Obviously, the value of these empirical probabilities depends crucially on the value of the setting~$\zv b$. 
However, one can demonstrate without much effort that (H$_{17}^\alpha$)'s failure to involve the setting~$\zv b$ on a hidden level, extends to the empirical level, if one assumes measurement independence: 
\begin{align}
\label{eq:hjp-ind-b}
P(\alpha \beta | a b) &=  \sum_\lambda P(\lambda|ab) P(\alpha \beta | a b \lambda) \stackrel{(\text{MI})}{=}  \sum_\lambda P(\lambda|ab^\prime) P(\alpha \beta | a b\lambda) \notag  \\ &\stackrel{(\text{H}^\alpha_{17})}{=} \sum_\lambda P(\lambda|ab^\prime) P(\alpha \beta | a b^\prime \lambda) =  P(\alpha \beta | a b^\prime) 
\end{align}
This demonstrates that according to (H$_{17}^\alpha$) all empirical probabilities $P(\alpha \beta | a b)$ that only differ by their value for the setting~$\zv b$ must equal another---which obviously contradicts \eqref{pcorr-1-1}.

In the same way, all other product forms that do not involve the setting $\zv b$ are in conflict with the perfect (anti\=/)correlations \eqref{pcorr-1-1},  
and, similarly, all product forms that fail to involve the setting $\zv a$ are in conflict with, for instance, the conditions
\begin{align}
\label{pcorr-1-3}
P(\alpha_\pm \beta_\pm | a_i b_i) &= \frac{1}{2}  &
P(\alpha_\pm \beta_\pm | a_{i_\bot} b_i) &= 0,  
\end{align}
which also belong to the perfect correlations and anti\-/correlations, respectively. 
\parbox{0cm}{} \hfill $\Box$

\subsubsection{Proof of \autoref{lemma-1.2}}

Condition~(ii), that the defining product form involves both settings and its first factor involves the distant outcome and at most one setting, is fulfilled by the classes (H$_{4}^\alpha$), (H$_{5}^\alpha$), (H$_{10}^\alpha$), (H$_{15}^\alpha$) and (H$_{16}^\alpha$). Here we have to show the inconsistency of these classes with the set of assumptions measurement independence, perfect correlations and perfect anti\-/correlations. 

We start with (H$^\alpha_{16}$) 
and proceed by reductio. {We assume perfect correlations and perfect anti\-/correlations}, 
\begin{align}%{10}
\label{perf-corr-0}
P(\alpha_\pm \beta_\mp|a_i b_i) = 0 && P(\alpha_\pm \beta_\mp|a_{i_\bot} b_{i_\bot}) = 0 \\
P(\alpha_\pm \beta_\pm|a_i b_{i_\bot}) = 0 && P(\alpha_\pm \beta_\pm|a_{i_\bot} b_i) = 0,
\end{align}
and, using measurement independence and (H$_{16}^\alpha$), we rewrite  the left hand side of these equations 
according to the scheme 
$P(\alpha \beta | a b)  = \sum_\lambda P(\lambda) P(\alpha | \beta  a \lambda) P(\beta |b \lambda)$ (assuming that all conditional probabilities are well-defined, especially $P(\beta  a \lambda) >0$).  
The resulting system of eight equations can be solved. 
Since probabilities are non-negative and since without loss of generality we can assume $P(\lambda) > 0$ for all $\lambda$, at least one of the two factors of the product form in each summand must be zero, i.e. for every $i$ and $\lambda$ 
{all of the following disjunctions must be true}: \\[-1.0cm]

\noindent 
\begin{minipage}[t]{7.2cm}
\begin{alignat}{10}
\label{case+-i}& P(\alpha_+ | \beta_-  a_i \lambda) &= \! 0  \;\; \vee  & \;\; P(\beta_-|b_i \lambda) & = \! 0
\end{alignat}
\end{minipage} 
\begin{minipage}[t]{7.2cm} 
\begin{alignat}{10}
\label{case-+i} & P(\alpha_- | \beta_+  a_i \lambda) &= \! 0  \;\; \vee  & \;\; P(\beta_+|b_i \lambda) &= \! 0 
\end{alignat}  
\end{minipage}\\[-0.4cm]
\noindent 
\begin{minipage}[t]{7.2cm}
\begin{alignat}{10}
\label{case+-ibot} P(\alpha_+ | \beta_-  a_{i_\bot} \lambda) &= \! 0 \; \vee  & \; P(\beta_-|b_{i_\bot} \lambda) &= \! 0 
\end{alignat}
\end{minipage} 
\begin{minipage}[t]{7.2cm} 
\begin{alignat}{10}
\label{case-+ibot} P(\alpha_- | \beta_+  a_{i_\bot} \lambda) &= \! 0 \; \vee  & \; P(\beta_+|b_{i_\bot} \lambda) &= \! 0 
\end{alignat}  
\end{minipage} \\[-0.4cm] 
\noindent 
\begin{minipage}[t]{7.2cm}
\begin{alignat}{10}
\label{case++i} P(\alpha_+ | \beta_+ a_i \lambda) &= \! 0 \;\; \vee  & \;\; P(\beta_+|b_{i_\bot} \lambda) &= \! 0
\end{alignat}
\end{minipage} 
\begin{minipage}[t]{7.2cm} 
\begin{alignat}{10}
\label{case--i}  P(\alpha_- | \beta_-  a_i \lambda) &= \! 0 \;\; \vee  & \;\; P(\beta_-|b_{i_\bot} \lambda) &= \! 0 
\end{alignat}  
\end{minipage} \\[-0.4cm] 
\noindent 
\begin{minipage}[t]{7.2cm}
\begin{alignat}{10}
\label{case++ibot} P(\alpha_+ | \beta_+ a_{i_\bot} \lambda) &= \! 0 \;\; \vee  & \;\; P(\beta_+|b_i \lambda) &= \! 0
\end{alignat}
\end{minipage} 
\begin{minipage}[t]{7.2cm} 
\begin{alignat}{10}
\label{case--ibot} P(\alpha_- | \beta_-  a_{i_\bot} \lambda) &= \! 0 \;\; \vee  & \;\; P(\beta_-|b_i \lambda) &= \! 0 
\end{alignat}  
\end{minipage}\\[0.25cm]

From these conditions one can infer that all involved probabilities must be 0~or~1 (determinism). More precisely, for every $i$ and $\lambda$ one of the following two cases holds: \\

\noindent \underline{Case 1: $P(\alpha_+ | \beta_-  a_i \lambda) = 0$} 
\vspace{-0.25cm}
\begin{alignat*}{4}
&\stackrel{(\text{C})}{\Rightarrow} \!\! P(\alpha_- | \beta_-  a_i \lambda) = 1 
&&\stackrel{\eqref{case--i}}{\Rightarrow} && P(\beta_- | b_{i_\bot} \lambda) = 0 
&& \stackrel{(\text{C})}{\Rightarrow}  P(\beta_+ | b_{i_\bot} \lambda) = 1  \\
&\overset{\eqref{case-+ibot}}{\underset{\eqref{case++i}}{\Rightarrow}}  P(\alpha_- | \beta_+  a_{i_\bot} \lambda)  = 0    && \enspace \wedge \enspace && P(\alpha_+ | \beta_+  a_i \lambda) = 0  
&& \stackrel{(\text{C})}{\Rightarrow} P(\alpha_+ | \beta_+  a_{i_\bot} \lambda)  = 1 \enspace \wedge \enspace P(\alpha_- | \beta_+  a_i \lambda) = 1  \\
&\overset{\eqref{case++ibot}}{\underset{\eqref{case-+i}}{\Rightarrow}}   P(\beta_+ | b_i \lambda) = 0  \qquad 
&&\stackrel{(\text{C})}{\Rightarrow} && P(\beta_- | b_i \lambda) = 1 
&& \overset{\eqref{case--ibot}}{\Rightarrow} P(\alpha_- | \beta_-  a_{i_\bot} \lambda) = 0 
\stackrel{(\text{C})}{\Rightarrow}  P(\alpha_+ | \beta_-  a_{i_\bot} \lambda) = 1
\end{alignat*}

\noindent \emph{NB}: (C) stands for the following theorem of probability theory: $P(A|B) + P(\bar{A}|B) =1$.

\vspace{0.25cm}
\noindent \underline{Case 2: $P(\alpha_+ | \beta_-  a_i \lambda) > 0$}
\vspace{-0.25cm}
\begin{alignat*}{4}
&\overset{\eqref{case+-i}}{\Rightarrow}   P(\beta_- | b_i \lambda) = 0  
&&\stackrel{(\text{C})}{\Rightarrow}  P(\beta_+ | b_i \lambda) = 1
&&\overset{\eqref{case-+i}}{\underset{\eqref{case++ibot}}{\Rightarrow}}   P(\alpha_- | \beta_+  a_i \lambda)  = 0  && \enspace \wedge \enspace  P(\alpha_+ | \beta_+  a_{i_\bot} \lambda) = 0  \\
&\stackrel{(\text{C})}{\Rightarrow} P(\alpha_+ | \beta_+  a_i \lambda)  = 1  
\enspace && \wedge \enspace P(\alpha_- | \beta_+  a_{i_\bot} \lambda) = 1  \enspace
&& \overset{\eqref{case++i}}{\underset{\eqref{case-+ibot}}{\Rightarrow}}   P(\beta_+ | b_{i_\bot} \lambda) = 0 
&&\stackrel{(\text{C})}{\Rightarrow} P(\beta_- | b_{i_\bot} \lambda) = 1 \\
& \underset{\eqref{case--i}}{\overset{\eqref{case+-ibot}}{\Rightarrow}}  P(\alpha_+ | \beta_-  a_{i_\bot} \lambda) = 0 
\enspace && \wedge \enspace  P(\alpha_- | \beta_-  a_i \lambda) = 0 
&&\stackrel{(\text{C})}{\Rightarrow}  P(\alpha_- | \beta_-  a_{i_\bot} \lambda) = 1 
&& \enspace \wedge \enspace  P(\alpha_+ | \beta_-  a_i \lambda) = 1
\end{alignat*}

Since in each case we have $P(\alpha | \beta_+  a_i \lambda) = P(\alpha | \beta_-  a_i \lambda)$, 
it is true that $\forall \alpha, \beta, a, \lambda: \quad P(\alpha | \beta  a \lambda) = P(\alpha | a \lambda)$. 
By this statistical independence the product form (H$^\alpha_{16}$), $P(\alpha \beta | a b \lambda) = P(\alpha | \beta a \lambda) P(\beta | b \lambda)$,  
loses its dependence on the outcome $\beta$ in the first factor, i.e. it now reads 
$P(\alpha \beta | a b \lambda) = P(\alpha | a \lambda) P(\beta | b \lambda)$,  
which is the well known local product form (local factorization), contradicting the assumption that we have the non-local product form (H$^\alpha_{16}$).%
\footnote{{} 
In this proof we have assumed $P(\beta a_i \lambda) > 0$. If, however, for some $i$ and $\lambda$, for instance we have $P(\beta_+ a_i \lambda) = 0$, $P(\alpha | \beta_+ a_i \lambda)$ is not well-defined and we cannot reason as above. 
In this case, however, it is even easier to show that  the outcomes cannot depend on another, making (H$^\alpha_{16}$) impossible: 
\begin{align}
\label{eq:path1}
P(\beta_+  a_i \lambda) &= 0 \quad  \Rightarrow \quad P(\beta_+ | a_i \lambda) = 0 \; \wedge \; P(\alpha \beta_+ | a_i \lambda) = 0  
\\ 
P(\alpha|\beta_- a_i \lambda) &= \frac{P(\alpha\beta_-| a_i \lambda)}{P(\beta_-| a_i \lambda)} = \frac{P(\alpha\beta_-| a_i \lambda) + 0}{1- P(\beta_+| a_i \lambda)} \stackrel{\eqref{eq:path1}}{=} \frac{P(\alpha\beta_-| a_i \lambda) + P(\alpha\beta_+ | a_i \lambda)}{1} = P(\alpha| a_i \lambda) \\
P(\alpha \beta_+ | a_i \lambda) &= 0 = P(\alpha | a_i \lambda) \underbrace{P(\beta_+ | a_i \lambda)}_{=0}  
\end{align}
}
This completes the proof for the inconsistency of (H$^\alpha_{16}$) with measurement independence, perfect correlations and perfect anti\-/correlations.%
\footnote{{}
Note that this proof makes essential use of the perfect correlations and perfect anti\-/correlations \eqref{perf-corr-0}.
If these conditions are only slightly relaxed, i.e. if any of the involved probabilities takes on a positive value, even if very small, the conclusion does not follow. }

Mutatis mutandis, also the classes (H$_{10}^\alpha$) and (H$_{15}^\alpha$) lead to a similar inconsistency. In each case the product form looses its dependence on the distant outcome in the first factor, i.e. (H$_{10}^\alpha$) reduces to (H$_{14}^\alpha$), whereas (H$_{15}^\alpha$) reduces to (H$_{22}^\alpha$).

The proofs against the classes (H$_{4}^\alpha$) and (H$_{5}^\alpha$) work in a similar way, but require a little more care due to an additional case differentiation. 
Let me shortly demonstrate this for class (H$_{5}^\alpha$). 
As for (H$_{16}^\alpha$) one starts with expressing the perfect (anti\=/)correlations in terms of the product form,  
\begin{align}
\label{H5-pcorr1}
P(\alpha_\pm \beta_\mp|a_i b_i) &= 0 = \sum_\lambda P(\lambda) P(\alpha_\pm | \beta_\mp  a_i \lambda) P(\beta_\mp|a_i b_i \lambda)\\
\label{H5-pcorr2}
P(\alpha_\pm \beta_\mp|a_{i_\bot} b_{i_\bot}) &= 0 = \sum_\lambda P(\lambda) P(\alpha_\pm | \beta_\mp  a_{i_\bot} \lambda) P(\beta_\mp|a_{i_\bot} b_{i_\bot} \lambda) \\
\label{H5-pcorr3}
P(\alpha_\pm \beta_\pm|a_i b_{i_\bot}) &= 0 = \sum_\lambda P(\lambda) P(\alpha_\pm | \beta_\pm a_i \lambda) P(\beta_\pm|a_i b_{i_\bot} \lambda) \\
\label{H5-pcorr4}
P(\alpha_\pm \beta_\pm|a_{i_\bot} b_i) &= 0 = \sum_\lambda P(\lambda) P(\alpha_\pm | \beta_\pm a_{i_\bot} \lambda) P(\beta_\pm|a_{i_\bot} b_i \lambda).
\end{align}
In the case of (H$_{16}^\alpha$) there were two cases, defined by $P(\alpha_+ | \beta_-  a_i \lambda) = 0$ or 
$P(\alpha_+ | \beta_-  a_i \lambda) > 0$, 
respectively, and all other probabilities followed from each of these defining probabilities. 
In the present case, however, when, accordingly, we 
consider each of these cases, 
only the factors of the product form 
on the right hand side of equations \eqref{H5-pcorr1} and \eqref{H5-pcorr3} are implied, i.e. the probabilities in \eqref{H5-pcorr2} and \eqref{H5-pcorr4} remain undetermined by these assumptions (due to the fact that there are \emph{two} settings in the second factor of the product form). The latter probabilities have to be determined by further assumptions, e.g. by setting $P(\alpha_+ | \beta_-  a_{i_\bot} \lambda) = 0$ or 
$P(\alpha_+ | \beta_-  a_{i_\bot} \lambda) > 0$, 
respectively. These assumptions introduce two new cases, that are logically independent of the former two. 
In total, this makes four cases (instead of two): 
\begin{align}
P(\alpha_+ | \beta_-  a_i \lambda) &= 0   \qquad \wedge  \qquad P(\alpha_+ | \beta_-  a_{i_\bot} \lambda) = 0 \\
P(\alpha_+ | \beta_-  a_i \lambda) &= 0   \qquad \wedge  \qquad P(\alpha_+ | \beta_-  a_{i_\bot} \lambda) > 0 \\
P(\alpha_+ | \beta_-  a_i \lambda) &> 0  \qquad \wedge  \qquad P(\alpha_+ | \beta_-  a_{i_\bot} \lambda) = 0 \\
P(\alpha_+ | \beta_-  a_i \lambda) &> 0  \qquad \wedge  \qquad P(\alpha_+ | \beta_-  a_{i_\bot} \lambda) > 0 
\end{align}
While this renders the proof slightly more complex, the crucial fact to mention here is that in all four cases we have 
\begin{equation}
\forall \alpha, \beta, a, \lambda : \quad P(\alpha|\beta a\lambda) = P(\alpha | a\lambda), 
\end{equation}
i.e. (H$_{5}^\alpha$) reduces to (H$_{12}^\alpha$). 
Similarly, one can show that  (H$_{4}^\alpha$) reduces to (H$_{11}^\alpha$). 
\parbox{0cm}{} \hfill $\Box$

\subsubsection{Proof of \autoref{lemma-1.3}}
\label{sec:proof-lemma1.3}

Condition~$\neg$(i) \emph{and}~$\neg$(ii), that the product form involves both settings and in case the distant outcome appears in the first factor, also both settings appear in that factor, is fulfilled by the product forms $\{(\text{H}_{1}^\alpha), \dots,(\text{H}_{14}^\alpha)\}\backslash\{(\text{H}_{4}^\alpha), (\text{H}_{5}^\alpha), (\text{H}_{10}^\alpha)\}$, (H$_{22}^\alpha$) and (H$_{29}^\alpha$). Here we have to show the consistency of these classes with the set of assumptions measurement independence, perfect correlations and perfect anti\-/correlations.

Since a class being inconsistent with certain assumptions means that every distribution of a class contradicts the assumptions, a class being consistent means that there is at least one probability distribution in that class which is compatible with the assumptions. 
Hence, in order to show consistency, we need to provide 
 one example of a probability distribution for each of the mentioned classes that respects the background assumptions. In fact, such examples are easy to construct. Let me demonstrate the procedure with one of the strongest classes in that group, (H$_{29}^\alpha$), whose product form is local factorization. 

Requiring just \emph{any} example we can presuppose a minimal setup, i.e. the hidden variable as well as each setting can be assumed to have only two possible values: $\zv \lambda = \lambda_1, \lambda_2$, $\zv a = a_i, a_{i_\bot}$ and $\zv b = b_i, b_{i_\bot}$ with $a_i = b_i$ and $a_{i_\bot}= b_{i_\bot}$. 
We assume perfect correlations and perfect anti\-/correlations: \vspace{-0.25cm}

\noindent 
\begin{minipage}[t]{5cm}
\begin{equation}
\label{calc1}
P(\alpha_\pm \beta_\mp|a_i b_i) = 0
\end{equation}
\end{minipage}
\hspace{2cm}
\begin{minipage}[t]{5cm}
\begin{equation}
\label{calc2}
P(\alpha_\pm \beta_\mp|a_{i_\bot} b_{i_\bot}) = 0 
\end{equation} 
\end{minipage} \\
\begin{minipage}[t]{5cm}
\begin{equation}
\label{calc3} 
P(\alpha_\pm \beta_\pm|a_{i_\bot} b_i) = 0 
\end{equation}  
\end{minipage}
\hspace{2cm}
\begin{minipage}[t]{5cm}
\begin{equation}
\label{calc4}
P(\alpha_\pm \beta_\pm|a_i b_{i_\bot}) = 0
\end{equation}  
\end{minipage} \\
\begin{minipage}[t]{5cm}
\begin{equation}
\label{calc5}
P(\alpha_\pm \beta_\pm|a_i b_i) = \frac{1}{2} 
\end{equation}
\end{minipage}
\hspace{2cm}
\begin{minipage}[t]{5cm}
\begin{equation}
\label{calc6}
P(\alpha_\pm \beta_\pm|a_{i_\bot} b_{i_\bot}) = \frac{1}{2}  \end{equation} 
\end{minipage} \\
\begin{minipage}[t]{5cm}
\begin{equation}
\label{calc7}
P(\alpha_\pm \beta_\mp|a_{i_\bot} b_i) = \frac{1}{2} 
\end{equation}
\end{minipage}
\hspace{2cm}
\begin{minipage}[t]{5cm}
\begin{equation}
\label{calc8}
P(\alpha_\pm \beta_\mp|a_i b_{i_\bot}) = \frac{1}{2}. 
\end{equation}
\end{minipage}
\\[0.25cm]

\noindent By measurement independence and the product form of (H$_{29}^\alpha$) one can rewrite the left hand side of these equations according to the scheme $P(\alpha \beta |a b) = \sum_\lambda P(\lambda) P(\alpha | a \lambda) P(\beta | b \lambda)$, 
which yields a system of 16 coupled equations. 

Then choose a value for any of the conditional probabilities that does not lead into inconsistencies, e.g. $P(\alpha_+|a_i\lambda_1) = 0$. 
By \eqref{calc1}--\eqref{calc4} this entails the following probabilities: 
\begin{alignat*}{23}
&\stackrel{(\text{C})}{\Rightarrow}  && P(\alpha_-|a_i\lambda_1) &&= 1 \enspace && \overset{\eqref{calc1}}{\underset{\eqref{calc4}}{\Rightarrow}}  && P(\beta_+|b_i\lambda_1) &&= 0  
&&\enspace \wedge \enspace &&P(\beta_-|b_{i_\bot}\lambda_1) &&= 0  \\
&\stackrel{(\text{C})}{\Rightarrow} && P(\beta_-|b_i\lambda_1) && = 1 
&& \enspace  \wedge \enspace && P(\beta_+|b_{i_\bot}\lambda_1) &&= 1 \enspace 
&& \overset{\eqref{calc3}}{\underset{\eqref{calc2}}{\Rightarrow}}  && P(\alpha_-|a_{i_\bot}\lambda_1) &&= 0 \enspace    
&&\stackrel{(\text{C})}{\Rightarrow} && P(\alpha_+|a_{i_\bot}\lambda_1)  &&= 1
\end{alignat*}

Similarly, choose a value for the corresponding probability conditional on $\lambda_2$, e.g. $P(\alpha_+|a_i\lambda_2) = 1$
and draw the appropriate consequences: 
\begin{alignat*}{24}
& \quad \overset{\eqref{calc1}}{\underset{\eqref{calc4}}{\Rightarrow}}  && P(\beta_-|b_i\lambda_2) &&= 0  
&&\enspace \wedge \enspace &&P(\beta_+|b_{i_\bot}\lambda_2) &&= 0  \enspace 
&&\stackrel{(\text{C})}{\Rightarrow} && P(\beta_+|b_i\lambda_2) && = 1 
&& \enspace \wedge \enspace &&P(\beta_-|b_{i_\bot}\lambda_2) &&= 1 \\
& \overset{\eqref{calc1},\eqref{calc4}}{\underset{\eqref{calc2},\eqref{calc3}}{\Rightarrow}}  \; &&  P(\alpha_-|a_i\lambda_2) &&= 0   && \enspace \wedge \enspace && P(\alpha_+|a_{i_\bot}\lambda_2) &&=0 &&\stackrel{(\text{C})}{\Rightarrow} && P(\alpha_-|a_{i_\bot}\lambda_2)  &&= 1 
\label{end-prob}
\end{alignat*}

These probabilities determine the values of the hidden joint probabilities consistently with equations \eqref{calc1}--\eqref{calc4}. Note that we have 
\begin{align}
&\forall \alpha, \lambda: \quad  P(\alpha | a_i \lambda) \neq P(\alpha | a_{i_\bot} \lambda)  && \forall \alpha, a: \quad  P(\alpha | a \lambda_1) \neq P(\alpha | a \lambda_2) \\
&\forall \beta, \lambda: \quad  P(\beta | b_i \lambda) \neq P(\beta | b_{i_\bot} \lambda)  && \forall \beta, b: \quad  P(\beta | b \lambda_1) \neq P(\beta | b \lambda_2), 
\end{align}
which means that the product form does not reduce to any other product form (i.e. the product form is consistent with the assumptions so far).

Inserting the determined values of the hidden joint probability into equations \eqref{calc5}--\eqref{calc8} yields $ P(\lambda_1) = \frac{1}{2}$ and $ P(\lambda_2) = \frac{1}{2}$. 
Finally we can freely choose, say, 
$P(a_i) = \frac{1}{2} = P(a_{i_\bot})$ and $  P(b_i) = \frac{1}{2} = P(b_{i_\bot})$, 
and by the formula $P(\alpha \beta a b \lambda) = P(\alpha | a \lambda) P(\beta | b \lambda) \cdot P(\lambda) P(a) P(b)$ 
we arrive at the following probability distribution: 
\vspace{-0.4cm}

\note{WV müsste man nicht abdrucken}{}
\begin{align}
 P(\alpha_+ \beta_+ a_i b_i \lambda_1) &= 0  & P(\alpha_+ \beta_- a_i b_i \lambda_1) &= 0 \notag 
\\  P(\alpha_- \beta_+ a_i b_i \lambda_1) &= 0 & P(\alpha_- \beta_- a_i b_i \lambda_1) &= \tfrac{1}{8} 
\\[0.1cm] P(\alpha_+ \beta_+ a_i b_{i_\bot} \lambda_1) &= 0  & P(\alpha_+ \beta_- a_i b_{i_\bot} \lambda_1) &= 0 \notag
\\ P(\alpha_- \beta_+ a_i b_{i_\bot} \lambda_1) &= \tfrac{1}{8} & P(\alpha_- \beta_- a_i b_{i_\bot} \lambda_1) &= 0 
\\[0.1cm] P(\alpha_+ \beta_+ a_{i_\bot} b_i \lambda_1) &= 0  & P(\alpha_+ \beta_- a_{i_\bot} b_i \lambda_1) &= \tfrac{1}{8} \notag
\\ P(\alpha_- \beta_+ a_{i_\bot} b_i \lambda_1) &= 0 & P(\alpha_- \beta_- a_{i_\bot} b_i \lambda_1) &= 0 
\\[0.1cm] P(\alpha_+ \beta_+ a_{i_\bot} b_{i_\bot} \lambda_1) &= \tfrac{1}{8} & P(\alpha_+ \beta_- a_{i_\bot} b_{i_\bot} \lambda_1) &= 0 \notag
\\ P(\alpha_- \beta_+ a_{i_\bot} b_{i_\bot} \lambda_1) &= 0  & P(\alpha_- \beta_- a_{i_\bot} b_{i_\bot} \lambda_1) &= 0 
\\[0.2cm] P(\alpha_+ \beta_+ a_i b_i \lambda_2) &= \tfrac{1}{8}   & P(\alpha_+ \beta_- a_i b_i \lambda_2) &= 0  \notag
\\ P(\alpha_- \beta_+ a_i b_i \lambda_2) &= 0  & P(\alpha_- \beta_- a_i b_i \lambda_2) &= 0
\\[0.1cm] P(\alpha_+ \beta_+ a_i b_{i_\bot} \lambda_2) &= 0  & P(\alpha_+ \beta_- a_i b_{i_\bot} \lambda_2) &= \tfrac{1}{8} \notag
\\ P(\alpha_- \beta_+ a_i b_{i_\bot} \lambda_2) &= 0 & P(\alpha_- \beta_- a_i b_{i_\bot} \lambda_2) &= 0
\\[0.1cm] P(\alpha_+ \beta_+ a_{i_\bot} b_i \lambda_2) &= 0  & P(\alpha_+ \beta_- a_{i_\bot} b_i \lambda_2) &= 0 \notag
\\ P(\alpha_- \beta_+ a_{i_\bot} b_i \lambda_2) &= \tfrac{1}{8} & P(\alpha_- \beta_- a_{i_\bot} b_i \lambda_2) &= 0
\\[0.1cm] P(\alpha_+ \beta_+ a_{i_\bot} b_{i_\bot} \lambda_2) &= 0  & P(\alpha_+ \beta_- a_{i_\bot} b_{i_\bot} \lambda_2) &= 0 \notag
\\ P(\alpha_- \beta_+ a_{i_\bot} b_{i_\bot} \lambda_2) &= 0 & P(\alpha_- \beta_- a_{i_\bot} b_{i_\bot} \lambda_2) &= \tfrac{1}{8}  
\end{align}
\vspace{-0.4cm}

This distribution is in accordance with the axioms of probability theory; by construction its hidden joint probability has the product form that is characteristic for class (H$_{29}^\alpha$), and it reproduces the perfect correlations and perfect anti\-/correlations.  This explicit example shows that class (H$_{29}^\alpha$) is consistent with the assumptions measurement independence, perfect correlations and perfect \mbox{(anti-)}cor\-relations.

In a similar way one can construct examples of probability distributions for the other classes fulfilling $\neg$(i) and $\neg$(ii). Since (H$^\alpha_{22}$) is symmetric to (H$^\alpha_{29}$) under interchanging the settings, it is clear that the constructed distribution for the latter class can easily be turned into an example for the former by swapping the values of the settings  in each total probability distribution. 
Furthermore, it is straightforward to modify the construction 
such that it yields distributions for the classes 
$\{(\text{H}_{1}^\alpha),…,(\text{H}_{14}^\alpha)\}\backslash\{(\text{H}_{4}^\alpha), (\text{H}_{5}^\alpha), (\text{H}_{10}^\alpha)\}$. 
Note that in these latter classes 
there are more degrees of freedom than in the presented example, so one might freely choose more values of probabilities. 
\parbox{0cm}{} \hfill $\Box$

\subsection{Proof of \autoref{th:dir-npcorr}%2
}
\label{sec:proof-theorem2.1}

\stepcounter{theoremcounter}

\theoremb*

\noindent 
The theorem is equivalent to the conjunction of the following claims: 

\begin{lemma}
\label{lemma-2.1}
A class of probability distributions forms an inconsistent set with measurement independence, nearly perfect correlations and nearly perfect anti\-/correlations if (i) its defining product form involves at most one of the settings. 
\end{lemma}
\begin{lemma}
\label{lemma-2.2}
A class of probability distributions forms a consistent set with  measurement independence, nearly perfect correlations and nearly perfect anti\-/correlations if $\neg$(i) its defining product form involves both settings. 
\end{lemma}

\subsubsection{Proof of  \autoref{lemma-2.1}}

Condition (i), that the product form involves at most one of the settings, is fulfilled by the classes $\{(\text{H}_{17}^\alpha),…,(\text{H}_{32}^\alpha)\}\backslash \{(\text{H}_{22}^\alpha),(\text{H}_{29}^\alpha)\}$. Here we have to show the inconsistency of these classes with the set of assumptions measurement independence, nearly perfect correlations and nearly perfect anti\-/correlations.

The proof runs very similar to our demonstration of \autoref{lemma-1.1}
(in \autoref{sec:proof-theorem1.1}).  
On the one hand, the nearly perfect correlations and nearly perfect anti\-/correlations involve dependences on each of the settings, e.g. the conditions  
\begin{align}
\label{npcorr-1-1}
P(\alpha_\pm \beta_\pm | a_i b_i) &= \frac{1}{2}-\delta_{ii}  &
P(\alpha_\pm \beta_\pm | a_i b_{i_\bot}) &= \delta_{ii_\bot}
\end{align}
reveal a dependence on the setting $\zv b$, while e.g. the conditions  
\begin{align}
\label{npacorr-1-1}
P(\alpha_\pm \beta_\pm | a_i b_i) &= \frac{1}{2}-\delta_{ii}  &
P(\alpha_\pm \beta_\pm | a_{i_\bot} b_i) &= \delta_{i_\bot i} 
\end{align}
show a dependence on the setting $\zv a$. On the other hand, any hidden joint probability that does not involve the setting $\zv b$ 
cannot account for changing values in the empirical joint probability with changing values of $\zv b$ (cf. Eq. \eqref{eq:hjp-ind-b}); so it necessarily contradicts the set of Equations \eqref{npcorr-1-1}. 
And similarly, hidden joint probabilities that are independent of $\zv a$ contradict the set of Equations \eqref{npacorr-1-1}. 

Note that condition (ii) from \autoref{th:dir-pcorr} is not a criterion for inconsistency according to \autoref{th:dir-npcorr}, because the inconsistency in question essentially relies on \emph{strictly} perfect (anti\=/)correlations, which are not assumed in \autoref{th:dir-npcorr}.  
\parbox{0cm}{} \hfill $\Box$

\subsubsection{Proof of \autoref{lemma-2.2}}
\label{sec:proof-lemma2.2} 

The classes fulfilling criterion~$\neg$(i) to involve both settings in their product forms are (H$_{1}^\alpha$), \dots, (H$_{16}^\alpha$), (H$_{22}^\alpha$) and (H$_{29}^\alpha$). 
Here we have to show the consistency of these classes with the set of assumptions measurement independence, nearly perfect correlations and nearly perfect anti\-/correlations.

As in the proof of \autoref{lemma-1.3} (\autoref{sec:proof-lemma1.3}) one can demonstrate the present claim 
by providing an example of a probability distribution for each class that is consistent with these assumptions. 
Since nearly perfect correlations are a weaker requirement than strictly perfect ones, it is clear that for all classes which we have shown to be consistent with the latter---viz. (H$_1^\alpha$), \dots, (H$_{14}^\alpha$)$\backslash\{(\text{H}_{4}^\alpha), (\text{H}_{5}^\alpha), (\text{H}_{10}^\alpha)\}$, (H$_{22}^\alpha$) and (H$_{29}^\alpha$)---are also consistent with the former. 
Therefore, what still needs to be proven here is that measurement independence and nearly perfect (anti\=/)correlations are consistent with those classes fulfilling criterion $(\neg\text{i})$ that are inconsistent with the strictly perfect ones (because they fulfill (ii)):  
$(\text{H}_{4}^\alpha)$, $(\text{H}_{5}^\alpha)$, $(\text{H}_{10}^\alpha)$, $(\text{H}_{15}^\alpha)$ and $(\text{H}_{16}^\alpha)$. 

Again, the best way to find examples of this kind is by constructing them such that the conditions are fulfilled. Here we show how to construct a distribution for class $(\text{H}_{10}^\alpha)$. 
The starting point are the equations for nearly perfect (anti\=/)correlations:
\begin{align}
P(\alpha_\pm \beta_\mp|a_i b_i) &= \delta_{ii} & P(\alpha_\pm \beta_\mp|a_{i_\bot} b_{i_\bot}) &= \delta_{i_\bot i_\bot}  
\\ P(\alpha_\pm \beta_\pm|a_i b_i) &= \tfrac{1}{2}-\delta_{ii} & P(\alpha_\pm \beta_\pm|a_{i_\bot} b_{i_\bot}) &= \tfrac{1}{2}- \delta_{i_\bot i_\bot} 
\\ P(\alpha_\pm \beta_\pm|a_{i_\bot} b_i) &= \delta_{i_\bot i} & P(\alpha_\pm \beta_\pm|a_i b_{i_\bot}) &= \delta_{ii_\bot}  
\\ P(\alpha_\pm \beta_\mp|a_{i_\bot} b_i) &= \tfrac{1}{2}-\delta_{i_\bot i} & P(\alpha_\pm \beta_\mp|a_i b_{i_\bot}) &= \tfrac{1}{2}-\delta_{ii_\bot} 
\end{align}

\noindent
Replacing the empirical probability on the left hand side of each equation by an equivalent expression involving hidden probabilities of the product form,  
$P(\alpha \beta | a b) = \sum_\lambda P(\lambda) \cdot P(\alpha | \beta \lambda) P(\beta | a b \lambda) $, 
yields a set of 16 equations, whose solutions determine probability distributions with the required features.

The $\delta$'s in these equations indicate the deviation from strictly perfect correlations. One might use realistic empirical values for them, but since the task here is merely a conceptual one, one might as well just stipulate any small, positive values. 
Due to the lacking perfectness, the resulting set of equations is more complicated than  that in \autoref{th:ind-pcorr}, and solutions are best determined by appropriate computer algorithms. 
Here, we shall present a solution for the special case $\delta_{i i} = \delta_{i_\bot i_\bot} = \delta_{i_\bot i} = \delta_{i i_\bot} =: \delta$, 
which reads:  
\begin{align}
\label{H14-mPcorr-pf1}
P(\lambda_1) & = \tfrac{1}{2} & P(\lambda_2) & = \tfrac{1}{2} 
\\ \label{H14-mPcorr-pf2} P(\alpha_+ | \beta_+ \lambda_1) &= 0 & P(\alpha_- | \beta_+ \lambda_1) &= 1 
\\ P(\alpha_+ | \beta_+ \lambda_2) &= 1-2 \delta & P(\alpha_- | \beta_+ \lambda_2) &= 2 \delta
\\ \label{H14-mPcorr-pf3}
P(\alpha_+ | \beta_- \lambda_1) &= 2 \delta & P(\alpha_- | \beta_- \lambda_1) &= 1-2 \delta
\\ P(\alpha_+ | \beta_- \lambda_2) &= 1 & P(\alpha_- | \beta_- \lambda_2) &= 0 
\\ \label{H14-mPcorr-pf4} 
P(\beta_+ | a_i b_i \lambda_1) &= 0 & P(\beta_- | a_i b_i \lambda_1) &= 1
\\ P(\beta_+ | a_i b_i \lambda_2) &= 1 & P(\beta_- | a_i b_i \lambda_2) &= 0 
\\ \label{H14-mPcorr-pf5} 
P(\beta_+ | a_i b_{i_\bot} \lambda_1) &= \tfrac{4\delta-1}{2\delta-1} & P(\beta_- | a_i b_{i_\bot} \lambda_1) &=  \tfrac{2\delta}{1-2\delta}
\\ P(\beta_+ | a_i b_{i_\bot} \lambda_2) &= \tfrac{2\delta}{1-2\delta} & P(\beta_- | a_i b_{i_\bot} \lambda_2) &= \tfrac{4\delta-1}{2\delta-1} 
\\ \label{H14-mPcorr-pf6} P(\beta_+ | a_{i_\bot} b_i \lambda_1) &= \tfrac{4\delta-1}{2\delta-1} & P(\beta_- | a_{i_\bot} b_i \lambda_1) &= \tfrac{2\delta}{1-2\delta}
\\ P(\beta_+ | a_{i_\bot} b_i \lambda_2) &= \tfrac{2\delta}{1-2\delta} & P(\beta_- | a_{i_\bot} b_{i_\bot} \lambda_2) &= \tfrac{4\delta-1}{2\delta-1} 
\\ \label{H14-mPcorr-pf7} P(\beta_+ | a_{i_\bot} b_{i_\bot} \lambda_1) &= 0 & P(\beta_- | a_{i_\bot} b_{i_\bot} \lambda_1) &= 1
\\ P(\beta_+ | a_{i_\bot} b_{i_\bot} \lambda_2) &= 1 & P(\beta_- | a_{i_\bot} b_{i_\bot} \lambda_2) &= 0 
\end{align}

\noindent Note that according to this solution all dependences of the product form (H$^\alpha_{10}$) are preserved, because, for instance, we have 
\begin{align}
P(\alpha_+ | \beta_+ \lambda_1) &\neq P(\alpha_+ | \beta_- \lambda_1)  & P(\alpha_+ | \beta_+ \lambda_1) &\neq P(\alpha_+ | \beta_+ \lambda_2) 
\\ P(\beta_+ | a_i b_i \lambda_1) &\neq P(\beta_+ | a_{i_\bot} b_i \lambda_1)  & P(\beta_+ | a_i b_i \lambda_1) &\neq P(\beta_+ | a_i b_{i_\bot} \lambda_1) 
\\ P(\beta_+ | a_i b_i \lambda_1) &\neq P(\beta_+ | a_i b_i \lambda_2) 
\end{align} 

\noindent Finally, when we further assume, say, $P(a_i) = \frac{1}{2}$, $P(a_{i_\bot}) = \frac{1}{2}$, $P(b_i) = \frac{1}{2}$ and $P(b_{i_\bot}) = \frac{1}{2}$,
by the equation $P(\alpha \beta a b \lambda) = P(\alpha | \beta \lambda) P(\beta | a b \lambda) P(\lambda) P(a) P(b)$
the results so far determine the values of the total probability distribution:
\note{WV nicht unbedingt abdrucken}{}

\begin{align}
P(\alpha_+ \beta_+ a_i b_i \lambda_1) &= 0 & P(\alpha_+ \beta_- a_i b_i \lambda_1) &= \tfrac{\delta}{4} \notag
\\ P(\alpha_- \beta_+ a_i b_i \lambda_1) &=0 & P(\alpha_- \beta_- a_i b_i \lambda_1) &= \tfrac{1-2\delta}{8} 
\\ P(\alpha_+ \beta_+ a_i b_{i_\bot} \lambda_1) &= 0 & P(\alpha_+ \beta_- a_i b_{i_\bot} \lambda_1) &= \tfrac{\delta^2}{2(1-2\delta)} \notag 
\\ P(\alpha_- \beta_+ a_i b_{i_\bot} \lambda_1) &= \tfrac{1-4\delta}{8(1-2\delta)} & P(\alpha_- \beta_- a_i b_{i_\bot} \lambda_1) &= \tfrac{\delta}{4} 
\\ P(\alpha_+ \beta_+ a_{i_\bot} b_i \lambda_1) &= 0  & P(\alpha_+ \beta_- a_{i_\bot} b_i \lambda_1) &= \tfrac{\delta^2}{2(1-2\delta)} \notag
\\ P(\alpha_- \beta_+ a_{i_\bot} b_i \lambda_1) &=  \tfrac{1-4\delta}{8(1-2\delta)} & P(\alpha_- \beta_- a_{i_\bot} b_i \lambda_1) &= \tfrac{\delta}{4} 
\\ P(\alpha_+ \beta_+ a_{i_\bot} b_{i_\bot} \lambda_1) &=0 & P(\alpha_+ \beta_- a_{i_\bot} b_{i_\bot} \lambda_1) &= \tfrac{\delta}{4} \notag
\\ P(\alpha_- \beta_+ a_{i_\bot} b_{i_\bot} \lambda_1) &= 0 & P(\alpha_- \beta_- a_{i_\bot} b_{i_\bot} \lambda_1) &=  \tfrac{1-2\delta}{8} 
\\[0.2cm]
P(\alpha_+ \beta_+ a_i b_i \lambda_2) &= \tfrac{1-2\delta}{8}   & P(\alpha_+ \beta_- a_i b_i \lambda_2) &= 0 \notag
\\  P(\alpha_- \beta_+ a_i b_i \lambda_2) &= \tfrac{\delta}{4} & P(\alpha_- \beta_- a_i b_i \lambda_2) &= 0 
\\ P(\alpha_+ \beta_+ a_i b_{i_\bot} \lambda_2) &= \tfrac{\delta}{4} & P(\alpha_+ \beta_- a_i b_{i_\bot} \lambda_2) &=\tfrac{1-4\delta}{8(1-2\delta)} \notag
\\  P(\alpha_- \beta_+ a_i b_{i_\bot} \lambda_2) &= \tfrac{\delta^2}{2(1-2\delta)} & P(\alpha_- \beta_- a_i b_{i_\bot} \lambda_2) &= 0 
\\ P(\alpha_+ \beta_+ a_{i_\bot} b_i \lambda_2) &= \tfrac{\delta}{4} & P(\alpha_+ \beta_- a_{i_\bot} b_i \lambda_2) &= \tfrac{1-4\delta}{8(1-2\delta)} \notag
\\ P(\alpha_- \beta_+ a_{i_\bot} b_i \lambda_2) &= \tfrac{\delta^2}{2(1-2\delta)} & P(\alpha_- \beta_- a_{i_\bot} b_i \lambda_2) &= 0 
\\ P(\alpha_+ \beta_+ a_{i_\bot} b_{i_\bot} \lambda_2) &= \tfrac{1-2\delta}{8} & P(\alpha_+ \beta_- a_{i_\bot} b_{i_\bot} \lambda_2) &= 0 \notag
\\ P(\alpha_- \beta_+ a_{i_\bot} b_{i_\bot} \lambda_2) &= \tfrac{\delta}{4} & P(\alpha_- \beta_- a_{i_\bot} b_{i_\bot} \lambda_2) &= 0 
\end{align}

By construction this distribution has the product form that is characteristic for class (H$_{10}^\alpha$), and it involves measurement independence, nearly perfect correlations for parallel settings and nearly perfect anti\-/correlations for perpendicular settings. This explicitly shows class (H$_{10}^\alpha$) to be consistent with these assumptions. 
In a similar way, one can find examples for classes $(\text{H}_{4}^\alpha)$, $(\text{H}_{5}^\alpha)$, $(\text{H}_{15}^\alpha)$ and $(\text{H}_{16}^\alpha)$ consistent with the mentioned assumptions. 
\parbox{0cm}{} \hfill $\Box$

\subsection{Proof of \autoref{th:ind-pcorr}%3
}
\label{sec:proof-theorem1.2}
\stepcounter{theoremcounter}

\theoremc*

\noindent 
The theorem is equivalent to the conjunction of the following claims: 
\begin{lemma}
\label{lemma-3.1}
Given measurement independence, perfect correlations and perfect anti\-/correlations, a consistent class (i.e. a class that fulfills $\neg$(i) and $\neg$(ii)) implies Bell inequalities if (iii) each factor of its defining product form involves at most one setting.  
\end{lemma}
\begin{lemma}
\label{lemma-3.2}
Given measurement independence, perfect correlations and perfect anti\-/correlations, a consistent class (i.e. a class that fulfills $\neg$(i) and $\neg$(ii)) does not imply Bell inequalities if $\neg$(iii) at least one factor of its defining product form involves both settings.   
\end{lemma}

\subsubsection{Proof of  \autoref{lemma-3.1}}

The set of classes fulfilling $\neg$(i), $\neg$(ii) and (iii) consists of (H$_{22}^\alpha$) and (H$_{29}^\alpha$). Here we have to show that, given measurement independence, perfect correlations and perfect (anti\=/)correlations, each of these classes implies Bell inequalities.  

By usual derivations of Wigner-Bell inequalities, it is well-known that local factorization (H$_{29}^\alpha$) implies Bell inequalities (given measurement independence and perfect correlations; cf. premise (P4) of the Bell argument above). 
Now, it is easy to see that in a very similar way one can use (H$_{22}^\alpha$) to derive Bell inequalities. For, as we have said, (H$_{22}^\alpha$) differs from local factorization only in that the settings in the product form are swapped: Instead of a dependence of each outcome on the local settings each factor involves a dependence on the \emph{distant} setting. Accordingly, the derivation from (H$_{22}^\alpha$) results from the usual one by interchanging the settings in each expression.  
\parbox{0cm}{} \hfill $\Box$

\subsubsection{Proof of  \autoref{lemma-3.2}} 

The classes fulfilling conditions $\neg$(i), $\neg$(ii) and $\neg$(iii)
are (H$_1^\alpha$), \dots,(H$_{14}^\alpha$)$\backslash\{(\text{H}_{4}^\alpha)$, $(\text{H}_{5}^\alpha)$, $(\text{H}_{10}^\alpha)\}$. 
Here we have to show that given the background assumptions measurement independence, perfect correlations and perfect anti\-/correlations, these classes do \emph{not} imply the Bell inequalities, i.e. that  
there is at least one distribution for each class that fulfills measurement independence, perfect correlations, perfect anti\-/correlations and violates the Bell inequalities. 

One way to find such examples is to look at existing hidden-variable theories that successfully explain the statistics of EPR/B experiments. 
In our overview of the classes we have seen that the de-Broglie-Bohm theory falls under different classes depending on which temporal order the experiment has, (H$^\alpha_6$), (H$^\alpha_{9}$) or (H$^\alpha_{12}$). For each of these classes, the probability distribution of the theory provides an example with the desired features. 
Moreover, the example for (H$^\alpha_{9}$) can be turned into one for (H$^\alpha_{8}$) by reversing the dependence on the settings. And similarly, the example for (H$^\alpha_{12}$) can be turned into one for (H$^\alpha_{11}$). 
Since (H$^\alpha_{1}$), (H$^\alpha_{2}$), (H$^\alpha_{3}$) and (H$^\alpha_{7}$) are weaker product forms (involve more dependences) than one or several of the classes (H$^\alpha_6$), (H$^\alpha_{8}$), (H$^\alpha_{9}$), (H$^\alpha_{11}$) or (H$^\alpha_{12}$), by small modifications of the available examples one can construct examples for these classes as well. 

It remains to find examples for classes (H$^\alpha_{13}$) and (H$^\alpha_{14}$). Since there are no theories available for these classes, here the construction has to be from scratch. 
Let me demonstrate how the construction works for class (H$^\alpha_{14}$). 
We first of all take into account the perfect correlations and perfect anti\-/correlations~\eqref{calc1}--\eqref{calc8}. 
This goes, mutatis mutandis, very similar to finding a probability distribution from class (H$_{29}^\alpha$) that is compatible with perfect (anti\=/)correlations (see proof of  \autoref{lemma-1.3}, \autoref{sec:proof-lemma1.3}). We substitute in each Equation \eqref{calc1}--\eqref{calc8} 
\begin{equation}
\label{eq:JP-hidden}
P(\alpha\beta|ab)=\sum_\lambda P(\alpha|\lambda)P(\beta|ab\lambda)P(\lambda). 
\end{equation} 
The first four of the resulting equations have two possible solutions for every $i$ and $\lambda$: 
 \\[-0.25cm]

\noindent \underline{Case~I}: 
\vspace{-0.25cm}
\begin{align}
P(\alpha_+|\lambda) &= 0 & P(\alpha_-|\lambda) &= 1 & 
\\ P(\beta_+|a_ib_i\lambda) &= 0 & P(\beta_-|a_ib_i\lambda) &= 1
& P(\beta_+|a_{i_\bot} b_{i_\bot} \lambda) &= 0 & P(\beta_-|a_{i_\bot} b_{i_\bot} \lambda) &= 1
\\ P(\beta_+|a_{i} b_{i_\bot} \lambda) &= 1   &   P(\beta_-|a_{i} b_{i_\bot} \lambda) &= 0
& P(\beta_+|a_{i_\bot} b_{i} \lambda) &= 1  & P(\beta_-|a_{i_\bot} b_{i} \lambda) &= 0
\end{align}

\noindent \underline{Case~II}: (replace all 0's in case~I by 1 and vice versa) \\[-0.25cm]

Requiring just \emph{any} example we can assume a toy model with only two possible hidden states ($\zv \lambda= \lambda_1, \lambda_2$). Then we might, for instance, choose case~I for $\lambda_1$ and case~II for $\lambda_2$ for all $i$'s. 
Then, by the rewritten equations \eqref{calc5}--\eqref{calc8} it follows $P(\lambda_1) = \frac{1}{2}$ and $P(\lambda_2) = \frac{1}{2}$. 
In this way we have accounted for the perfect correlations as well as for the perfect anti\-/correlations. 

Now it remains to reproduce the EPR/B correlations for non-parallel and non-per\-pen\-dicular settings. A minimal set of such probabilities, which can violate the Bell inequalities (both the usual ones as well as the Wigner-Bell inequalities), can be found if each of the settings $\zv a$ and $\zv b$ has two possible values, e.g. $a_1=0^\circ$, $a_2=30^\circ$, $b_1 = 30^\circ$ and $b_2=60^\circ$. Measuring the quantum state $\psi_0 = (|++\rangle + |--\rangle)/\sqrt{2}$ at these settings yields  the following observable probabilities: 
\begin{align}
\label{emp-prob-H14-PCorr-1}
P(\alpha_\pm \beta_\pm | a_1 b_1) &= \tfrac{3}{8} &  P(\alpha_\pm \beta_\mp | a_1 b_1) &= \tfrac{1}{8} & P(\alpha_\pm \beta_\pm | a_1 b_2) &= \tfrac{1}{8} &  P(\alpha_\pm \beta_\mp | a_1 b_2) & = \tfrac{3}{8} 
\\ \label{emp-prob-H14-PCorr-2} P(\alpha_\pm \beta_\pm | a_2 b_1) &= \tfrac{1}{2} &  P(\alpha_\pm \beta_\mp | a_2 b_1) &= 0  & P(\alpha_\pm \beta_\pm | a_2 b_2) &= \tfrac{3}{8} &  P(\alpha_\pm \beta_\mp | a_2 b_2) &= \tfrac{1}{8}
\end{align}

\noindent These are sixteen equations, and any of the probabilities on their left hand sides can be expressed by the product form of the hidden joint probability \eqref{eq:JP-hidden}. 
As we have derived above, $P(\lambda)$ and $P(\alpha|\lambda)$ are already completely determined by the perfect \mbox{(anti-)}cor\-re\-la\-tions, $P(\beta | a b \lambda)$ partly so (namely only for the parallel settings $a_2=b_1$).  
Inserting these values consistently in \eqref{eq:JP-hidden} 
yields the following values for the missing probabilities $P(\beta|ab\lambda)$:
\begin{align}
P(\beta_+ | a_1b_1\lambda_1) &= \tfrac{1}{4} & P(\beta_- | a_1 b_1\lambda_1) &= \tfrac{3}{4} & P(\beta_+ | a_1 b_1\lambda_2) &= \tfrac{3}{4} & P(\beta_- | a_1 b_1\lambda_2) &= \tfrac{1}{4}
\\ P(\beta_+ | a_1b_2\lambda_1) &= \tfrac{3}{4} & P(\beta_- | a_1 b_2\lambda_1) &= \tfrac{1}{4} & P(\beta_+ | a_1 b_2\lambda_2) &= \tfrac{1}{4} & P(\beta_- | a_1 b_1\lambda_2) &= \tfrac{3}{4}
\\ P(\beta_+ | a_2 b_2\lambda_1) &= \tfrac{1}{4} & P(\beta_- | a_2 b_2\lambda_1) &= \tfrac{3}{4} & P(\beta_+ | a_2 b_2\lambda_2) &= \tfrac{3}{4} & P(\beta_- | a_2 b_2\lambda_2) &= \tfrac{1}{4}
\end{align}

\noindent Finally, choosing, say, $P(a_i) = \frac{1}{2}$, $  P(a_{i_\bot}) = \frac{1}{2}$, $P(b_i) = \frac{1}{2}$ and $P(b_{i_\bot})  = \frac{1}{2}$, 
the formula 
$P(\alpha \beta a b \lambda) = P(\alpha | \lambda) P(\beta | a b \lambda) P(\lambda) P(a) P(b)$
entails the following total probabilities, which constitute the searched for probability distribution: \note{WV evtl. nicht drucken}{}
\vspace{-0.4cm}

\begin{align}%{7}
P(\alpha_+ \beta_+ a_1 b_1 \lambda_1) &= 0  & P(\alpha_+ \beta_- a_1 b_1 \lambda_1) &= 0  \notag
\\ P(\alpha_- \beta_+ a_1 b_1 \lambda_1) &= \tfrac{1}{32} & P(\alpha_- \beta_- a_1 b_1 \lambda_1) &= \tfrac{3}{32} 
\\ P(\alpha_+ \beta_+ a_1 b_2 \lambda_1) &= 0  & P(\alpha_+ \beta_- a_1 b_2 \lambda_1) &= 0 \notag
\\ P(\alpha_- \beta_+ a_1 b_2 \lambda_1) &= \tfrac{3}{32} & P(\alpha_- \beta_- a_1 b_2 \lambda_1) &= \tfrac{1}{32} 
\\ P(\alpha_+ \beta_+ a_2 b_1 \lambda_1) &= 0  & P(\alpha_+ \beta_- a_2 b_1 \lambda_1) &= 0 \notag
\\ P(\alpha_- \beta_+ a_2 b_1 \lambda_1) &= 0 & P(\alpha_- \beta_- a_2 b_1 \lambda_1) &= \tfrac{1}{8} 
\\ P(\alpha_+ \beta_+ a_2 b_2 \lambda_1) &= 0 & P(\alpha_+ \beta_- a_2 b_2 \lambda_1) &= 0 \notag 
\\ P(\alpha_- \beta_+ a_2 b_2 \lambda_1) &= \tfrac{1}{32} & P(\alpha_- \beta_- a_2 b_2 \lambda_1) &= \tfrac{3}{32} 
\\[0.1cm] P(\alpha_+ \beta_+ a_1 b_1 \lambda_2) &= \tfrac{3}{32} & P(\alpha_+ \beta_- a_1 b_1 \lambda_2) &= \tfrac{1}{32}  \notag
\\ P(\alpha_- \beta_+ a_1 b_1 \lambda_2) &= 0 & P(\alpha_- \beta_- a_1 b_1 \lambda_2) &= 0
\\ P(\alpha_+ \beta_+ a_1 b_2 \lambda_2) &= \tfrac{1}{32} & P(\alpha_+ \beta_- a_1 b_2 \lambda_2) &= \tfrac{3}{32} \notag
\\ P(\alpha_- \beta_+ a_1 b_2 \lambda_2) &= 0 & P(\alpha_- \beta_- a_1 b_2 \lambda_2) &= 0
\\ P(\alpha_+ \beta_+ a_2 b_1 \lambda_2) &= \tfrac{1}{8} & P(\alpha_+ \beta_- a_2 b_1 \lambda_2) &= 0 \notag
\\ P(\alpha_- \beta_+ a_2 b_1 \lambda_2) &= 0 & P(\alpha_- \beta_- a_2 b_1 \lambda_2) &= 0
\\ P(\alpha_+ \beta_+ a_2 b_2 \lambda_2) &= \tfrac{3}{32} & P(\alpha_+ \beta_- a_2 b_2 \lambda_2) &= \tfrac{1}{32} \notag
\\ P(\alpha_- \beta_+ a_2 b_2 \lambda_2) &= 0 & P(\alpha_- \beta_- a_2 b_2 \lambda_2) &= 0 
\end{align} 

\vspace{-0.4cm}
\noindent Note that here we have not explicitly noted the probabilities for parallel or perpendicular settings, but by constructing the distribution in the indicated way we have implicitly taken account of the perfect (anti\=/)correlations at these settings, and it is straight forward  to extent the distribution to include these settings as well (the distribution just becomes much longer, when for each measurement setting at one side one includes a parallel and a perpendicular setting at the other side).  

This completes our construction of a distribution from class (H$^\alpha_{14}$) which respects, measurement independence, perfect correlations, perfect anti\-/correlations and violates the Bell inequalities. In a similar way, one can construct an example for class (H$^\alpha_{13}$), which differs from (H$^\alpha_{14}$) just in that the dependence on both settings is not in the second but in the first factor of its product form. 
\parbox{0cm}{} \hfill $\Box$

\subsection{Proof of \autoref{th:ind-npcorr}%4
}
\label{sec:proof-theorem2.2}
\stepcounter{theoremcounter} 

\theoremd*

\noindent 
The theorem is equivalent to the conjunction of the following claims: 
\begin{lemma}
\label{lemma-4.1}
Given measurement independence, nearly perfect correlations and nearly perfect anti\-/correlations, a consistent class (i.e. a class that fulfills $\neg$(i)) 
implies Bell inequalities if (iii) each factor of its defining product form involves at most one setting. 
\end{lemma}
\begin{lemma} 
\label{lemma-4.2}
Given measurement independence, nearly perfect correlations and nearly perfect anti\-/correlations, a consistent class (i.e. a class that fulfills $\neg$(i)) 
does not imply Bell inequalities if $\neg$(iii) at least one factor of its defining product form involves both settings.
\end{lemma}

\subsubsection{Proof of \autoref{lemma-4.1}}

The set of classes fulfilling $\neg$(i) and (iii) consists of (H$_{15}^\alpha$), (H$_{16}^\alpha$), (H$_{22}^\alpha$) and (H$_{29}^\alpha$). 
It has to be shown that given measurement independence, nearly perfect correlations and nearly perfect anti\-/correlations, each of these classes implies Bell inequalities. 

It suffices to show that under these conditions $(\text{H}_{16}^\alpha)$ implies Bell inequalities, because then it is clear that the other classes imply the inequalities as well: 
$(\text{H}_{15}^\alpha)$ only differs from $(\text{H}_{16}^\alpha)$ in that the settings are swapped in the product form,  
local factorization $(\text{H}_{29}^\alpha)$ is a stronger product form than $(\text{H}_{16}^\alpha)$, and 
$(\text{H}_{22}^\alpha)$ is a stronger form than $(\text{H}_{15}^\alpha)$.
So we now demonstrate the derivation for class $(\text{H}_{16}^\alpha)$.

We define two partitions for the values of $\zv \lambda$: \\

\setlength{\tabcolsep}{5mm}
\renewcommand{\arraystretch}{1.5}
\begin{tabular}{ll}
\underline{Partition $\prescript{\alpha\!}{}{\Lambda}{^i}$} 
& \underline{Partition $\prescript{\beta\!}{}{\Lambda}{^j}$} \\
$\prescript{\alpha \!}{}{\Lambda}{_1^i} :=\{\lambda|0 ≤P(\alpha_-|\beta_+a_i\lambda)≤\epsilon_1\}$  
& $\prescript{\beta\!}{}{\Lambda}{_1^j}:=\{\lambda|0 ≤P(\beta_+|b_j\lambda)≤\epsilon_1\}$ \\
$\prescript{\alpha\!}{}{\Lambda}{_2^i}:=\{\lambda|\epsilon_1<P(\alpha_-|\beta_+a_i\lambda)<1-\epsilon_2\}$
& $\prescript{\beta\!}{}{\Lambda}{_2^j}:=\{\lambda|\epsilon_1<P(\beta_+|b_j\lambda)<1-\epsilon_2\}$\\
$\prescript{\alpha\!}{}{\Lambda}{_3^i}:=\{\lambda| 1-\epsilon_2 \leq P(\alpha_-|\beta_+a_i\lambda) ≤1 \}$
& $\prescript{\beta\!}{}{\Lambda}{_3^j}:=\{\lambda| 1-\epsilon_2 \leq P(\beta_+|b_j\lambda) ≤1 \}$
\end{tabular}\\[0.25cm]
The allowed values for the parameters of the partitions are $\epsilon_1 ≤ \frac{1}{2}$ and $\epsilon_2 < 1-\epsilon_1$.  Note that each value $i$ or $j$, respectively, defines a different partition. The two partitions give rise to a more fine-grained partition with nine elements ($\prescript{\alpha\!}{}{\Lambda}{_1^i} \cap \prescript{\beta\!}{}{\Lambda}{_1^j}$, $\prescript{\alpha\!}{}{\Lambda}{_1^i} \cap \prescript{\beta\!}{}{\Lambda}{_2^j}$, …), which, by the product form $(\text{H}_{16}^\alpha)$, imply restrictions for the hidden joint probability $P(\alpha_- \beta_+ | a_i b_j \lambda)$.  We introduce the shorthand $\Lambda^{ij}_{kl} := \prescript{\alpha\!}{}{\Lambda}{_k^i} \cap \prescript{\beta\!}{}{\Lambda}{_l^j}$ for the intersection  between elements of the two partitions. 

Using the product form $(\text{H}_{16}^\alpha)$ as well as (some of) the conditions for nearly perfect anti\hyp{}correlations, 
\begin{align}
 P(\alpha_- \beta_+ | a_i b_i) = \delta_{ii} &&
P(\alpha_- \beta_+ | a_i b_{i_\bot}) = \tfrac{1}{2} - \delta_{i i_\bot}
\end{align} 
(where {$\delta_{ii}$ and $\delta_{ii_\bot}$ are positive and small 
and for the sake of simplicity, we assume $\delta_{ii}=\delta_{ii_\bot} =: \delta$ in the following), we can derive constraints on the weights of the elements of the partitions:
\begin{align}
\tag{L1} \frac{P(\alpha_- \beta_+ | a_i b_j)-\epsilon_1}{1-\epsilon_1} \stackrel{\text{(a)}}{≤} & P([\prescript{\alpha\!}{}{\Lambda}{_2^i} \cup \prescript{\alpha\!}{}{\Lambda}{_3^i}] \cap  [\prescript{\beta\!}{}{\Lambda}{_2^j} \cup \prescript{\beta\!}{}{\Lambda}{_3^j}])\stackrel{\text{(b)}}{≤} \frac{P(\alpha_- \beta_+ | a_i b_j)}{\epsilon_1^2}
\\
\tag{L2} 1-\frac{P(\alpha_- \beta_+ | a_i b_j)}{\epsilon_1^2} \stackrel{\text{(a)}}{≤}  &P(\prescript{\alpha\!}{}{\Lambda}{_1^i} \cup \prescript{\beta\!}{}{\Lambda}{_1^j}) \stackrel{\text{(b)}}{≤}  \frac{1-P(\alpha_- \beta_+ | a_i b_j)}{1-\epsilon_1} 
\\
\tag{L3} \frac{1}{2}-\frac{2\delta+\epsilon_1}{2 (1-\epsilon_1)} \stackrel{(a)}{≤} & P(\prescript{x\!}{}{\Lambda}{_2^i} \cup \prescript{x\!}{}{\Lambda}{_3^i}) \stackrel{(b)}{≤} \frac{1}{2}+\frac{\delta}{\epsilon_1^2}+\frac{2\delta+\epsilon_1}{2 (1-\epsilon_1)} \quad (x= \alpha, \beta) 
\\
\tag{L4} \frac{1}{2}-\frac{\delta}{\epsilon_1^2}-\frac{2\delta+\epsilon_1}{2 (1-\epsilon_1)} \stackrel{(a)}{≤} & P(\prescript{x\!}{}{\Lambda}{_1^i})  \stackrel{(b)}{≤} \frac{1}{2}+\frac{2\delta+\epsilon_1}{2(1-\epsilon_1)} \quad (x= \alpha, \beta)
\\
\tag{L5} \frac{1}{2}+\frac{2\delta +\epsilon_1}{2(1-\epsilon_1)}-\frac{2\delta+\epsilon_1}{\epsilon_2} &- \delta \frac{1-\epsilon_1-\epsilon_2}{\epsilon_1^2 \epsilon_2} \stackrel{\text{(a)}}{≤} P(\prescript{x\!}{}{\Lambda}{_3^i}) \stackrel{\text{(b)}}{≤} \frac{1}{2} + \frac{\delta}{\epsilon_1^2} + \frac{2\delta+\epsilon_1}{2 (1-\epsilon_1)} 
\\
\tag{L6} & P(\prescript{x\!}{}{\Lambda}{_2^i})  ≤ \frac{2\delta +\epsilon_1}{\epsilon_2} + \frac{\delta (1-\epsilon_1)}{\epsilon_1^2 \epsilon_2} \quad (x= \alpha, \beta)
\\
\tag{L7} & P(\Lambda_{33}^{ij})  ≤ \frac{P(\alpha_- \beta_+ | a_i b_j)}{(1-\epsilon_2)^2}
\\
\tag{L8} & P(\Lambda_{22}^{ij} \cup \Lambda_{23}^{ij} \cup \Lambda_{32}^{ij}) ≤ 2 (\frac{2\delta +\epsilon_1}{\epsilon_2} + \frac{\delta (1-\epsilon_1)}{\epsilon_1^2 \epsilon_2})
\end{align}
 
In order for (L1)–(L8) to yield relevant restrictions, $\delta$, $\epsilon_1$ and $\epsilon_2$ must be small, but $\epsilon_1$ and $\epsilon_2$ must be considerably greater than $\delta$ such that $\frac{\delta}{\epsilon_1^2 \epsilon_2}$ (see L5a) is still small. Furthermore, the relevance of some of the expressions crucially depends on the settings: for (L1a) and (L2b) to be interesting statements, i.e. to exclude logically possible probability values, it must be the case that $P(\alpha_- \beta_+ | a_i b_j) > \epsilon_1$, i.e. $|i-j| \gg 0$; likewise, (L1b) and (L2a), respectively, require $P(\alpha_- \beta_+ | a_i b_j)< \epsilon_1^2$, i.e. $i \approx j$. (Note that since $P(\alpha_- \beta_+ | a_i b_j)  ≤ \frac{1}{2}$ for all $i,j$, (L7) already gives a relevant supremum if $\epsilon_2$ is small.)  

If these conditions are met, the above constraints roughly say that for every $i$, the values of $\lambda$ divide as follows: $P(\prescript{x\!}{}{\Lambda}{_1^i}) \approx 1/2$ (L4), $P(\prescript{x\!}{}{\Lambda}{_2^i}) \approx 0$ (L6) and $P(\prescript{x\!}{}{\Lambda}{_3^i}) \approx 1/2$ (L5). Since $\prescript{x\!}{}{\Lambda}{_2^i}$ is small for any setting $i$, the intersections involving it — $\Lambda_{23}^{ij}$, $\Lambda_{32}^{ij}$, $\Lambda_{21}^{ij}$, $\Lambda_{12}^{ij}$, $\Lambda_{22}^{ij}$ — are small and cannot contribute substantially to the value of $P(\alpha_- \beta_+ | a_i b_i)$. The value of the probability rather crucially depends on the weights of the intersections $\Lambda_{33}^{ij}$, $\Lambda_{31}^{ij}$, $\Lambda_{13}^{ij}$, $\Lambda_{11}^{ij}$, which is best illustrated with the extreme cases of parallel and perpendicular settings: If $i=j$, $P(\Lambda_{33}^{ii}) \approx 0$  (L7), hence, since $P(\prescript{x\!}{}{\Lambda}{_3^i}) \approx 1/2$ (L5), we have $P(\Lambda_{31}^{ii}) \approx 1/2$ and $P(\Lambda_{13}^{ii})\approx 1/2$; therefore $P(\Lambda_{11}^{ii}) \approx 0$. Vice versa, if $j=i_\bot$,  $P(\Lambda_{33}^{ii_\bot}) \approx 1/2$  (L7), hence, since $P(\prescript{x\!}{}{\Lambda}{_3^i}) \approx 1/2$ (L5),  we have $P(\Lambda_{31}^{ii_\bot}) \approx 0$ and $P(\Lambda_{13}^{ii_\bot})\approx 0$; therefore $P(\Lambda_{11}^{ii_\bot})\approx 1/2$. For the settings in between parallel and perpendicular there is a smooth transition between these extreme cases. This particular behaviour of the partitions for $\lambda$ noted in (L1)–(L8) can be used to derive a Bell inequality. Before we shall demonstrate this, we  sketch their proofs: 
\\

\noindent \underline{Proof of (L1a)}: We proceed indirectly. Supposing one would have 
\begin{equation}
P([\prescript{\alpha\!}{}{\Lambda}{_2^i} \cup \prescript{\alpha\!}{}{\Lambda}{_3^i}] \cap  [\prescript{\beta\!}{}{\Lambda}{_2^j} \cup \prescript{\beta\!}{}{\Lambda}{_3^j}]) = \frac{P(\alpha_- \beta_+ | a_i b_j)-\epsilon_1}{1-\epsilon_1} - \mu 
\end{equation}
with $\mu>0$, 
one can derive a contradiction: 
\begin{align} 
P(\alpha_- \beta_+ | a_i b_j) &= \sum_{\lambda \in \Lambda} P(\lambda) P(\alpha_- \beta_+ | a_i b_j \lambda) \notag
\\
 &= \sum_{\lambda \in (\prescript{\alpha\!}{}{\Lambda}{_1^i} \cup \prescript{\beta\!}{}{\Lambda}{_1^j})}  P(\lambda) \underbrace{P(\alpha_- | \beta_+ a_i  \lambda) P(\beta_+ | b_j \lambda)}_{≤\epsilon_1} 
\notag\\
& \hphantom{= \;} + \sum_{\lambda \in ([\prescript{\alpha\!}{}{\Lambda}{_2^i} \cup \prescript{\alpha\!}{}{\Lambda}{_3^i}] \cap  [\prescript{\beta\!}{}{\Lambda}{_2^j} \cup \prescript{\beta\!}{}{\Lambda}{_3^j}])} P(\lambda) \underbrace{P(\alpha_- | \beta_+ a_i  \lambda)}_{≤ 1} \underbrace{P(\beta_+ | b_j  \lambda)}_{≤ 1} 
\notag\\
&≤ \epsilon_1 \underbrace{\sum_{\lambda \in (\prescript{\alpha\!}{}{\Lambda}{_1^i} \cup \prescript{\beta\!}{}{\Lambda}{_1^j})}  P(\lambda)}_{=1- (\frac{P(\alpha_- \beta_+ | a_i b_j)-\epsilon_1}{1-\epsilon_1} - \mu)} + \underbrace{\sum_{\lambda \in ([\prescript{\alpha\!}{}{\Lambda}{_2^i} \cup \prescript{\alpha\!}{}{\Lambda}{_3^i}] \cap  [\prescript{\beta\!}{}{\Lambda}{_2^j} \cup \prescript{\beta\!}{}{\Lambda}{_3^j}])} P(\lambda)}_{=\frac{P(\alpha_- \beta_+ | a_i b_j)-\epsilon_1}{1-\epsilon_1} - \mu}  
\notag\\
& = P(\alpha_- \beta_+ | a_i b_j) - \mu (1-\epsilon_1) < P(\alpha_- \beta_+ | a_i b_j).
\end{align}

\begin{align}
NB: %\hspace{4cm}
&& \prescript{\alpha\!}{}{\Lambda}{_1^i} \cup \prescript{\beta\!}{}{\Lambda}{_1^j} &= \Lambda_{11}^{ij} \cup \Lambda_{12}^{ij} \cup \Lambda_{13}^{ij} \cup \Lambda_{21}^{ij} \cup \Lambda_{31}^{ij} \\
&& [\prescript{\alpha\!}{}{\Lambda}{_2^i} \cup \prescript{\alpha\!}{}{\Lambda}{_3^i}] \cap  [\prescript{\beta\!}{}{\Lambda}{_2^j} \cup \prescript{\beta\!}{}{\Lambda}{_3^j}] &= \Lambda_{22}^{ij} \cup \Lambda_{23}^{ij} \cup \Lambda_{32}^{ij} \cup \Lambda_{33}^{ij} 
\end{align}\\

\noindent \underline{Proof of (L1b)}: Indirectly. Assuming $P([\prescript{\alpha\!}{}{\Lambda}{_2^i} \cup \prescript{\alpha\!}{}{\Lambda}{_3^i}] \cap  [\prescript{\beta\!}{}{\Lambda}{_2^j} \cup \prescript{\beta\!}{}{\Lambda}{_3^j}]) > \frac{P(\alpha_- \beta_+ | a_i b_j)}{\epsilon_1^2}$ yields a contradiction:
\begin{align}
P(\alpha_- \beta_+ | a_i b_j) &= \sum_{\lambda \in \Lambda} P(\lambda) P(\alpha_- \beta_+ | a_i b_j \lambda) 
\notag\\
&= \sum_{\lambda \in (\prescript{\alpha\!}{}{\Lambda}{_1^i} \cup \prescript{\beta\!}{}{\Lambda}{_1^j})} P(\lambda) \underbrace{P(\alpha_- | \beta_+ a_i  \lambda) P(\beta_+ | b_j \lambda)}_{≥0} 
\notag \\
& \hphantom{= \;} + \sum_{\lambda \in ([\prescript{\alpha\!}{}{\Lambda}{_2^i} \cup \prescript{\alpha\!}{}{\Lambda}{_3^i}] \cap  [\prescript{\beta\!}{}{\Lambda}{_2^j} \cup \prescript{\beta\!}{}{\Lambda}{_3^j}])} P(\lambda) \underbrace{P(\alpha_- | \beta_+ a_i  \lambda)}_{> \epsilon_1} \underbrace{P(\beta_+ | b_j  \lambda)}_{> \epsilon_1} 
\notag\\   
& ≥ \epsilon_1^2 \underbrace{\sum_{\lambda \in  ([\prescript{\alpha\!}{}{\Lambda}{_2^i} \cup \prescript{\alpha\!}{}{\Lambda}{_3^i}] \cap  [\prescript{\beta\!}{}{\Lambda}{_2^j} \cup \prescript{\beta\!}{}{\Lambda}{_3^j}])} P(\lambda)}_{>\frac{P(\alpha_- \beta_+ | a_i b_j)}{\epsilon_1^2}} > P(\alpha_- \beta_+ | a_i b_j).
\end{align}\\

\noindent \underline{Proof of (L2)}: The claim follows immediately from (L1) and the fact
\begin{equation}
P(\prescript{\alpha\!}{}{\Lambda}{_1^i} \cup \prescript{\beta\!}{}{\Lambda}{_1^j})   = P(\overline{[\prescript{\alpha\!}{}{\Lambda}{_2^i} \cup \prescript{\alpha\!}{}{\Lambda}{_3^i}] \cap  [\prescript{\beta\!}{}{\Lambda}{_2^j} \cup \prescript{\beta\!}{}{\Lambda}{_3^j}]}) = 1- P([\prescript{\alpha\!}{}{\Lambda}{_2^i} \cup \prescript{\alpha\!}{}{\Lambda}{_3^i}] \cap  [\prescript{\beta\!}{}{\Lambda}{_2^j} \cup \prescript{\beta\!}{}{\Lambda}{_3^j}]).
\end{equation}\\

\noindent \underline{Proof of (L3a)}: We show the claim for $x=\alpha$ (it can be demonstrated mutatis mutandis for $x=\beta$). We proceed indirectly. Assuming $P(\prescript{\alpha\!}{}{\Lambda}{_2^i} \cup \prescript{\alpha\!}{}{\Lambda}{_3^i}) = \frac{1}{2}-\frac{2\delta+\epsilon}{2 (1-\epsilon)} - \mu$ 
with $\mu > 0$, yields a contradiction: 
\begin{align}
\frac{1}{2} - \delta &\stackrel{(66)}{=} P(\alpha_- \beta_+ | a_i b_{i_\bot}) = \sum_{\lambda \in \Lambda} P(\lambda) P(\alpha_- \beta_+ | a_i b_{i_\bot} \lambda)
\notag\\
&= \sum_{\lambda \in \prescript{\alpha\!}{}{\Lambda}{_1^i}} P(\lambda) \underbrace{P(\alpha_- | \beta_+ | a_i  \lambda)}_{≤\epsilon_1} \underbrace{P(\beta_+ | b_{i_\bot} \lambda)}_{≤1} + \sum_{\lambda \in (\prescript{\alpha\!}{}{\Lambda}{_2^i} \cup \prescript{\alpha\!}{}{\Lambda}{_3^i})} P(\lambda) \underbrace{P(\alpha_- | \beta_+ | a_i  \lambda)}_{≤1} \underbrace{P(\beta_+ | b_{i_\bot}  \lambda)}_{≤1}
\notag\\
&≤ \epsilon_1 \underbrace{\sum_{\lambda \in \prescript{\alpha\!}{}{\Lambda}{_1^i}} P(\lambda)}_{=1-(\frac{1}{2}-\frac{2\delta+\epsilon_1}{2 (1-\epsilon_1)}-\mu)} + \underbrace{\sum_{\lambda \in (\prescript{\alpha\!}{}{\Lambda}{_2^i} \cup \prescript{\alpha\!}{}{\Lambda}{_3^i})} P(\lambda)}_{=\frac{1}{2}-\frac{2\delta+\epsilon_1}{2 (1-\epsilon_1)}-\mu}  = \frac{1}{2}-\delta -(1-\epsilon_1)\mu< \frac{1}{2}-\delta. 
\end{align}\\

\noindent \underline{Proof of (L3b)}: We show the claim for $x=\alpha$ (it can be demonstrated mutatis mutandis for $x=\beta$). We proceed indirectly. Suppose one would have $P(\prescript{\alpha\!}{}{\Lambda}{_2^i} \cup \prescript{\alpha\!}{}{\Lambda}{_3^i})  > \frac{1}{2}+\frac{\delta}{\epsilon_1^2}+\frac{2\delta+\epsilon_1}{2 (1-\epsilon_1)}$. By (L3a) (for $x=\beta$) we know that $P(\prescript{\beta\!}{}{\Lambda}{_2^i} \cup \prescript{\beta\!}{}{\Lambda}{_3^i}) ≥ \frac{1}{2}-\frac{2\delta+\epsilon_1}{2 (1-\epsilon_1)}$. Using the generally valid expression 
\begin{equation}
P(A \cap B) = P(A) + P(B) -P(A \cup B) ≥ P(A) + P(B) - 1
\end{equation}
one can derive a contradiction:
\begin{equation}
\frac{\delta}{\epsilon_1^2} \stackrel{\text{(L1b)}}{≥} P([\prescript{\alpha\!}{}{\Lambda}{_2^i} \cup \prescript{\alpha\!}{}{\Lambda}{_3^i}] \cap  [\prescript{\beta\!}{}{\Lambda}{_2^i} \cup \prescript{\beta\!}{}{\Lambda}{_3^i}]) ≥ P(\prescript{\alpha\!}{}{\Lambda}{_2^i} \cup \prescript{\alpha\!}{}{\Lambda}{_3^i}) + P(\prescript{\beta\!}{}{\Lambda}{_2^i} \cup \prescript{\beta\!}{}{\Lambda}{_3^i}) -1 > \frac{\delta}{\epsilon_1^2}
\end{equation}\\

 \noindent \underline{Proof of (L4)}: The claim follows from (L3) and the fact $P(\prescript{x\!}{}{\Lambda}{_1^i}) = 1-P(\prescript{x\!}{}{\Lambda}{_2^i} \cup \prescript{x\!}{}{\Lambda}{_3^i})$. \\

\noindent \underline{Proof of (L5a)}: We show the claim for $x=\alpha$ (it can be demonstrated mutatis mutandis for $x=\beta$). Let us define $S:= \frac{1}{2}+\frac{2\delta +\epsilon_1}{2(1-\epsilon_1)}-\frac{2\delta+\epsilon_1}{\epsilon_2}- \delta \frac{1-\epsilon_1-\epsilon_2}{\epsilon_1^2 \epsilon_2}$ and $T:=\frac{1}{2}+\frac{\delta}{\epsilon_1^2}+\frac{2\delta+\epsilon_1}{2 (1-\epsilon_1)}$. We proceed indirectly. Suppose one would have $P(\prescript{\alpha\!}{}{\Lambda}{_3^i}) = S - \mu$ with $\mu > 0$. Even when we allow that $P(\prescript{\alpha\!}{}{\Lambda}{_2^i} \cup \prescript{\alpha\!}{}{\Lambda}{_3^i}) = T$ is maximal (L3b), and hence  $P(\prescript{\alpha\!}{}{\Lambda}{_2^i}) = T - (S - \mu)$ and $P(\prescript{\alpha\!}{}{\Lambda}{_1^i}) = 1-  T$, a contradiction follows:
\begin{align}
\frac{1}{2} - \delta &\stackrel{(66)}{=}   P(\alpha_- \beta_+ | a_i b_{i_\bot}) = \sum_{\lambda \in \Lambda} P(\lambda) P(\alpha_- \beta_+ | a_i b_{i_\bot} \lambda) 
\notag\\
 & =  \sum_{\lambda \in \prescript{\alpha\!}{}{\Lambda}{_1^i}} P(\lambda) \underbrace{P(\alpha_- | \beta_+ | a_i  \lambda)}_{≤\epsilon_1} \underbrace{P(\beta_+ | b_{i_\bot} \lambda)}_{≤1} + \sum_{\lambda \in  \prescript{\alpha\!}{}{\Lambda}{_2^i}} P(\lambda) \underbrace{P(\alpha_- | \beta_+ | a_i  \lambda)}_{<1-\epsilon_2} \underbrace{P(\beta_+ | b_{i_\bot} \lambda)}_{≤1} 
\notag \\
& \hphantom{= \;} +  \sum_{\lambda \in \prescript{\alpha\!}{}{\Lambda}{_3^i}} P(\lambda) \underbrace{P(\alpha_- | \beta_+ | a_i  \lambda)}_{≤1} \underbrace{P(\beta_+ | b_{i_\bot}  \lambda)}_{≤1}
\notag\\
& ≤   \epsilon_1 \underbrace{\sum_{\lambda \in \prescript{\alpha\!}{}{\Lambda}{_1^i}} P(\lambda)}_{=1-T} + (1-\epsilon_2) \underbrace{\sum_{\lambda \in \prescript{\alpha\!}{}{\Lambda}{_2^i}} P(\lambda)}_{=T-S+\mu} + \underbrace{\sum_{\lambda \in \prescript{\alpha\!}{}{\Lambda}{_3^i}} P(\lambda)}_{=S-\mu} = \frac{1}{2}-\delta -\epsilon_2 \mu < \frac{1}{2}-\delta.
\end{align}\\

\noindent \underline{Proof of (L5b)}: The claim follows directly from (L3b) and the fact \mbox{$P(\prescript{x\!}{}{\Lambda}{_3^i}) ≤ P(\prescript{x\!}{}{\Lambda}{_2^i} \cup \prescript{x\!}{}{\Lambda}{_3^i})$}. \\

\noindent \underline{Proof of (L6)}: $P(\prescript{x\!}{}{\Lambda}{_2^i}) = P(\prescript{x\!}{}{\Lambda}{_2^i} \cup \prescript{x\!}{}{\Lambda}{_3^i})-P(\prescript{x\!}{}{\Lambda}{_3^i}) + \underbrace{P(\prescript{x\!}{}{\Lambda}{_2^i} \cap \prescript{x\!}{}{\Lambda}{_3^i})}_{=0} \stackrel{\text{(L3b),(L5a)}}{≤} \frac{2\delta +\epsilon_1}{\epsilon_2} + \frac{\delta (1-\epsilon_1)}{\epsilon_1^2 \epsilon_2}$. \\

\noindent \underline{Proof of (L7)}: We proceed indirectly. Supposing $P(\Lambda_{33}^{ij}) > \frac{P(\alpha_- \beta_+ | a_i b_j)}{(1-\epsilon_2)^2}$, one can derive the following contradiction: 
\begin{align}
P(\alpha_- \beta_+ | a_i b_j) & ≥ \sum_{\lambda \in \Lambda_{33}^{ij}} P(\lambda) \underbrace{P(\alpha_- \beta_+ | a_i b_j  \lambda)}_{≥ (1-\epsilon_2)^2} \notag \\
& ≥ (1-\epsilon_2)^2 \underbrace{\sum_{\lambda \in \Lambda_{33}^{ij}} P(\lambda)}_{> \frac{P(\alpha_- \beta_+ | a_i b_j)}{(1-\epsilon_2)^2}} > P(\alpha_- \beta_+ | a_i b_j).
\end{align}\\ 

\noindent \underline{Proof of (L8)}: 
\begin{equation}
P(\Lambda_{22}^{13} \cup \Lambda_{23}^{13} \cup \Lambda_{32}^{13}) = \underbrace{\sum_{\Lambda_{22}^{13}  \cup \Lambda_{23}^{13}} P(\lambda)}_{≤P(\prescript{\alpha\!}{}{\Lambda}{_2^1})} + \underbrace{\sum_{\Lambda_{32}^{13}} P(\lambda)}_{≤P(\prescript{\beta\!}{}{\Lambda}{_2^3})} \stackrel{\text{(L6)}}{≤} 2(\frac{2\delta +\epsilon_1}{\epsilon_2} + \frac{\delta (1-\epsilon_1)}{\epsilon_1^2 \epsilon_2}).
\end{equation}
\\

\noindent This completes the proofs of (L1)–(L8). 

Given the estimations for the weights of the elements of the partitions (L1)–(L8)
one can derive a generalised Wigner-Bell inequality. 
Consider the inequality 
\begin{equation}
P(A \cap D) ≤ P(A \cap B) + P(C \cap D) + P(\overline{B \cup C}),
\end{equation}
which in general holds for any events {$A, B, C, D$} of a measurable space, as can easily be seen  by rewriting and estimating the probability on the left hand side: 
\begin{align}
P(A \cap D) &=P(A \cap D \cap [(B \cup C) \cup (\overline{B \cup C})]) \notag\\
&= P(A \cap D \cap (B \cup C)) + P(A \cap D \cap (\overline{B \cup C})) \notag\\
& ≤ P(A \cap D \cap B) + P(A \cap D \cap C) + P(A \cap D \cap (\overline{B \cup C})) \notag\\
&≤ P(A \cap B) + P(C \cap D) + P(\overline{B \cup C}) 
\end{align}

Assuming 
 $A = \prescript{\alpha\!}{}{\Lambda}{_3^1}$, $B=\prescript{\beta\!}{}{\Lambda}{_3^2}$, $C=\prescript{\alpha\!}{}{\Lambda}{_3^2}$, $D=\prescript{\beta\!}{}{\Lambda}{_3^3}$
gives the inequality 

\begin{equation}
\label{WB-ineq-basic}
P(\Lambda_{33}^{13}) ≤ P(\Lambda_{33}^{12}) + P(\Lambda_{33}^{23}) + P(\overline{\prescript{\alpha\!}{}{\Lambda}{_3^2} \cup \prescript{\beta\!}{}{\Lambda}{_3^2}}). 
\end{equation}
This inequality can be transformed to yield a generalized Wigner-Bell inequality. We have to rewrite the inequality such that it involves only empirically accessible probabilities, i.e. probabilities that do not involve the hidden state $\zv \lambda$, and this can be done by using the estimates for the 
weights of the elements of the partitions (L1)–(L8). 
Especially, we have to find a lower estimate for the left hand side of the inequality and an upper estimate for its right hand side. We start by deriving the former: 

\begin{align}
P(\Lambda_{33}^{13}) & \stackrel{(\sigma\text{-additivity})}{=} \sum_{\lambda \in \Lambda_{33}^{13}} P(\lambda) ≥ \sum_{\lambda \in \Lambda_{33}^{13}} P(\lambda) P(\alpha_- \beta_+ | a_1 b_3 \lambda) \notag \\
&=  \sum_{\lambda \in \Lambda} P(\lambda) P(\alpha_- \beta_+ | a_1 b_3 \lambda) - \sum_{\lambda \in \overline{\Lambda_{33}^{13}}} P(\lambda) P(\alpha_- \beta_+ | a_1 b_3 \lambda) \notag\\
& = P(\alpha_- \beta_+ | a_1 b_3) - \sum_{\lambda \in (\prescript{\alpha\!}{}{\Lambda}{_1^1} \cup \prescript{\beta\!}{}{\Lambda}{_1^3})} P(\lambda) \underbrace{P(\alpha_- \beta_+ | a_1 b_3 \lambda)}_{≤\epsilon_1} \\
& \hphantom{= \;} - \sum_{\lambda \in (\Lambda_{22}^{13} \cup \Lambda_{23}^{13} \cup \Lambda_{32}^{13})} P(\lambda) \underbrace{P(\alpha_- \beta_+ | a_1 b_3 \lambda)}_{≤(1-\epsilon_2)} \notag\\
& ≥ P(\alpha_- \beta_+ | a_1 b_3) - \epsilon_1 \underbrace{\sum_{\lambda \in (\prescript{\alpha\!}{}{\Lambda}{_1^1}{\Lambda} \cup \prescript{\beta\!}{}{\Lambda}{_1^3})} P(\lambda)}_{\stackrel{\text{(L2b)}}{≤} \frac{1-P(\alpha_- \beta_+ | a_1 b_3)}{1-\epsilon_1}} - (1-\epsilon_2) \underbrace{\sum_{\lambda \in (\Lambda_{22}^{13} \cup \Lambda_{23}^{13} \cup \Lambda_{32}^{13})} P(\lambda)}_{\stackrel{\text{(L8)}}{≤} 2 (\frac{2\delta +\epsilon_1}{\epsilon_2} + \frac{\delta (1-\epsilon_1)}{\epsilon_1^2 \epsilon_2})}  \notag \\
& ≥ \frac{P(\alpha_- \beta_+ | a_1 b_3)-\epsilon_1}{1-\epsilon_1} - 2 (1-\epsilon_2) (\frac{2\delta +\epsilon_1}{\epsilon_2}-\frac{\delta}{\epsilon_1\epsilon_2}+ \frac{\delta}{\epsilon_1^2 \epsilon_2})
\end{align}

A supremum for the right hand side of \eqref{WB-ineq-basic} can be caculated in two steps. Here is the estimation for the first two  terms: 
\begin{align}
P(\Lambda_{33}^{12}) + P(\Lambda_{33}^{23}) & = \sum_{\lambda \in \Lambda_{33}^{12}} P(\lambda) + \sum_{\lambda \in \Lambda_{33}^{23}} P(\lambda) \notag\\
& ≤ \sum_{\lambda \in \Lambda_{33}^{12}} P(\lambda) \frac{P(\alpha_- \beta_+ | a_1 b_2 \lambda)}{(1-\epsilon_2)^2} + \sum_{\lambda \in \Lambda_{33}^{23}} P(\lambda) \frac{P(\alpha_- \beta_+ | a_2 b_3 \lambda)}{(1-\epsilon_2)^2} \notag\\
& ≤ \frac{P(\alpha_- \beta_+ | a_1 b_2 \lambda) + P(\alpha_- \beta_+ | a_2 b_3 \lambda)}{(1-\epsilon_2)^2} 
\end{align}
And here is the estimation for the third term:
\begin{align}
P(\overline{\prescript{\alpha\!}{}{\Lambda}{_3^2} \cup \prescript{\beta\!}{}{\Lambda}{_3^2}}) &= 1-[P(\prescript{\alpha\!}{}{\Lambda}{_3^2}) + {P(\prescript{\beta\!}{}{\Lambda}{_3^2})} - P(\Lambda_{33}^{22})] \notag\\
& \stackrel{\text{(L5a),(L7)}}{≤} 2\frac{2\delta+\epsilon_1}{\epsilon_2}- \frac{2\delta +\epsilon_1}{1-\epsilon_1} +2 \delta \frac{1-\epsilon_1-\epsilon_2}{\epsilon_1^2 \epsilon_2} +\frac{\delta}{(1-\epsilon_2)^2}
\end{align}

The resulting  inequality, 
\begin{align} 
\frac{P(\alpha_- \beta_+ | a_1 b_3)}{1-\epsilon_1}  & ≤ \frac{P(\alpha_- \beta_+ | a_1 b_2 \lambda) + P(\alpha_- \beta_+ | a_2 b_3 \lambda)}{(1-\epsilon_2)^2} \notag\\
&\hphantom{= \;} +  \delta \left (\frac{2}{\epsilon_1}+\frac{8}{\epsilon_2} + 4 \frac{1-\epsilon_1-\epsilon_2}{\epsilon_1^2 \epsilon_2} + \frac{1}{(1-\epsilon_2)^2} -\frac{2}{1-\epsilon_1} -4 \right ) \notag \\
&\hphantom{= \;}  + 2\epsilon_1 \left (\frac{2}{\epsilon_2}-1 \right), 
\end{align}
is the Wigner-Bell inequality we have been looking for. 
It 
generalizes usual Wigner-Bell inequalities such as 
\begin{equation}
P(\alpha_-\beta_+|a_1 b_3)  \leq  P(\alpha_-\beta_+|a_1b_2) + P(\alpha_-\beta_+|a_2b_3)
\end{equation} 
in that it introduces correction terms with the deviation from perfect correlations $\delta$ and the parameters of the partitions $\epsilon_1$ and $\epsilon_2$. If the correlations are perfect ($\delta \rightarrow 0$) one can choose $\epsilon_1 \rightarrow 0$ and $\epsilon_2 \rightarrow 0$, which yields the original inequality.
One can check numerically that the new inequality is violated by 
the empirical measurement results 
\begin{align}
\label{eq:emp-data}
P(\alpha_-\beta_+|a_1 b_3) = 0.375, && P(\alpha_-\beta_+|a_1 b_2) = 0.125,  && P(\alpha_-\beta_+|a_2 b_3) = 0.125, 
\end{align} 
(which are a maximal violation of the usual Wigner-Bell inequality and occur e.g. for the measurement settings being chosen as $1=0^\circ$, $2=30^\circ$, $3=60^\circ$ given the quantum state $\psi_0$), 
if  
 $\delta < 7.42 \cdot 10^{-7}$ (and the  values for the parameters of the partition are chosen appropriately, viz. $\epsilon_1= 3.97\cdot 10^{-2}, \epsilon_2=5.55\cdot 10^{-2}$).
Hence, for the generalized Wigner-Bell inequality still to be violated 
at least 99.9999258\% of the photons must be perfectly correlated and anti-correlated.%
\footnote{{} Note that in this calculation we have assumed that the empirical measurement results \eqref{eq:emp-data} agree with the theoretical predictions, which is approximately the case 
when the measurement settings are not close to being aligned or perpendicular. Of course, if there are consistent deviations from the theoretical values (though small), the allowed maximal deviation from perfect correlations $\delta$ (for Bell inequalities still to be violated) might be slightly higher than indicated here.} 
\parbox{0cm}{} \hfill $\Box$

\subsubsection{Proof of \autoref{lemma-4.2}} 

The classes fulfilling condition $\neg$(i) and $\neg$(iii) are (H$_1^\alpha$), \dots, (H$_{14}^\alpha$). 
Here we have to show that given the background assumptions measurement independence, nearly perfect correlations and nearly perfect anti\-/correlations, these classes do \emph{not} imply the Bell inequalities, i.e. that 
there is at least one distribution for each class that fulfills 
the background assumptions and violates the Bell inequalities. 

We know already from \autoref{th:ind-pcorr} that the classes (H$_1^\alpha$), \dots, (H$_{14}^\alpha$)$\backslash\{(\text{H}_{4}^\alpha), (\text{H}_{5}^\alpha), (\text{H}_{10}^\alpha)\}$ can violate the inequalities given the assumptions of measurement independence and {strictly} perfect (anti-)correlations. Since the latter are a stronger condition than \emph{nearly} perfect correlations, it is clear that these classes can violate the Bell inequalities also in the present case.
It remains to show that the classes $(\text{H}_{4}^\alpha), (\text{H}_{5}^\alpha), (\text{H}_{10}^\alpha)$ can violate the inequalities under the given assumptions. 
Here we explicitly construct an example for class  $(\text{H}_{10}^\alpha)$. 

In the proof of \autoref{lemma-2.2} (\autoref{sec:proof-lemma2.2}) 
we constructed a toy example of a probability distribution for this class that is compatible with measurement independence and nearly perfect (anti\=/)correlations. When, for any setting $i$, we use the resulting probabilities \eqref{H14-mPcorr-pf1}--\eqref{H14-mPcorr-pf7} we can be sure that the distribution we are about to construct is consistent with the nearly perfect (anti\=/)correlations. 
What remains to be done is to reproduce the EPR/B correlations for non-parallel and non-per\-pen\-dicular settings. We again choose the settings $a_1=0^\circ$, $a_2=30^\circ$, $b_1 = 30^\circ$ and $b_2=60^\circ$ as well as the  quantum state $\psi = (|++\rangle + |--\rangle)/\sqrt{2}$. Then the observable probabilities read: 
\begin{align}
\label{empr-prob-H14-nPCorr-I}
P(\alpha_\pm \beta_\pm | a_1 b_1) &= \tfrac{3}{8} &  P(\alpha_\pm \beta_\mp | a_1 b_1) &= \tfrac{1}{8} 
\\ P(\alpha_\pm \beta_\pm | a_1 b_2) &= \tfrac{1}{8} &  P(\alpha_\pm \beta_\mp | a_1 b_2) & = \tfrac{3}{8} 
\\  P(\alpha_\pm \beta_\pm | a_2 b_1) &= \tfrac{1}{2}-\delta & P(\alpha_\pm \beta_\mp | a_2 b_1) &= \delta 
\\ \label{empr-prob-H14-nPCorr-II} P(\alpha_\pm \beta_\pm | a_2 b_2) &= \tfrac{3}{8} &  P(\alpha_\pm \beta_\mp | a_2 b_2) &= \tfrac{1}{8}
\end{align} 

\noindent (Note the difference to the probabilities with the same settings and quantum state in \eqref{emp-prob-H14-PCorr-1}--\eqref{emp-prob-H14-PCorr-2}, which involve \emph{strictly} perfect anti\-/correlations for parallel settings $a_2=b_1$, i.e. $\delta = 0$.

\noindent
These are sixteen equations, and any of the probabilities on their left hand sides can be expressed by the product form of the hidden joint probability, $P(\alpha \beta | a b ) = \sum_\lambda P(\lambda) \cdot P(\alpha|\beta\lambda) P(\beta|a b \lambda)$. 
$P(\lambda)$ and $P(\alpha|\beta \lambda)$ are completely determined by the requirements following from the perfect \mbox{(anti-)}cor\-re\-la\-tions \eqref{H14-mPcorr-pf1}--\eqref{H14-mPcorr-pf3},  $P(\beta | a b \lambda)$ partly so (namely only for parallel settings, \eqref{H14-mPcorr-pf4} or \eqref{H14-mPcorr-pf7}). 

Inserting these predetermined probabilities into equations \eqref{empr-prob-H14-nPCorr-I}--\eqref{empr-prob-H14-nPCorr-II} yields the following consistent values for the missing probabilities $P(\beta|ab\lambda)$:
\begin{align}
P(\beta_+ | a_1b_1\lambda_1) &= \tfrac{1-8\delta}{4(1-2\delta)} & P(\beta_- | a_1 b_1\lambda_1) &= \tfrac{3}{4(1-2\delta)}  
\\ P(\beta_+ | a_1 b_1\lambda_2) &= \tfrac{3}{4(1-2\delta)} & P(\beta_- | a_1 b_1\lambda_2) &= \tfrac{1-8\delta}{4(1-2\delta)}
\\ P(\beta_+ | a_1b_2\lambda_1) &= \tfrac{3-8\delta}{4(1-2\delta)} & P(\beta_- | a_1 b_2\lambda_1) &= \tfrac{1}{4(1-2\delta)} 
\\ P(\beta_+ | a_1 b_2\lambda_2) &= \tfrac{1}{4(1-2\delta)} & P(\beta_- | a_1 b_1\lambda_2) &= \tfrac{3-8\delta}{4(1-2\delta)}
\\ P(\beta_+ | a_2 b_2\lambda_1) &=\tfrac{1-8\delta}{4(1-2\delta)} & P(\beta_- | a_2 b_2\lambda_1) &= \tfrac{3}{4(1-2\delta)} 
\\ P(\beta_+ | a_2 b_2\lambda_2) &= \tfrac{3}{4(1-2\delta)} & P(\beta_- | a_2 b_2\lambda_2) &= \tfrac{1-8\delta}{4(1-2\delta)} 
\end{align} 

\noindent Finally, choosing, say, $P(a_i) = \frac{1}{2}$, $P(a_{i_\bot}) = \frac{1}{2}$, $P(b_i) = \frac{1}{2}$ and $ P(b_{i_\bot}) = \frac{1}{2}$, 
the formula $P(\alpha \beta a b \lambda) = P(\alpha | \lambda) P(\beta | a b \lambda) P(\lambda) P(a) P(b)$ 
entails the following total probabilities: 
\note{evtl. nicht abdrucken}{}
\begin{align}
P(\alpha_+ \beta_+ a_1 b_1 \lambda_1) &= 0 & P(\alpha_+ \beta_- a_1 b_1 \lambda_1) &= \tfrac{3\delta}{16(1-2\delta)} 
\\ P(\alpha_- \beta_+ a_1 b_1 \lambda_1) &= \tfrac{1-8\delta}{32(1-2\delta)} & P(\alpha_- \beta_- a_1 b_1 \lambda_1) &= \tfrac{3}{32} \notag
\\ P(\alpha_+ \beta_+ a_1 b_2 \lambda_1) &= 0  & P(\alpha_+ \beta_- a_1 b_2 \lambda_1) &= \tfrac{\delta}{16(1-2\delta)} 
\\ P(\alpha_- \beta_+ a_1 b_2 \lambda_1) &= \tfrac{3-8\delta}{32(1-2\delta)} & P(\alpha_- \beta_- a_1 b_2 \lambda_1) &= \tfrac{1}{32} \notag
\\ P(\alpha_+ \beta_+ a_2 b_1 \lambda_1) &= 0  & P(\alpha_+ \beta_- a_2 b_1 \lambda_1) &= \tfrac{\delta}{4} 
\\ P(\alpha_- \beta_+ a_2 b_1 \lambda_1) &= 0 & P(\alpha_- \beta_- a_2 b_1 \lambda_1) &= \tfrac{1-2\delta}{8} \notag
\\ P(\alpha_+ \beta_+ a_2 b_2 \lambda_1) &= 0  & P(\alpha_+ \beta_- a_2 b_2 \lambda_1) &= \tfrac{3\delta}{16(1-2\delta)} 
\\ P(\alpha_- \beta_+ a_2 b_2 \lambda_1) &= \tfrac{1-8\delta}{32(1-2\delta)} & P(\alpha_- \beta_- a_2 b_2 \lambda_1) &= \tfrac{3}{32} \notag 
\\[0.2cm] 
P(\alpha_+ \beta_+ a_1 b_1 \lambda_2) &= \tfrac{3}{32}  & P(\alpha_+ \beta_- a_1 b_1 \lambda_2) &= \tfrac{1-8\delta}{32(1-2\delta)}  
\\ P(\alpha_- \beta_+ a_1 b_1 \lambda_2) &= \tfrac{3\delta}{16(1-2\delta)} & P(\alpha_- \beta_- a_1 b_1 \lambda_2) &= 0
\\ P(\alpha_+ \beta_+ a_1 b_2 \lambda_2) &= \tfrac{1}{32}  & P(\alpha_+ \beta_- a_1 b_2 \lambda_2) &= \tfrac{3-8\delta}{32(1-2\delta)} 
\\ P(\alpha_- \beta_+ a_1 b_2 \lambda_2) &= \tfrac{\delta}{16(1-2\delta)} & P(\alpha_- \beta_- a_1 b_2 \lambda_2) &= 0
\\ P(\alpha_+ \beta_+ a_2 b_1 \lambda_2) &= \tfrac{1-2\delta}{8}  & P(\alpha_+ \beta_- a_2 b_1 \lambda_2) &= 0 
\\ P(\alpha_- \beta_+ a_2 b_1 \lambda_2) &= \tfrac{\delta}{4} & P(\alpha_- \beta_- a_2 b_1 \lambda_2) &= 0
\\ P(\alpha_+ \beta_+ a_2 b_2 \lambda_2) &= \tfrac{3}{32}  & P(\alpha_+ \beta_- a_2 b_2 \lambda_2) &= \tfrac{1-8\delta}{32(1-2\delta)}  
\\ P(\alpha_- \beta_+ a_2 b_2 \lambda_2) &= \tfrac{3\delta}{16(1-2\delta)} & P(\alpha_- \beta_- a_2 b_2 \lambda_2) &= 0
\end{align}

\noindent This completes our construction of a distribution from class (H$^\alpha_{10}$) which respects, auto\-no\-my, nearly perfect correlations, nearly perfect anti\-/correlations and violates the Bell inequalities. Similarly, one can construct examples of distributions for class (H$^\alpha_{4}$) and (H$^\alpha_{5}$). 
\parbox{0cm}{} \hfill $\Box$

\subsection{Proof of \autoref{th:combi}}%5
\label{sec:proof-theorem-combi} 
\stepcounter{theoremcounter}

\theoreme*

\noindent The theorem is equivalent to the conjunction of the following claims: 

\begin{lemma}
\label{lemma-5.1}
Each class (H$_i^\alpha$) is inconsistent with those classes (H$_j^\beta$) that are not indicated in column~X of \autoref{table-classes}. 
\end{lemma} 
\begin{lemma}
\label{lemma-5.2}
Each class (H$_i^\alpha$) is consistent with those classes (H$_j^\beta$) that are indicated in column~X of \autoref{table-classes}. 
\end{lemma}

\subsubsection{Proof of \autoref{lemma-5.1}}

One can find inconsistent classes $(H_j^\beta)$ for a class $(H_i^\alpha)$ by analyzing $(H_i^\alpha)$ in terms of pairwise (in)dependences (see \autoref{table-independences} for the definitions of the independences and \autoref{th:analysis} in \autoref{sect-analysis-classes} for the analysis)
and by checking which of the (in-)dependences that are relevant for defining classes $(H_j^\beta)$ are implied by the defining (in-)dependences of $(H_i^\alpha)$: All classes $(H_j^\beta)$ that contradict the inferred independences are inconsistent. 
The following logical relations are relevant for this purpose (for $x, y \in \{\alpha, \beta\}$ with $x≠y$): 

\noindent
\begin{minipage}{0.45\linewidth} 
\begin{equation} 
\label{LR1}
(\text{OI}_1) \wedge (\text{PI}_2^x) \Leftrightarrow (\text{OI}_1) \wedge (\text{PI}_1^x) 
 \end{equation} 
\end{minipage} 
\hspace{0.5cm} 
\begin{minipage}{0.45\linewidth} 
\begin{equation} 
\label{LR2}
 (\text{OI}_1) \wedge (\text{LPI}_2^x) \Leftrightarrow (\text{OI}_1) \wedge (\text{LPI}_1^x) \end{equation} 
\end{minipage}
\\[-0.35cm]
\parbox{0cm}{} \hspace{0.0cm}
\begin{minipage}{0.935\linewidth}
\begin{flalign}
\label{LR3} (\text{LPI}_1^x) \wedge (\text{PI}_2^y) \Leftrightarrow (\text{LPI}_1^x) \wedge (\text{PI}_1^y) \Leftrightarrow (\text{LPI}_2^x) \wedge (\text{PI}_1^y) &&
\end{flalign}
\end{minipage}
\\[-0.35cm]
\begin{minipage}{0.45\linewidth} 
\begin{equation} 
\label{LR4} 
(\text{PI}_1^x) \wedge (\text{PI}_2^x) \Rightarrow (\text{LPI}_2^y) \vee (\text{OI}_1) \end{equation} 
\end{minipage} 
\hspace{0.5cm} 
\begin{minipage}{0.45\linewidth} 
\begin{equation}
\label{LR5} 
(\text{LPI}_1^x) \wedge (\text{LPI}_2^x) \Rightarrow (\text{PI}_2^y) \vee (\text{OI}_1)
 \end{equation} 
\end{minipage} \\[0.25cm]

Relations \eqref{LR1}--\eqref{LR3} follow straightforwardly from the graphoid axioms  \citep[that provably hold for conditional independences, see][p. 11]{Pearl2000}, while \eqref{LR4} for $x=\alpha$ and $y = \beta$ can be proven as follows (and mutatis mutandis for $x=\beta$ and $y = \alpha$; also \eqref{LR5} can be proven similarly): 
\begin{align}
(\text{PI}_2^\alpha)  \Leftrightarrow  & \; I(\zv \alpha, \zv b| \zv a \zv \lambda) \\
\Leftrightarrow  & \;  P(\alpha|b a \lambda) = P(\alpha|b^\prime a \lambda) \\
\Leftrightarrow  & \; \sum_{\beta=\pm} P(\alpha|b a \lambda\beta) P(\beta|b a \lambda)  = \sum_{\beta=\pm} P(\alpha|b^\prime a \lambda \beta) P(\beta|b^\prime a \lambda) \\ 
 \overset{I(\zv \alpha, \zv b|\zv \beta \zv a \zv \lambda)}{\Longrightarrow} &
 \sum_{\beta=\pm} P(\alpha|b a \lambda \beta) P(\beta|b a \lambda)  = \sum_{\beta=\pm} P(\alpha|b a \lambda\beta) P(\beta|b^\prime a \lambda) \\
\begin{split}
\Leftrightarrow & \; P(\alpha|b a \lambda \beta_+) P(\beta_+|b a \lambda) + P(\alpha|b a \lambda \beta_-) \underbrace{P(\beta_-|b a \lambda)}_{1-P(\beta_+|b a \lambda)}   \\
& \qquad \qquad \qquad  = P(\alpha|b a \lambda \beta_+) P(\beta_+|b^\prime a \lambda) + P(\alpha|b a \lambda \beta_-) \underbrace{P(\beta_-|b^\prime a \lambda)}_{1-P(\beta_+|b^\prime a \lambda)} \end{split}\\
 \Leftrightarrow & \; \left(P(\alpha|b a \lambda\beta_+)-P(\alpha|b a \lambda \beta_-)\right) \left(P(\beta_+|b a \lambda)-P(\beta_+|b^\prime a \lambda)\right) =0\\
\Leftrightarrow & \; \underbrace{P(\alpha|b a \lambda \beta_+)=P(\alpha|b a \lambda \beta_-)}_{%
(\text{OI}_1)} \quad  \vee \quad \underbrace{P(\beta_+|b a \lambda)=P(\beta_+|b^\prime a \lambda)}_{%
(\text{LPI}^\beta_2)}
\end{align}

Here is an example how one can apply these relations between independences in order to rule out inconsistent classes. 
\autoref{table-classes} says that class $(\text{H}^\alpha_{9})$, defined by the product form $P(\alpha \beta | a b \lambda) = P(\alpha | a b \lambda) P(\beta | b \lambda)$, is inconsistent with all classes $(\text{H}^\beta_j)$ except $(\text{H}^\beta_{12})$, whose defining product form is identical to that of $(\text{H}^\alpha_{9})$. The inconsistencies can be seen as follows: The analysis of the former class is
$(\text{H}^\alpha_{9}) \Leftrightarrow (\text{OI}_1)  \wedge \neg(\text{PI}_1^\alpha) \wedge \neg(\text{LPI}_1^\alpha)  \wedge (\text{PI}_2^\beta) \wedge \neg(\text{LPI}_2^\beta) $
which implies $(\text{PI}_1^\beta)$ and $\neg(\text{PI}_2^\alpha)$ by \eqref{LR1} and  $\neg(\text{LPI}_1^\beta)$ and $\neg(\text{PI}_2^\alpha)$ by \eqref{LR2}. Jointly with $(\text{OI}_1)$ these are the defining (in-)dependences of $(\text{H}^\beta_{12})$, so all other classes are impossible. 

Or, as a slightly more complex example, consider 
$(\text{H}^\alpha_{7}) \Leftrightarrow \neg(\text{OI}_1)  \wedge \neg(\text{PI}_1^\alpha) \wedge \neg(\text{LPI}_1^\alpha)  \wedge (\text{PI}_2^\beta) \wedge (\text{LPI}_2^\beta)$,
which entails that the following expressions cannot be true: $\neg(\text{LPI}_2^\beta)  \wedge (\text{PI}_1^\alpha)$ (by \eqref{LR3} with $x=\alpha$, $y=\beta$), $\neg(\text{LPI}_1^\beta)  \wedge (\text{PI}_2^\alpha)$ (by \eqref{LR3} with $x=\beta$, $y=\alpha$), $(\text{PI}_1^\beta)  \wedge (\text{LPI}_2^\alpha)$ (by \eqref{LR4} with $x=\beta$, $y=\alpha$) and $\neg(\text{LPI}_1^\beta)  \wedge \neg(\text{PI}_2^\alpha)$ (by \eqref{LR5} with $x=\beta$, $y=\alpha$). 
By some basic logical reasoning one can show that these requirements exclude all classes $(\text{H}^\beta_{j})$ except
 $(\text{H}^\beta_{1})$, $(\text{H}^\beta_{2})$, $(\text{H}^\beta_{3})$ and $(\text{H}^\beta_{7})$. 

In a similar way, one can prove inconsistency between each class $(\text{H}^\alpha_{i})$ and every class $(\text{H}^\beta_{j})$ that is not mentioned in the respective line of column~X. 

\subsubsection{Proof of \autoref{lemma-5.2}}

A class $(\text{H}^\alpha_i)$ can be shown to be consistent with a class $(\text{H}^\beta_j)$ by providing an example of a probability distribution that falls under both classes. Since any example suffices, one can assume a toy model according to which all variables are two-valued; the total probability distribution $P(\alpha \beta a b \lambda)$ then has 64 values that need to be determined such that the characteristic pattern of \mbox{(in-)}dependences for both classes holds. Finding appropriate values for the total probability distribution is straightforward by construction: Factorize the total probability distribution according to class $(\text{H}^\alpha_i)$, which for, say, $(\text{H}^\alpha_7)$ reads
$P(\alpha \beta a b \lambda) = P(\alpha| \beta b a \lambda) P(\beta|b\lambda) P(a) P(b) P(\lambda)$. 
Then, for each combination of variable values choose a value for each probability on the right hand side (these probability values need to be consistent with the axioms of probability theory, i.e. must be between 0 and 1 and must sum to 1, but can be arbitrary otherwise); from these probability values calculate the values of the total probability distribution, and by construction the total probability distribution will have exactly the  characteristic \mbox{(in-)}dependences of class $(\text{H}^\alpha_i)$. 

In general (i.e. for almost any choice of values), the probability distribution will have the \mbox{(in-)}dependences that are characteristic of the weakest class $(\text{H}^\beta_j)$ (i.e. the class involving most variables) that is consistent with $(\text{H}^\alpha_i)$, e.g. $(\text{H}^\beta_1)$ for a construction of $(\text{H}^\alpha_7)$. Combinations with weaker classes (if there are any consistent ones) can be constructed by imposing additional constraints on the probabilities whose values have been chosen. For instance, in order to construct a probability distribution from classes $(\text{H}^\alpha_7)$ and $(\text{H}^\beta_3)$ (instead of $(\text{H}^\beta_1)$), one needs to express the additionally required independence, $\alpha$-parameter independence$_2$, in terms of the probabilities of the factorization form of  $(\text{H}^\alpha_7)$: 
\begin{align}
(\text{PI}^\alpha_2) \Leftrightarrow && P(\alpha | b_1 a \lambda) &= P(\alpha | b_2 a \lambda) \\
\Leftrightarrow && \frac{\sum_{\beta} P(\alpha \beta b_1 a \lambda)}{\sum_{\alpha,\beta} P(\alpha \beta b_1 a \lambda)} &= \frac{\sum_{\beta} P(\alpha \beta b_2 a \lambda)}{\sum_{\alpha,\beta} P(\alpha \beta b_2 a \lambda)} \\
\Leftrightarrow && \frac{\sum_{\beta} P(\alpha | \beta b_1 a \lambda) P(\beta | b_1 \lambda)}{\sum_{\alpha,\beta} P(\alpha |\beta b_1 a \lambda) P(\beta | b_1 \lambda)} &= \frac{\sum_{\beta} P(\alpha | \beta b_2 a \lambda) P(\beta | b_2 \lambda)}{\sum_{\alpha,\beta} P(\alpha | \beta b_2 a \lambda) P(\beta | b_2 \lambda)}.
\end{align}
The latter expression involves only those probabilities whose values are chosen arbitrarily at construction; it yields one equation for each set of variables $\alpha, a, \lambda$ and thus reduces the number 
of probabilities that need to be chosen. 
If these additional conditions are respected the resulting total probability distribution will have the required \mbox{(in-)}dependences. 

By constructing probability distributions in this way one can find examples of probability distributions from each class $(\text{H}^\alpha_i)$ and jointly from any class $(\text{H}^\beta_j)$ that is listed in column~X of \autoref{table-classes}.

\subsection{Proof of \autoref{th:analysis}%6
}
\label{sec:proof-theorem3}
\stepcounter{theoremcounter}

\theoremf*

Formally, \autoref{th:analysis} can be written as $(\text{H}^\alpha_i) \Leftrightarrow \text{Ind}(\text{H}^\alpha_i) \wedge \text{Dep}(\text{H}^\alpha_i)$. 
Since a class $(\text{H}^\alpha_i)$ is defined as the conjunction of a factorization condition $(\text{F}^\alpha_i)$ and the claim that this factorization is minimal 
($\text{Min}(\text{F}^\alpha_i)$),%
\footnote{{}
We have said in \autoref{sec:classes} that a factorization condition that holds for a (set of) probability distribution(s)  is minimal if and only if it is the factorization condition involving fewest possible variables according to this (set of) disribution(s). 
Precisely, minimality of, say, $(\text{F}^\alpha_{24})$ is defined as the conjunction of the following expressions: 
\begin{align} 
\label{eq:H24-strich2}
\tag{Min($F^\alpha_{24})_1$} 
& \exists \; \alpha, \beta, a, b, \lambda:  &P(\alpha \beta | a b \lambda) &\neq P(\alpha | \lambda)P(\beta | b \lambda) 
\\
\label{eq:H24-strich3}
\tag{Min($F^\alpha_{24})_2$} 
& \exists \; \alpha, \beta, a, b, \lambda:  &P(\alpha \beta | a b \lambda) &\neq P(\alpha | \beta \lambda)P(\beta | \lambda) 
\end{align}
}
more explicitly \autoref{th:analysis} reads: $(\text{F}^\alpha_i) \wedge \text{Min}(\text{F}^\alpha_i) \Leftrightarrow \text{Ind}(\text{H}^\alpha_i) \wedge \text{Dep}(\text{H}^\alpha_i)$. 
This theorem holds if the conjunction of the following claims is true:%
\footnote{{} 
The conjunction of the lemmas implies the theorem; they are \emph{not} equivalent.  
}

\begin{lemma}
\label{lemma-6.1}
The factorization condition of a class  (H$_i^\alpha$) is equivalent to the conjunction of $(\text{H}_i^\alpha)$'s specific pattern of independences: 
$(\text{F}^\alpha_i) \Leftrightarrow \text{Ind}(\text{H}^\alpha_i)$. 
\end{lemma}

\begin{lemma}
\label{lemma-6.2}
The factorization condition of a class (H$_i^\alpha$) and the corresponding minimality condition imply the conjunction of $(\text{H}_i^\alpha)$'s specific pattern of dependences: 
$(\text{F}^\alpha_i) \wedge \text{Min}(\text{F}^\alpha_i) \Rightarrow \text{Dep}(\text{H}^\alpha_i)$.
\end{lemma}

\begin{lemma}
\label{lemma-6.3}
The specific pattern of dependences of a class (H$_i^\alpha$) implies the minimality condition for its factorization condition: 
$\text{Dep}(\text{H}^\alpha_i) \Rightarrow  \text{Min}(\text{F}^\alpha_i)$. 
\end{lemma}

\subsubsection{Proof of \autoref{lemma-6.1}}
\label{sec:lemma-5.1}

I demonstrate the lemma for
$(\text{F}_{24}^\alpha) \;\Leftrightarrow (\text{PI}^\alpha_1) \wedge (\text{LPI}^\alpha_1) \wedge (\text{PI}^\beta_2)$ (in case that $(\text{F}_{24}^\alpha)$ is well-defined for all values),  
and my proof can be easily adapted to cases of partial definition as well as to other classes. 
We first of all note suitable formal definitions for the involved expressions: 
\begin{align}
\label{eq:H24-strich1}
\tag{$\text{F}^\alpha_{24}$} 
& 
\forall \, \alpha, \beta, a, b, \lambda:  &P(\alpha \beta | a b \lambda) &= P(\alpha | \beta \lambda)P(\beta | b \lambda)\\
 \label{eq:OD_1-strich}
\tag*{$\neg(\text{OI}_1)$} 
&  \exists \; \alpha, \beta, a, b, \lambda:  &P(\alpha | \beta  b a \lambda) &\neq P(\alpha | b a \lambda)\\
\label{eq:PI-alpha_1-strich}
\tag{$\text{PI}^\alpha_1$}
 &  \forall \, \alpha, \beta, a, b, \lambda:  &P(\alpha | \beta  b a \lambda) &= P(\alpha | \beta a \lambda)\\
\label{eq:lPI-alpha_1-strich}
\tag{$\text{LPI}^\alpha_1$} 
& \forall \, \alpha, \beta, a, b, \lambda:  &P(\alpha | \beta  b a \lambda) &= P(\alpha | \beta b \lambda)\\
\label{eq:PI-beta_2-strich}  
\tag{$\text{PI}^\beta_2$} 
 & \forall \, \beta, a, b, \lambda:  &P(\beta | a b  \lambda) &= P(\beta | b \lambda) \\
\label{eq:lPI-beta_2-strich}
\tag*{$\neg(\text{LPI}^\beta_2)$} 
 & \exists \, \beta, a, b, \lambda:  &P(\beta | a b  \lambda) &\neq P(\beta | a \lambda) 
\end{align}

\noindent \fbox{$\leftarrow$} \hspace{0.25cm} 
Replacing the expressions on the right hand side of the generally valid equation 
\begin{equation}
\label{eq:H24-gen-fact}
\forall \alpha, \beta, a, b, \lambda: \quad 
P(\alpha \beta | a b \lambda) = P(\alpha | \beta b a \lambda)P(\beta | a b \lambda)
\end{equation}
by the expressions for the pairwise independences \eqref{eq:PI-alpha_1-strich}, \eqref{eq:lPI-alpha_1-strich} and \eqref{eq:PI-beta_2-strich} straightforwardly yields the product form \eqref{eq:H24-strich1}. \\

\noindent \fbox{$\rightarrow$} \hspace{0.25cm} 
Summing over $\alpha$ on both sides of \eqref{eq:H24-strich1} and omitting quantification over non-appearing variables we get
\begin{align}
\label{eq:lemma5.1-1}
\forall \beta, a, b, \lambda: \quad  
P(\beta | a b \lambda) &= P(\beta | b \lambda), 
\end{align}
which is $(\text{PI}^\beta_2)$. 
 
Rewriting the left hand side of \eqref{eq:H24-strich1} by \eqref{eq:H24-gen-fact} and its right hand side by 
\eqref{eq:lemma5.1-1} 
one can cancel the second factors on each side, which yields: 
\begin{align}
\label{eq:lemma5.1-2}
\forall \alpha, \beta, a, b, \lambda:  \quad P(\alpha | \beta b a \lambda) = P(\alpha | \beta \lambda). 
\end{align}

This result can be used to derive $(\text{LPI}^\alpha_1)$ (and mutatis mutandis for $(\text{PI}^\alpha_1)$):
\begin{multline}
\forall \alpha, \beta, a, b, \lambda:  \quad  P(\alpha | \beta b \lambda) = \sum_{a^\prime} P(\alpha a^\prime | \beta b \lambda) = \sum_{a^\prime} P(\alpha | a^\prime \beta b \lambda) P(a^\prime | \beta b \lambda)  = \\
\stackrel{\eqref{eq:lemma5.1-2}}{=} P(\alpha | \beta  \lambda) \underbrace{\sum_{a^\prime} P(a^\prime | \beta b \lambda)}_{=1} \stackrel{\eqref{eq:lemma5.1-2}}{=} P(\alpha | \beta a b \lambda). 
\end{multline}
\parbox{0cm}{} \hfill $\Box$

\subsubsection{Proof of \autoref{lemma-6.2}}
\label{sec:lemma-5.2}

By Lemma~5.1 we know that Ind($\text{H}^\alpha_i$) hold. 
We then proceed by reductio: Suppose one of the dependences in $\text{Dep}(\text{H}^\alpha_i)$ would \emph{not} hold. Then there would be an additional independence, which by Lemma~5.1 
would yield a factorization condition involving fewer variables, contradicting the 
assumption that $(\text{F}^\alpha_i)$ is minimal. 
Hence all dependences $\text{Dep}(\text{H}^\alpha_i)$ hold. 
\parbox{0cm}{} \hfill $\Box$

\subsubsection{Proof of \autoref{lemma-6.3}}
\label{sec:lemma-5.3}
Again we proceed by reductio. Suppose $(\text{F}^\alpha_i)$ were not minimal, i.e. another factorization condition $(\text{F}^\alpha_j)$ involving fewer variables holds. By Lemma5.1 this would imply $\text{Ind}(\text{H}^\alpha_i)$ plus at least one further independence, 
contradicting the assumption of $\text{Dep}(\text{H}^\alpha_i)$. 
Hence $(\text{F}^\alpha_i)$ must be minimal. 
\parbox{0cm}{} \hfill $\Box$

\end{document}